\documentclass{jaa}
\usepackage{natbib}
\usepackage{graphicx}
\usepackage[english]{babel}

\usepackage{amsmath}

\usepackage[colorlinks=true, allcolors=blue]{hyperref}
\usepackage{subcaption}

\usepackage{soul}

\begin{document}\sloppy

%%paper title
%%For line breaks \\ can be used within title
\title{On-axis afocal telescopes as framework for Cubesat based \\astronomical imagers and slit-less spectrographs }

%%author names are separated by comma (,)
%%use \and before the last author name
%%use a * along with the number separated by comma
%% for the  author for correspondence
%%\textsuperscript{number} is used for affiliation
%%\affilOne, \affilTwo etc., upto \affilTwentyfive is possible
%%Please note the first letter after \affil is capitalised in the command
%%

\author{Anwesh Kumar Mishra\textsuperscript{1,*}, Gourav Banerjee\textsuperscript{1,2}, Rekhesh Mohan\textsuperscript{1} and Maheswar Gopinathan\textsuperscript{1} }
\affilOne{\textsuperscript{1}Indian Institute of Astrophysics, Koramangala, Bangalore,560034, India \\}
 \affilTwo{\textsuperscript{2} Vainu Bappu Observatory, Kavalur, 635701, Tamilnadu,India.}

%%escape two column mode for title, affiliation and abstract
%%by giving \twocolumn command as shown

\twocolumn[{

\maketitle

%%include \corres to print the corresponding author Email id
\corres{mishraanwesh@gmail.com}

%%include \msinfo for
%%manuscript information such as
%%received, revised and accepted dates
%%
\msinfo{1 January 2015}{1 January 2015}

%%abstract
\begin{abstract}

Cubesats present unique opportunities for observational astronomy in the modern era. They are useful in observing difficult-to-access wavelength regions and  long-term monitoring of interesting astronomical sources. However, conventional telescope designs are not necessarily the best fit for restricted envelope of a Cubesat. Additionally, fine-pointing stability on these platforms is difficult due to the low mass of the spacecraft and special allocations within the optical design are needed to achieve stable pointing. We propose afocal telescope designs as the framework to realise imagers and low-resolution spectrographs on Cubesat platforms. These designs help reduce the number of components in the optical chain and aim to improve throughput and sensitivity compared to conventional designs. Additionally, they also provide a fine steering mechanism within a collimated beam section. Fine beam steering within the collimated beam section avoids issues of image degradation due to out-of-plane rotation of the image plane or offset in the rotation axis of the mirror. This permits the use of simple and  mostly off-the-shelf tip-tilt mirrors for beam steering. The designs discussed here also allow for a standard telescope design to be used in many instrument types; thus reducing the complexity as well as the development time and cost. The optical design, performance and SNR estimations of these designs along with some interesting science cases are discussed. A number of practical aspects in implementation such as guiding, tolerancing, choice of detectors, vibration analysis and laboratory test setups are also presented. 
\end{abstract}

%%insert keywords separated by 3 hyphens using \keywords{words}

}]
%%close the twocolumn escape here

%%include \doinum{number}for the DOI number in the header
%%include \volnum{number} for the volume number in the header
%%include \year{yyyy} for  year of publication in the header
%%include \pgrange{num--num} page range of article in the header
%%include \artcitid{num} for the article citation id
%%include \lp to print last page of the article
%%include \setcounter{page}{pagenum} for the exact starting page of the article

\doinum{12.3456/s78910-011-012-3}
\artcitid{\#\#\#\#}
\volnum{000}
\year{0000}
\pgrange{1--}
\setcounter{page}{1}
\lp{1}

\section{Introduction: astronomy with Cubesats }

In recent years, there has been remarkable progress in the technology and application of small satellites for both commercial and scientific use (\cite{poghosyan2017cubesat}). Particularly, the Cubesat format and standardisation have been quite popular in terms of the number of satellites launched and planned (\cite{liddle2020space}). Cubesat based observations from a Low Earth Orbit can complement ground based observations (\cite{shkolnik2018verge}) in terms of uninterrupted wavelength coverage (e.g. 150 nm-2500 nm) and good temporal coverage. This would be particularly interesting for monitoring of variable stars, such as pulsating or flaring types. Similarly, science cases such as asteroseismology (\cite{nowak2016reaching}) and exoplanet studies (\cite{pong2018orbit})  which require very long and uninterrupted observing runs can greatly benefit from Cubesat missions operating in sun-synchronous orbits.\\

At the same time, Cubesats can achieve these goals with significantly smaller mission costs and shorter development time compared to larger space-based missions (\cite{serjeant2020future}). This is promising for making space-based astronomical observations accessible to universities and research institutions alike. Additionally, Cubesats may also be used as a platform for technology demonstration and space qualification of critical components and systems and hence serve as a testing platform for bigger astronomical missions (\cite{morgan2019mems}). However, implementations based on Cubesats have to navigate around the following practical issues:\\

\begin{itemize}
    \item \textbf{Pointing stability limitations:}\\    
       Astronomical telescopes typically require arcsecond level pointing accuracy in the line-of-sight direction and must maintain this pointing with minimal deviation for the complete duration of observation. Space borne astronomical telescopes must achieve this while in presence of various sources of micro-vibrations such as disturbances resulting from  magnetic torques, solar radiation pressure, gravity gradient torques and several other factors (\cite{paluszek2023adcs}). Conventional Attitude Determination and Control Systems (ADCS) are typically designed for larger platforms with higher mass (and hence inertia) ratios between the reaction wheel and the platform and  are not sufficient for arcsecond-level attitude stabilisation on Cubesat type platforms (\cite{douglas2021practical}). Typically, a separate finer control loop is required to correct for the disturbance from within the the optical chain of the telescope itself (\cite{peri2023high}).
       
    \item \textbf{Limited aperture and focal lengths:}\\    
        The Cubesat specification defines the exact envelopes within which the complete spacecraft (including the telescope, instruments and avionics) must fit into\footnote{https://www.cubesat.org/}. Typically, the largest of Cubesats allow for 12U volume in 2UX2UX3U format where 1U is typically considered a cube with sides of 10 cm. The largest symmetrical telescope aperture that can fit within this volume  is  about 200 mm in diameter (practically about 185-190 mm might be more realistic). On smaller Cubesats, e.g. 6U (1UX2UX3U), 100 mm aperture is more typical.  There have been some attempts to maximize the available collecting area: such as by using a rectangular aperture (e.g. CUTE,\cite{fleming2017colorado} or an off-axis primary mirror (e.g. CUBESPEC, \cite{raskin2022cubespec}). This restriction in aperture highlights the necessity of being efficient with the gathered photons and to be creative with the science programs.

        Cubesat based platforms have the potential to be diffraction limited. However, the smaller aperture and limited space envelope also limit the focal length of Cubesat based telescopes. 
        Invariably, this leads to a fairly high plate-scale(of the order of hundreds of arcseconds/per mm) at the image plane; and fairly small pixel size (about 2-4 microns) is required to adequately sample this PSF. Detectors arrays with such small pixels are typically not aimed for scientific observation and therefore (depending on the nature of the science being attempted) might need dedicated facilities for their characterisation before launch. At the same time, use of larger pixel size will cause the PSF to be undersampled.
        
    \item \textbf{Size, complexity and reliability:}\\
         Although smaller in size, Cubesats must operate in a self-sufficient manner in terms of power, communication, onboard computing, battery backup, coarse pointing and other miscellaneous subsystems. Fortunately, there has been significant efforts from commercial companies\footnote{https://www.bluecanyontech.com/}\footnote{https://hex20.in/} towards availability of standardised Cubesat buses. Similarly, progress is being made towards miniaturizing mission critical components such as reaction wheels, magneto-torquers and star sensors (\cite{johnston2011arc},\cite{candini2012miniaturized}). Still, the complexity of Cubesats (that are capable of useful astronomy) remains a challenge and will probably benefit from further standardisation of components and systems as well as collaborative work between the astronomical community and the industry.
    
\end{itemize}

Keeping these aspects in mind, we introduce a framework of astronomical telescopes based on on-axis afocal optics. The general design considerations are aimed towards suitability of use in a compact Cubesat platform. We discuss advantages and flexibility of such a framework along with a number of example designs and some illustrative science cases.
\section{General Principle of Afocal telescopes:}
\begin{figure*}[!ht]
		\centering

        \begin{subfigure}{0.55\textwidth}
			\includegraphics[width=1.05\linewidth]{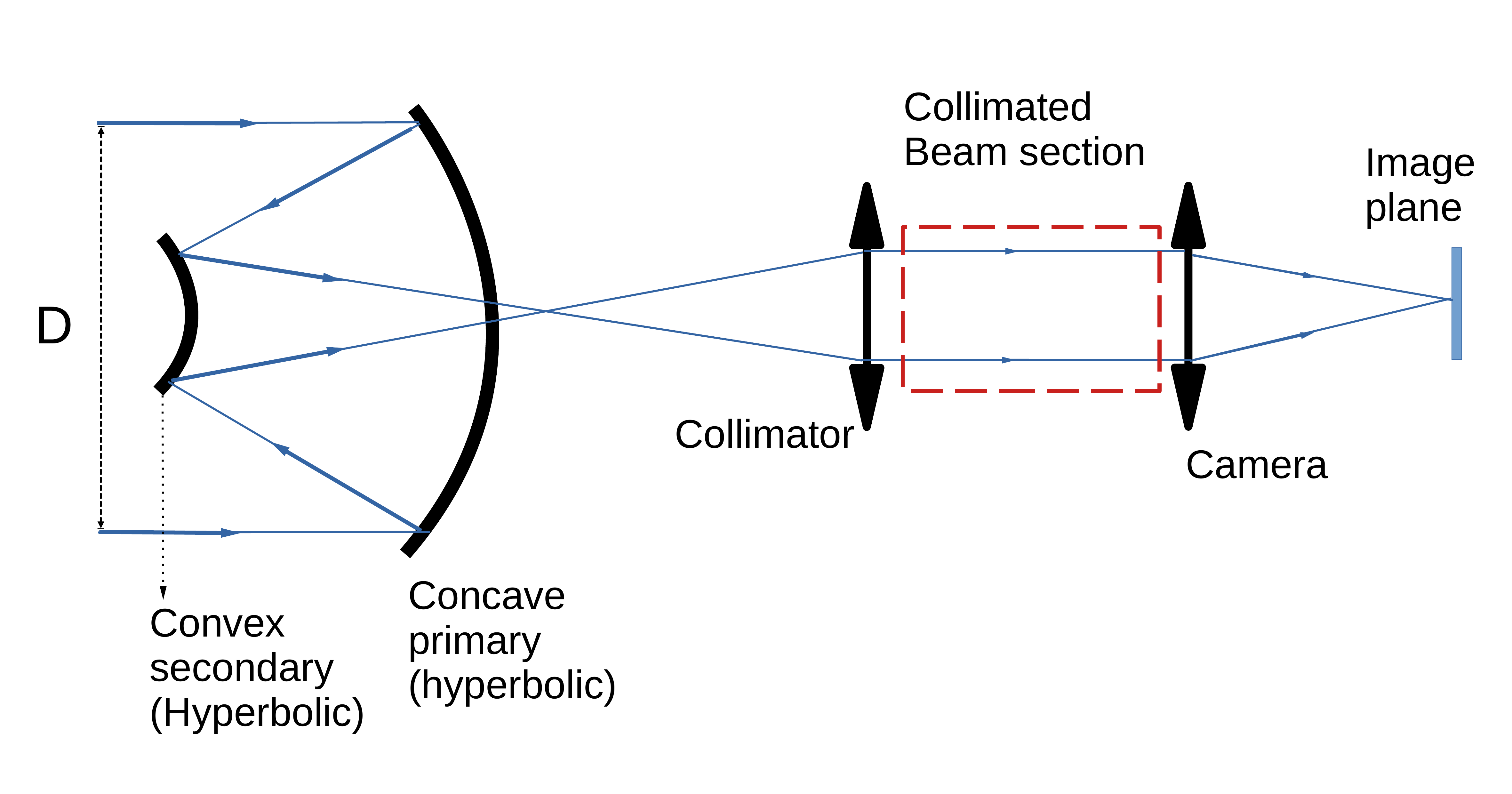}
			\caption{Conventional telescope with collimator camera type optics}\label{fig:conv_lay}
		\end{subfigure}
        \bigskip

        \begin{subfigure}{0.45\textwidth}
			\includegraphics[width=1.05\linewidth]{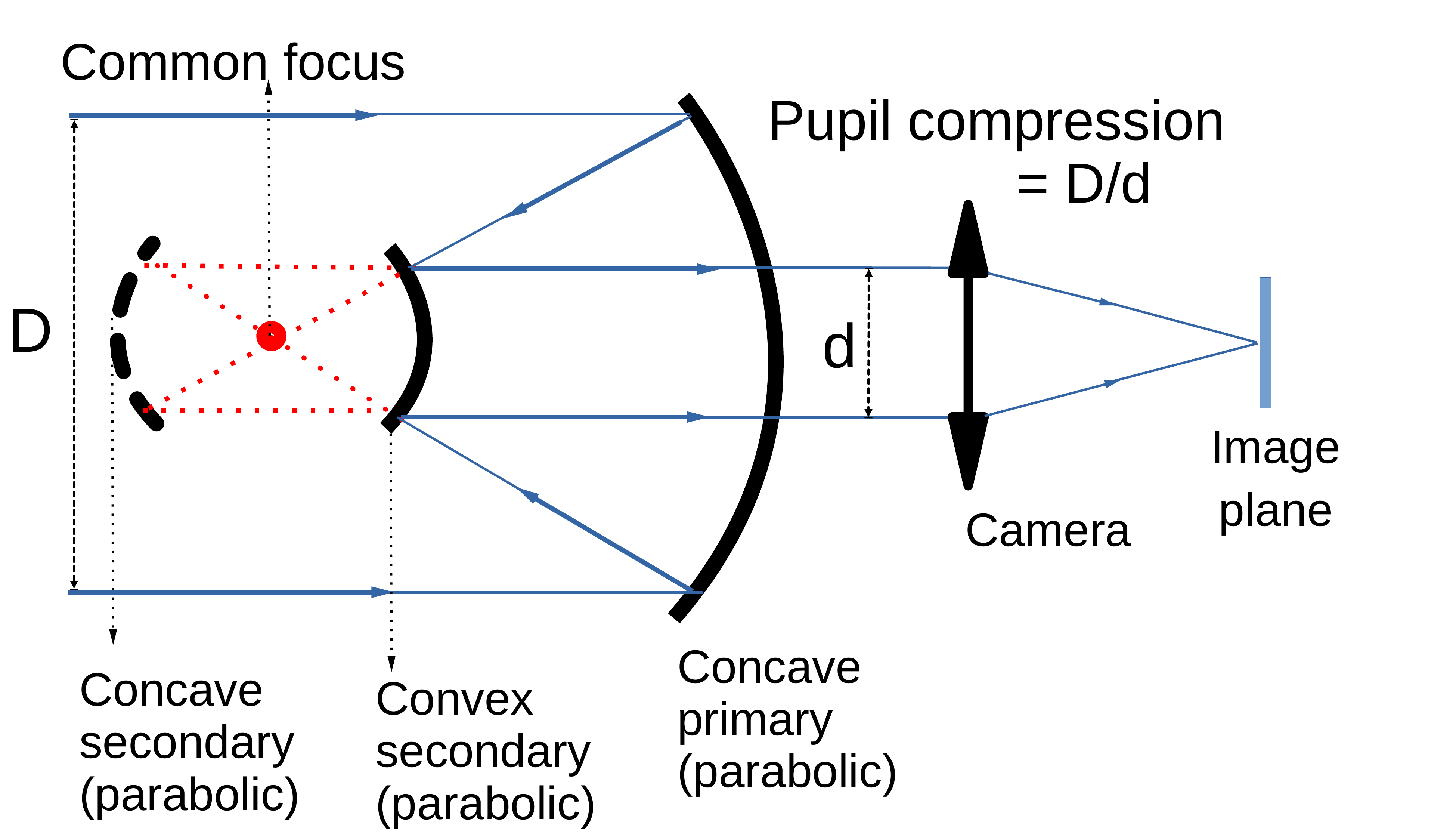}
			\caption{Schematic of an afocal telescope}\label{fig:afocal_layout}
		\end{subfigure}
		\begin{subfigure}{0.45\textwidth}
			\includegraphics[width=1.05\linewidth]{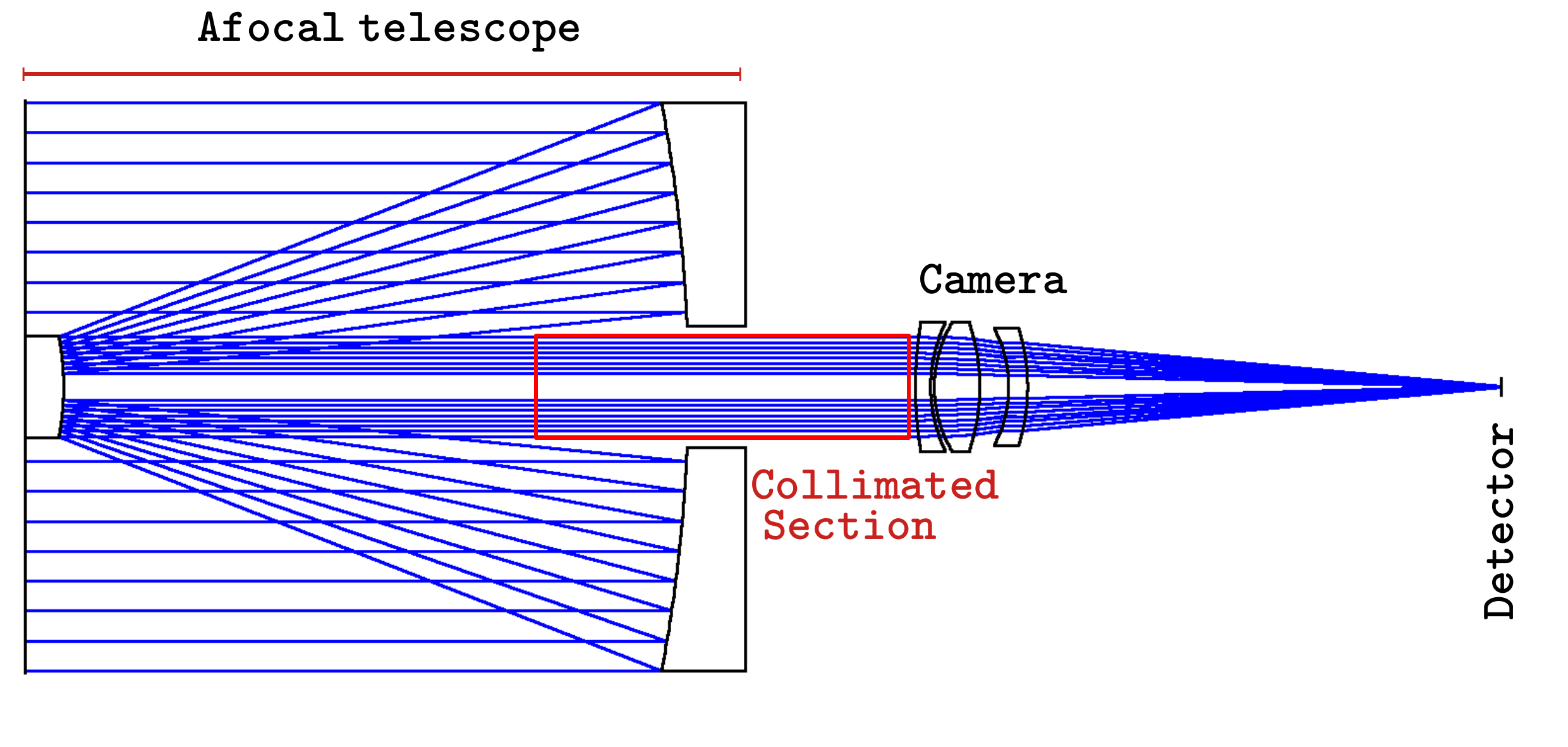}
			\caption{Ray trace of an afocal telescope}\label{fig:afocal_trace}
		\end{subfigure}
		\bigskip

     \caption{Layout of an afocal telescope with a camera: an afocal system in (b) is compared to a conventional designs in (a). In case of the conventional design the telescope produces a real image at the Cassegrain focus. This image may be re-collimated by a collimator optics and then re-imaged by a camera optics. The collimated beam section is used for placing filters, GRISMS etc. However, for the afocal telescope shown in (b), the combination of primary and secondary produces a collimated beam output that is only pupil compressed. This collimated beam is immediately available for placing components such as filters, GRISMS or for the purpose of beam steering. The final telescope F-number and platescale is determined by the camera optics. In (c) a real raytrace of an afocal system is shown to highlight the compactness of the system. }
		\label{fig:afocal_lay}  
	\end{figure*}

Afocal optical systems are defined as those systems which do not have a finite focal length associated with them. These typically have zero power and need a minimum of two optical components to be implemented (\cite{hazra2022foundations}). Practical examples of such systems include laser beam expanders (\cite{mahajan1998optical}) and guide star launchers for adaptive optics (\cite{clermont2020afocal}) as well as the classical Mersenne telescope. A detailed analysis of aberrations cancelling properties of the afocal telescope is presented in \cite{baker1969improving} and few more example can be found in \cite{terebizh2019survey}. In recent times, the COROT mission used an off-axis afocal telescope (\cite{auvergne2009corot}).

  In modern astronomy, there is often need for a collimated beam section to accommodate gratings, GRISMs, filters or polarization components. Typically, a collimated beam section is made available by means of a collimator camera type optics such as in figure \ref{fig:conv_lay}. However, an alternate method to produce a collimated beam section is shown in figure \ref{fig:afocal_layout} by means of an afocal system realized by means of two confocal paraboloids. A paraboloid has two conjugate points between which it achieves aberration free imaging. One of these points is at the focus of the parabola while the other point is at infinity. The primary mirror images the star(located at infinite object distance) onto its focus. However, this point is also the common focus for the secondary mirror.  Therefore the secondary mirror re-images the focus to its other conjugate point -- which is at infinity. The output is a collimated beam that has been pupil compressed. The ratio of compression can be expressed as:\\
\begin{equation}
 C= \frac{D}{d} = \frac{F_p}{F_s}
\end{equation}
Where D and d are the diameters of the primary and secondary mirror respectively and $F_p$ and $F_s$ are the focal lengths of the primary and secondary respectively. Since confocal parabolas satisfy the on-axis conjugate imaging condition, the beam compression is aberration free for the central field. In later sections, we demonstrate that it is possible to achieve good imaging performance to about 20-30 arcminutes in these type of systems.\\

 A camera is needed after the afocal telescope as shown in figure \ref{fig:afocal_trace}. The effective focal length of the complete system as well as the field of view (FOV) is related to that of the camera by:
\begin{equation}
    F_{telescope} = C \times F_{camera}
\end{equation}
and:
\begin{equation}
    FOV_{telescope} = \frac{FOV_{camera}}{C}
\end{equation}
 Where, $FOV_{camera}$ is the "apparent" FOV of the camera optics. The camera optics works at a smaller aperture where more flexibility is available. If required, strongly aspheric and free-form optics may be utilised. The afocal telescope is primarily responsible for providing a beam compression; the final image quality and plate-scale is largely dependent on this camera optics. It creates a unique opportunity of standardising the primary and secondary mirrors across a range of output F-numbers and wavelength ranges. A number of illustrative examples are presented at later sections(4.1 and 4.2).

\subsection*{Limitations of afocal designs}

The afocal design is not necessarily the most compact implementation possible.  Depending on specific output F-numbers required (typically for fairly short F-numbers - $<$ F6) it is possible to realize catadioptric systems that are shorter than afocal systems. One such example is shown later (section 4.1). For longer focal lengths, the afocal design is usually a fairly compact implementation.\\

Compared to the collimator-camera type implementation, the afocal design does not have an exit pupil in an accessible place. For a convex parabolic secondary the exit pupil is actually formed behind the secondary mirror and is not accessible. Off-axis designs using a concave secondary can bring this exit pupil to a more accessible spot (e.g COROT, \cite{auvergne2009corot}). However, this implementation is difficult to scale down to the permissible volumes of a Cubesat. Similarly, since there is no intermediate focus, there is also no place to include a slit and therefore the present work focuses only on imagers and slit-less spectrographs.

\subsection*{Advantages of Afocal designs}

\begin{itemize}
    \item Standardised testing and characterisation

       Same primary and secondary can be used across a range of different type of instruments; thus simplifying the process of testing and characterisation.            
   
   \item Flexibility in the choice of detector 
   
   The final platescale is only determined by the camera which can be optimised for a specific detector in mind; while keeping the primary and secondary mirror configuration fixed. This allows for use of off-the-shelf or commercial detector modules whenever possible.

   \item Lower part count and SNR advantage

     By eliminating the collimator group from a standard layout, the part count is reduced and thereby a simpler and higher throughput system maybe realized.

    \item Ease of fine steering and pointing stability :
    
      An unique advantage of the afocal design is the immediate availability of the collimated beam section. Beam steering from within a collimated beam section has significant advantages (discussed in the next section); and allows for simpler implementations of fine steering control loops.

\end{itemize}

\section{Beam-steering for fine pointing stability}

Astronomical telescopes require line-of-sight pointing accuracy at arcsecond levels or better.  Although it is possible for some larger observatories to make use of reaction wheels and magneto-torquers to achieve high line of sight stability (e.g. Hubble (\cite{beals1988hubble}) and COROT (\cite{auvergne2009corot})  ); this is much more difficult for smaller spacecrafts such as those based on Cubesat formats. This is partly due to the presence of more micro-disturbances in the low Earth orbits typical of Cubesats as well as an unfavourable ratio of typical reaction wheel mass to the spacecraft inertia (\cite{douglas2021practical}). There have been efforts to miniaturize reaction wheels for Cubesat applications (e.g. \cite{johnston2011arc}) ; and commercial Cubesat buses nowadays advertise stability of about an arcminute or even less \footnote{https://www.bluecanyontech.com/} , still this by itself is not sufficient for astronomical applications.\\

For astronomical usage of Cubesats, it is necessary to implement a separate fine control loop(e.g. in CUBESPEC mission \cite{peri2023high}) in addition to the spacecraft's pointing control. This loop is implemented by fine guidance sensors that can sense the deviation in pointing direction at a high enough resolution and some beam steering mechanism to correct for the deviation. The fine guidance sensor can be realized either using a fibre optic gyroscope ( \cite{sanders2012fiber}, \cite{jin2018orbit}) or a fine guiding camera (\cite{rowlands2004jwst}). Due to space constraints of the Cubesat type platforms we will focus mostly on the fine guiding camera. The guiding information is utilised by a fine steering mechanism to correct for pointing error. The different mechanisms for the same can be classified into three broad categories: (for a more comprehensive review of different actuators and mechanisms, \cite{milavsevivcius2023review} is  a good source.)

\begin{figure*}[!ht]
		\centering

        \begin{subfigure}{0.4\textwidth}
			\includegraphics[width=0.95\linewidth]{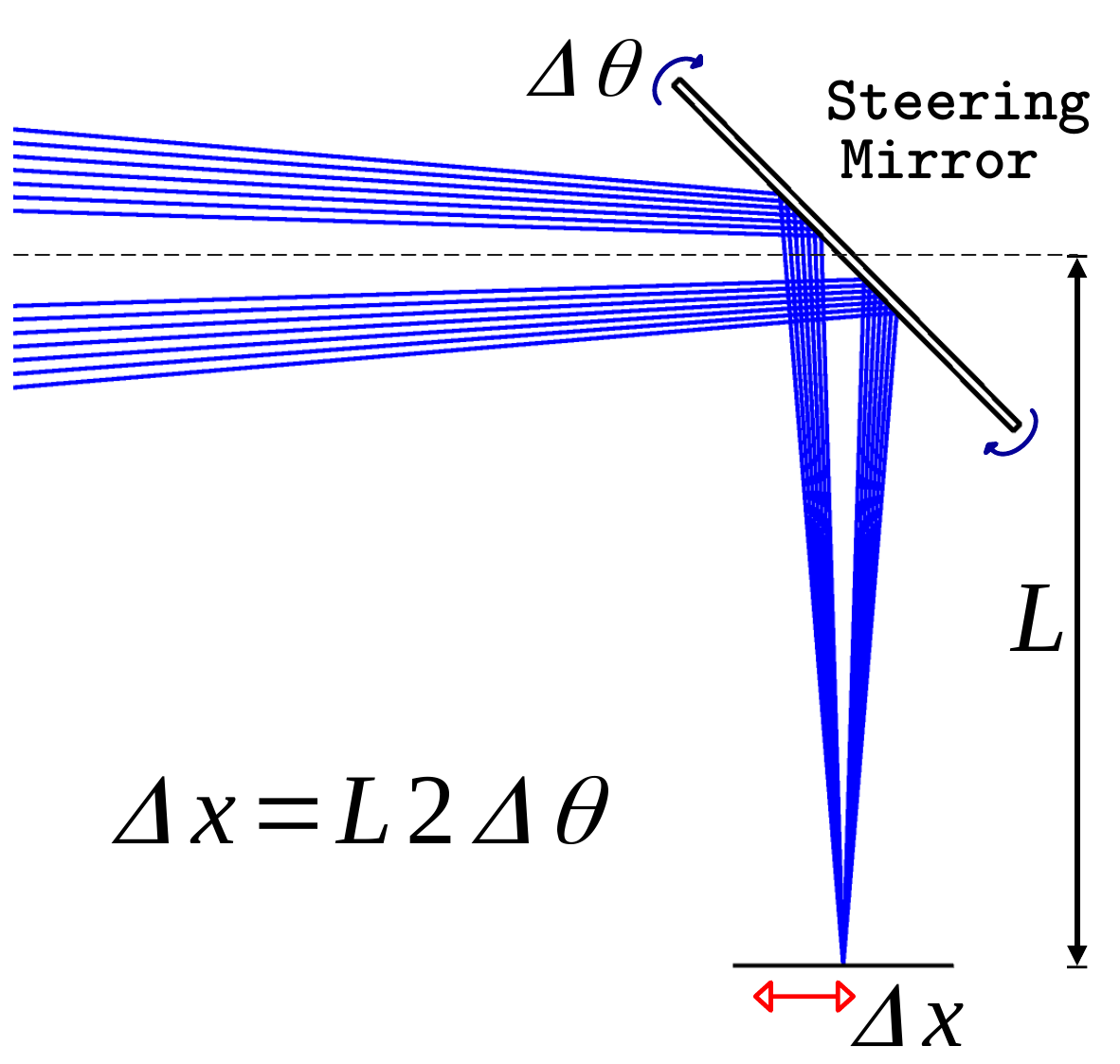}
			\caption{Basic beam steering }\label{fig:steera}
		\end{subfigure}
		\begin{subfigure}{0.42
        \textwidth}
			\includegraphics[width=0.95\linewidth]{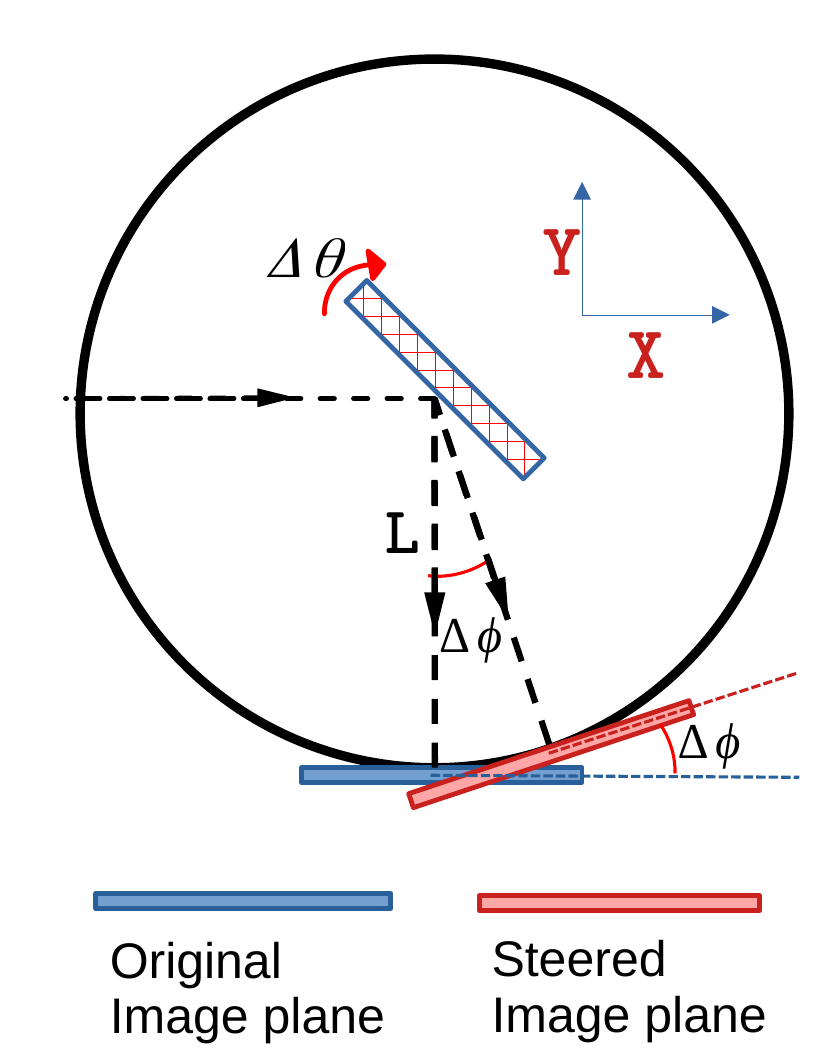}
			\caption{Parasitic movements with steering}\label{fig:steerb}
		\end{subfigure}
		\bigskip

  %     \begin{subfigure}{0.8\textwidth}
		% 	\includegraphics[width=0.95\linewidth]{steering_circle.pdf}
		% 	\caption{The steering circle}\label{fig:steerc}
		% \end{subfigure}
		% \bigskip
		
     \caption{Beam steering with a flat mirror: the basic principle of beam steering with a flat mirror is shown in (a). Angular motion of the steering mirror produces a "linear" displacement of the image at the image plane. This motions is amplified by having the image plane at a larger distance from the steering mirror. For small steering angles, the motion at the image plane can be approximated as a linear translation of the PSF. However, at larger angles, the parasitic motions are non-negligible. The general case is shown in (b). The beam steering in this case is along a circle centred around the mirror centre. In this case, a larger steering circle is preferable. }
		\label{fig:steer}  
	\end{figure*}

\begin{figure*}
		\centering

        \begin{subfigure}{0.8\textwidth}
			\includegraphics[width=1.\linewidth]{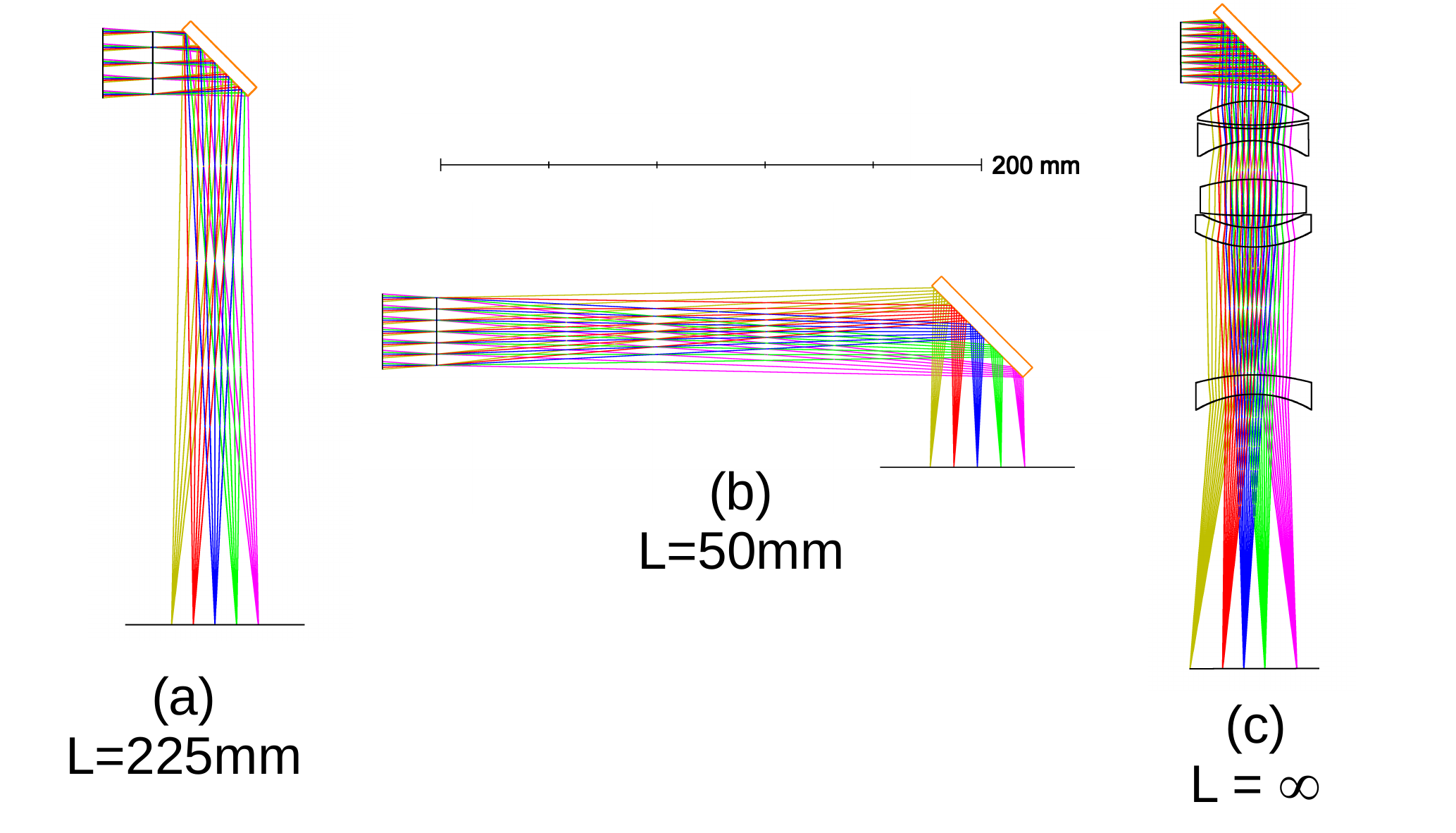}
			\caption{Different steering mechanisms}\label{fig:simul_lay}
		\end{subfigure}
        \bigskip

        \begin{subfigure}{0.45\textwidth}
			\includegraphics[width=1\linewidth]{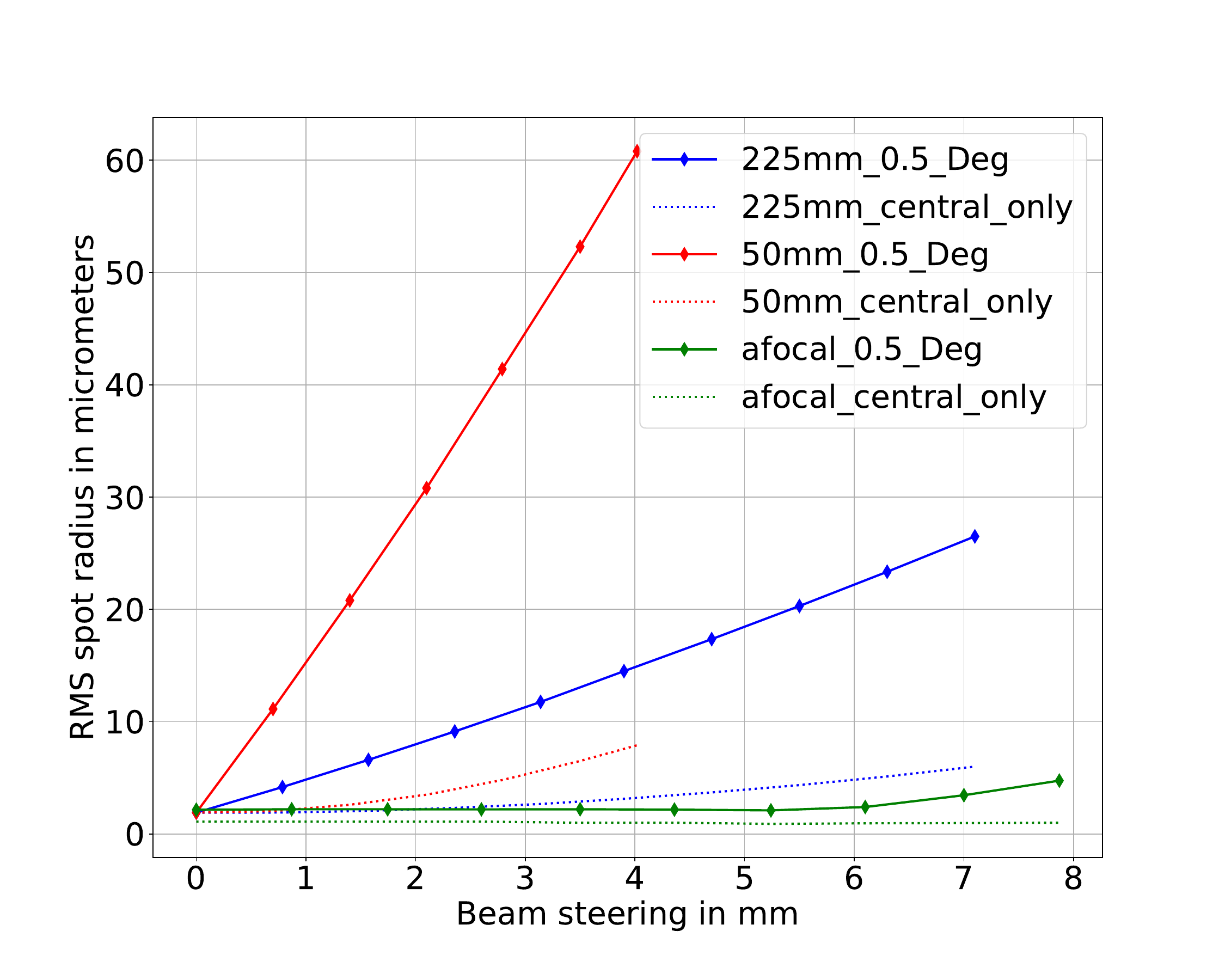}
			\caption{RMS spot degradation with steering angles}\label{fig:simul_RMS_tilt}
		\end{subfigure}
		\begin{subfigure}{0.4
        \textwidth}
			\includegraphics[width=1\linewidth]{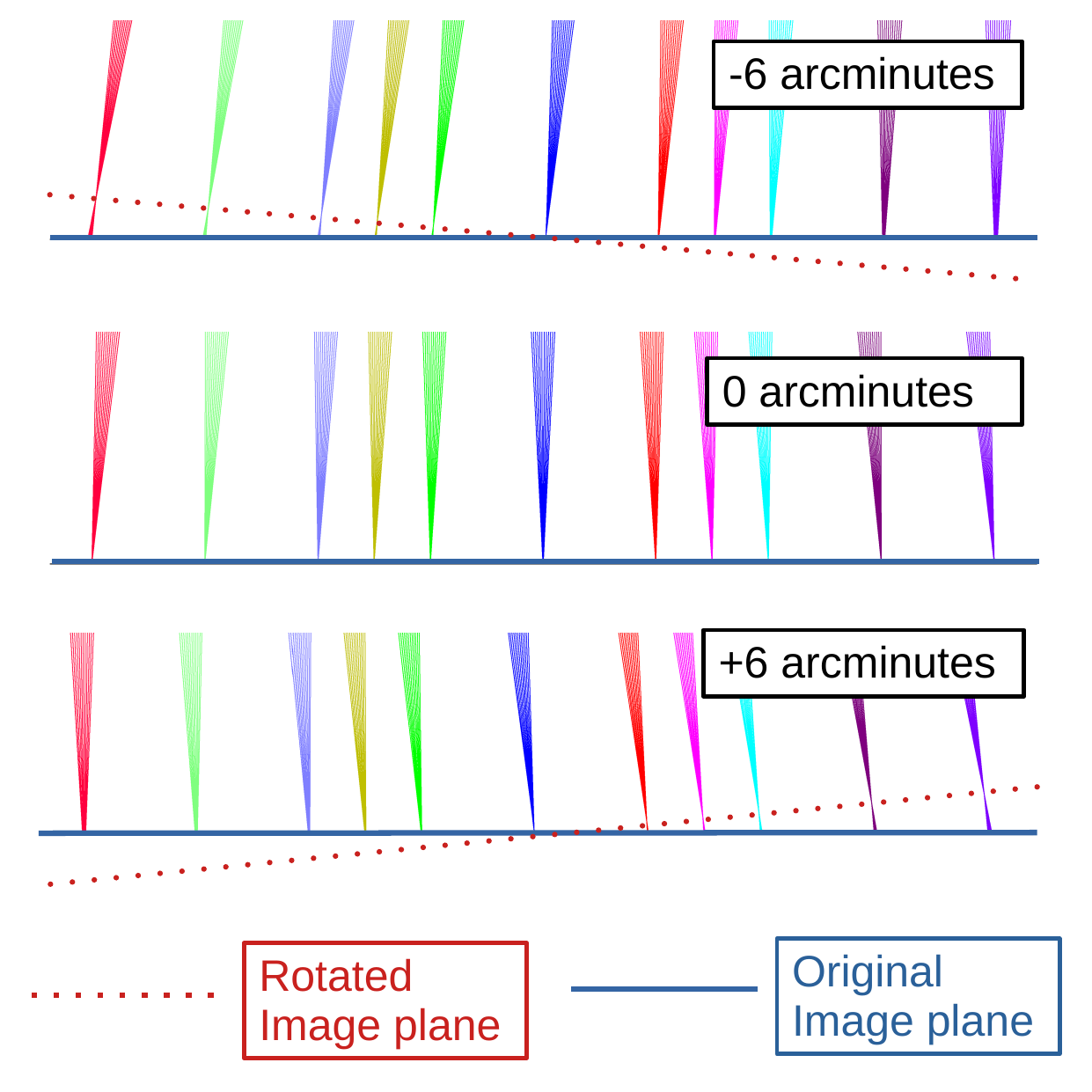}
			\caption{Illustration of out-of-plane rotation}\label{fig:simul_out_plane}
		\end{subfigure}
		\bigskip

     \caption{Comparison of efficacy of different steering mechanisms: layout of three different steering arrangements are shown in (a). The first two steer the beam in a converging beam at 225 mm and 50 mm from the image plane respectively. In the third one the beam is steered in a collimated section. All three systems have same output (F10) and a similar aperture. The degradation of RMS spot v steering angle is shown in (c). Steering from a shorter distance produces a tighter steering circle and hence significantly more degradation. The primary source of this degradation is an out-of-plane rotation of the image plane illustrated in (d). Steering in the collimated beam section produces the least amount of degradation.}
		\label{fig:steer_simulation}  
	\end{figure*}

\begin{figure*}
		\centering

        \begin{subfigure}{0.49\textwidth}
			\includegraphics[width=0.99\linewidth]{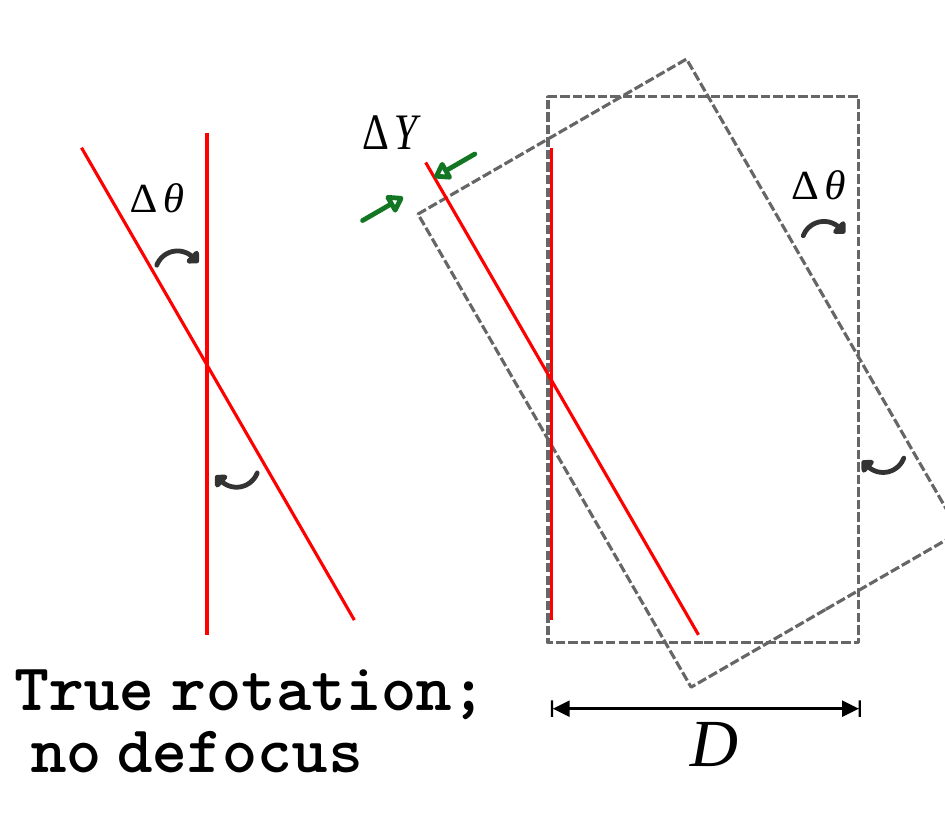}
			\caption{Defocus due to rotation offset }\label{fig:offseta}
		\end{subfigure}
		\begin{subfigure}{0.49
        \textwidth}
			\includegraphics[width=0.99\linewidth]{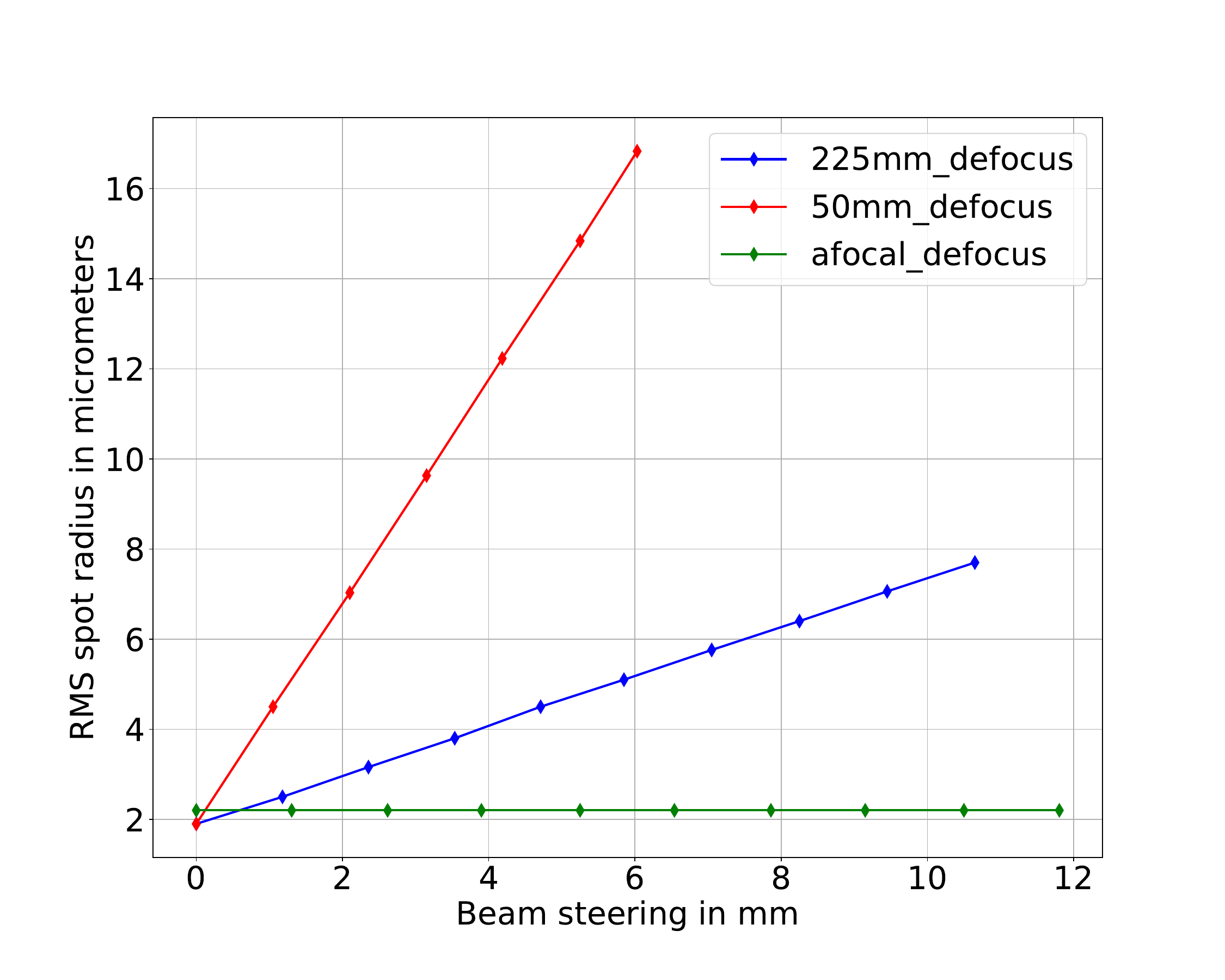}
			\caption{Defocus due to rotation offset}\label{fig:offsetb}
		\end{subfigure}
		\bigskip

  %     \begin{subfigure}{0.8\textwidth}
		% 	\includegraphics[width=0.95\linewidth]{steering_circle.pdf}
		% 	\caption{The steering circle}\label{fig:steerc}
		% \end{subfigure}
		% \bigskip
		
     \caption{Defocus by rotation of a real steering mirror: If a mirror is rotated about a centre that is not coincident with the reflective surface, an angle dependent defocus will occur, as shown in figure (a). The extent of this defocus depends on the steering angle and the thickness of the mirror. The required steering angle is also larger in case of a smaller steering circle. The spot degradation at the image plane is shown in (b) for a mirror thickness of 5 mm. The plot shows amount of image defocus for the same amount of linear steering at the image plane. The effect of longer steering distance is obvious and it is also seen that beam steering from within a collimated section is largely unaffected by offset in rotation axis of the mirror.}
		\label{fig:steer_offset}  
	\end{figure*}

\begin{itemize}
\item \textbf{Micro-thrusters }
   
Precise amount of correcting torque can be applied to the complete spacecraft uising micro-thrusters (\cite{flinois2022microthruster}, \cite{mandic2018habex},). A practical example of this is the Euclid observatory (\cite{mellier2024euclid}). This approach may hold promise for Cubesats in the future (\cite{dennehy2020application} and \cite{levchenko2018space}), but at present is limited by extra weight of fuel and limited lifetime.

\item \textbf{X-Y stage at the detector plane}
   
Let us consider a telescope of focal length $F$, if there is a pointing inaccuracy of $\Delta \alpha$ (in arcsecond units) the corresponding image motion at the detector plane is 
         \begin{equation}
               \Delta x = \frac{F \Delta \alpha}{206265} 
               \label{one}
         \end{equation}
 (the factor 206265 can be omitted if $\Delta \alpha$ is expressed in radians)  This error can be cancelled by translating the the detector via a X-Y stage. This method was attempted for both the PICSAT ( \cite{nowak2016reaching}) and ASTERIA missions. The ASTERIA mission (\cite{pong2018orbit}) has successfully demonstrated this principle is capable of achieving arc-second level accuracy. The typical piezo-driven X-Y stages used in this method have a range limitation ( $<$ 1 mm) so are typically suitable for short focal length imagers.
         
\item \textbf{Fine steering mirror}

The motion required at the detector plane is higher as longer focal length systems are to be accommodated or bigger errors are to be corrected. In such cases, fine steering mirrors are a general purpose solution for beam steering as they can accommodate spectrographs as well as basic imagers. The steering mirror acts as a lever and amplifies the motion. The basic principle is shown in figure \ref{fig:steera}. The translation $\Delta X$ at the image plane is given by 
       \begin{equation}
           \Delta X = \frac{2L\Delta \theta}{206265}
           \label{two}
       \end{equation}
       where $L$ is the steering distance and $\Delta\theta$ is the steering angle (again expressed in arcseconds) of the mirror from its nominal position. To correct for the pointing drift the steering motion must be equal to it magnitude and opposite in direction, therefore:
       \begin{equation*}
         \Delta x = -\Delta X
         \end{equation*}
       
The negative sign may be ignored if one is only concerned about the magnitude of steering angle; which can be rewritten as
       \begin{equation}
           \Delta \theta = \frac{F\Delta \alpha}{2L}
           \label{three}
       \end{equation}

 This method is used in JWST \cite{ostaszewski2007fine}, Herschel (SPIRE instrument \cite{pain2003spire}) as well for the planned mission CUBESPEC \cite{de2024high} as well as a number of satellite communication missions.
        
\end{itemize}

\subsection{Aberrations arising from beam steering}

An ideal beam steering solution is one that allows for controlled large amplitude motion in the image plane without degrading the PSF. However, X-Y stage or FSM based steering modify the optical layout in some manner and hence introduce steering related aberrations. For a X-Y stage, this aberration is mostly due to small deviations within beam-path as the pointing direction "wanders". This can be reduced by over-sizing the optics and optimising for a larger FOV than is imaged on to the detector. For example, let's consider a camera that images 10 arcminutes onto a detector. If the expected coarse pointing stability is 2(+/-1) arcminutes, then the optics must be designed for 12 arcminutes.

 FSM based beam steering is more complex as there are significant additional aberrations introduced by the steering mirror.  When a converging beam is steered by a flat mirror the optical path length is conserved along a circle (figure \ref{fig:steerb}). Therefore, beam steering (in 2 dimensions) by a flat mirror produces the best image on a sphere rather than a flat surface (\cite{bagnasco2007overview}). The centre of the sphere is coincident to the steering mirror. In reference to the detector this produces three separate motions. The first motions is along the X-axis expressed as:
\begin{equation}
    \Delta X = 2L \Delta \theta \times cos\Delta\phi
    \label{four}
\end{equation}

This is the desired steering motion and simplifies to equation \ref{two} when $\Delta \phi $ is small. The second is a defocus along the Y-axis expressed as :
\begin{equation}
    \Delta Y = 2L \Delta \theta \times sin\Delta\phi
    \label{five}
\end{equation}

This defocus can be made small by keeping the steering angle small. The third motion is an out-of-plane rotation in the image plane.
\begin{equation}
    \Delta \phi = 2\Delta \theta
    \label{six}
\end{equation}

This produces a defocus at different field points of the image plane. Particularly the defocus is positive for one half of the image and negative in the other half. The maximum defocus scales with the size of the image plane and can be written as:
\begin{equation}
    \Delta Y' =\pm \frac{W\Delta \phi}{2}
     \label{seven}
\end{equation}

Where, W is the size of the image plane. \\

JWST deals with the steering related aberrations by matching the field curvature of the intermediate telescope focus to that of the steering sphere. (\cite{bagnasco2007overview}). This method can not be scaled down to Cubesats where available space is very limited.

One additional source of image degradation  arises from off-centre steering of the FSM. Typically steering actuators are placed behind the mirror. The steering axis of the mirror is offset from the optical surface  by a distance equal to the substrate thickness $D$. A corresponding defocus of:
\begin{equation}
    \Delta Y = 2D sin(\Delta \theta)
    \label{nine}
\end{equation}

is seen at all field points of the image plane.

A comparative simulation of three separate steering arrangements is presented in figure \ref{fig:simul_lay}. In the first two arrangements, steering is done in an F10 converging beam with a steering distance of 225 mm and 50 mm respectively. The third arrangement steers the beam in a collimated beam section followed by an F10 camera.

In case of beam steering in a collimated beam, the motion at the image plane can be expressed as 
\begin{equation}
    \Delta X = 2f_a \Delta \theta
    \label{eight}
\end{equation}

where $f_a$ is the focal length of the imaging assembly following the steering mirror. 

In all three systems, the output is steered by means of tilting the steering mirror from its nominal position. The RMS spot radius is recorded for a number of different field for each incremental tilt of the steering mirror. The degradation of RMS spot radius is shown in figure \ref{fig:simul_RMS_tilt}. It is evident that degradation is inversely proportional to the steering distance. The major contributor for such degradation is the out-of-plane rotation of the image plane which introduces a positive defocus on half of the image plane and negative defocus for the other half. In case of the shortest beam steering distance, the out-of-plane rotation of the image plane is significant enough to be visible near the focal plane (figure \ref{fig:simul_out_plane}).  It makes sense to allow for a large $L$ to minimize the degradation of the spot.  In this aspect, the afocal system has a particular advantage; since from the perspective of the imaging system the object is at infinity, the optical distance between the image plane and the steering mirror can also be considered infinite. Therefore, a reasonably well optimized system is expected to show little or no degradation by beam steering from within a collimated beam section.

The spot degradations arising from this defocus is shown in figure \ref{fig:offsetb}. Since defocus is uniform across the field of view, data is plotted only for one field. Once again it is seen that setting the steering distance very short introduces the most amount of RMS spot degradation. This is related to equation \ref{two}, since the steering distance $L$ is small, a larger steering angle $\Delta\theta$ is required and hence a larger amount of defocus is observed. The afocal system shows no degradation in this scenario; this is  expected since small amounts of defocus in a collimated beam introduces no change in the image location. \\

We have presented a general description of the complexities related to beam steering by means of a FSM. The results are summarised as following:
\begin{itemize}
    \item The degradation is inversely proportional to the steering distance -- distance from the image plane to the centre of the steering mirror.
    \item The degradation is directly proportional to the physical size of the image plane.    
   \item  A defocus proportional to the thickness of the steering mirror is also seen.
   \item Beam steering within a collimated beam section avoids these aberrations.
   
\end{itemize}

We have highlighted that beam steering  within a collimated beam section has inherent advantages. Afocal telescopes conveniently produce a collimated section without the need for extra collimator optics. As such, afocal designs are a natural fit for Cubesat based observatories looking to utilise FSM based pointing stabilisation.

\section{Cubesat based observatory templates}

\begin{figure*}
		\centering
    
        \begin{subfigure}{0.5\linewidth}
			\includegraphics[width=1\linewidth]{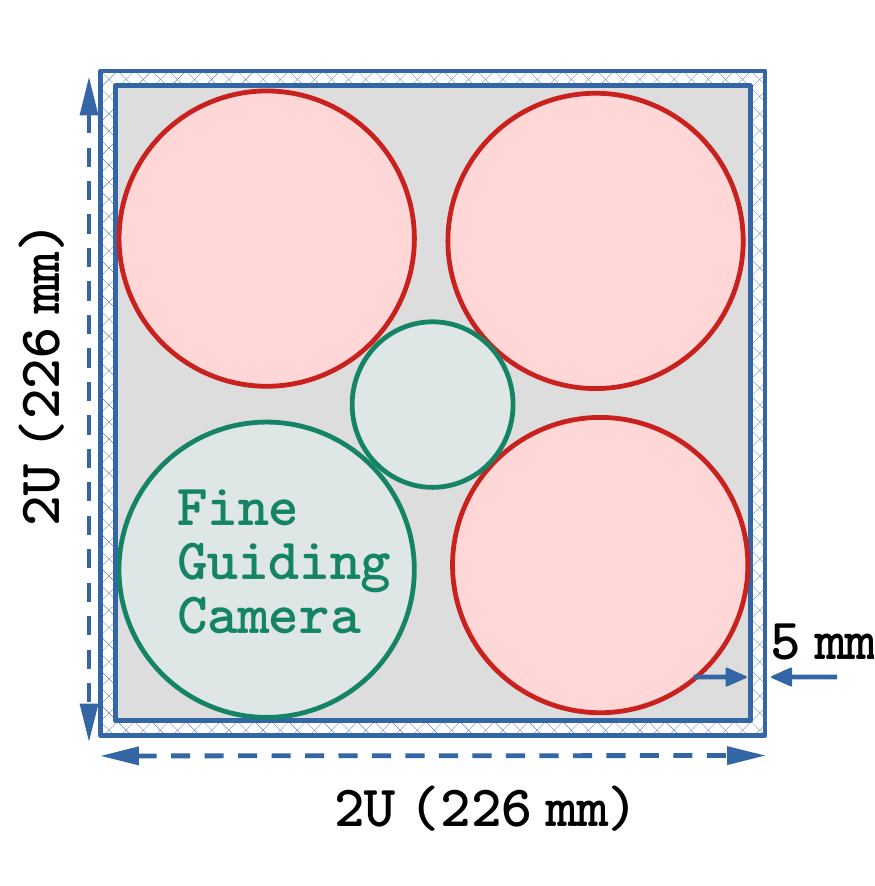}
			\caption{Top view}\label{fig:1U_topview}
		\end{subfigure}
		 \begin{subfigure}{0.49\linewidth}
			\includegraphics[width=1\linewidth]{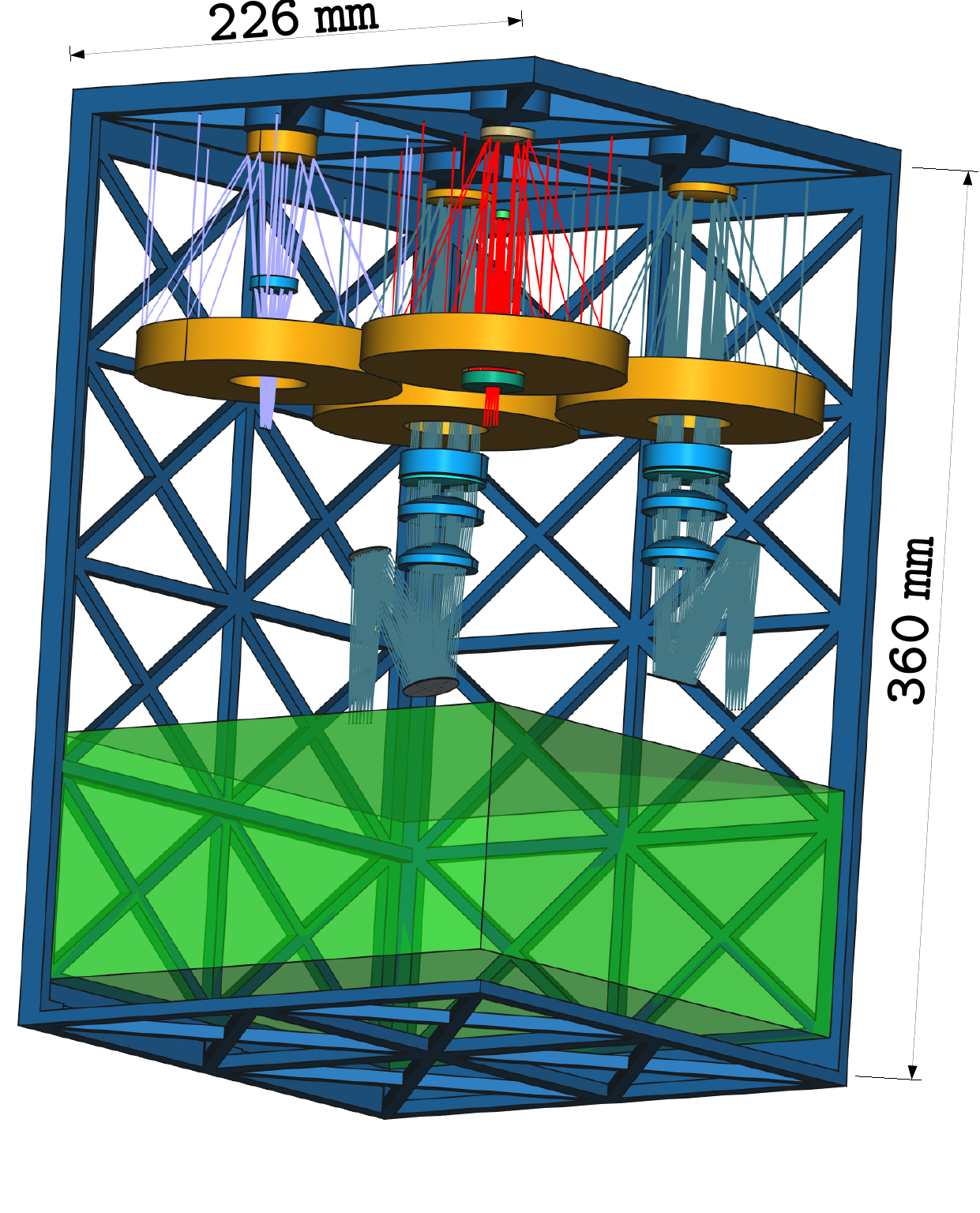}
			\caption{Side view }\label{fig:1u_3d}
		\end{subfigure}
        
		\begin{subfigure}{0.8\linewidth}
			\includegraphics[width=1\linewidth]{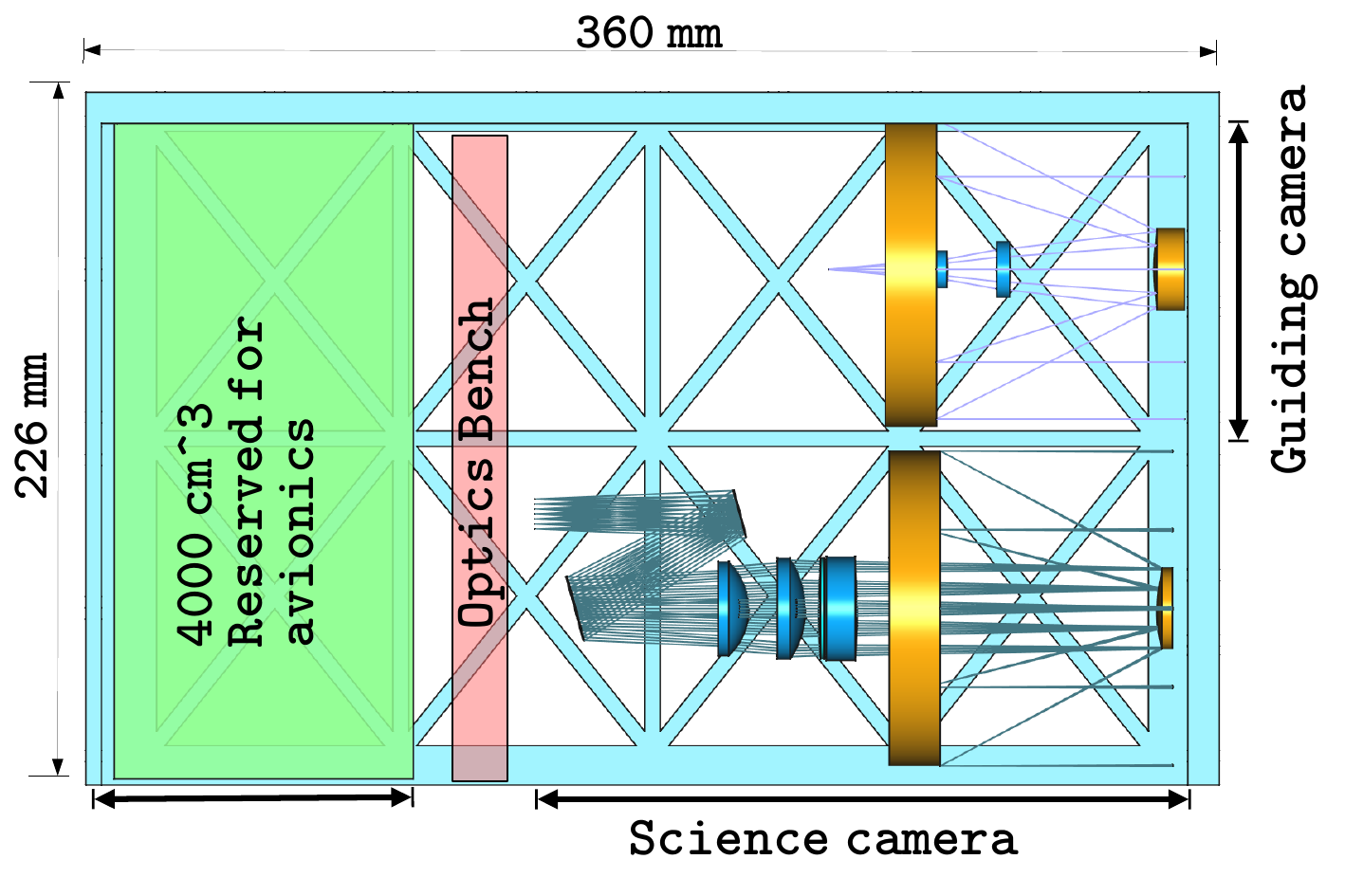}
			\caption{Side view }\label{fig:1u_sideview}
		\end{subfigure}

     \caption{ Plan for a Cubesat observatory utilising a four separate 10 cm aperture telescopes. The entire observatory fits within a 12U volume, of which about 6-8U is reserved for the optical telescope and components. The aperture distribution is shown in (a). The space allocation within 12U volume is shown in (b) and (c). Suitable fine guiding cameras for this template are shown in  figure \ref{fig:1U_guiders} and science focused optical designs  are shown in figure \ref{fig:1U refractive designs} and \ref{fig:1U_reflective designs}. }
		\label{fig:1U_observatory}  
	\end{figure*}

\begin{figure*}
		\centering

        \begin{subfigure}{0.44\textwidth}
			\includegraphics[width=1.\linewidth]{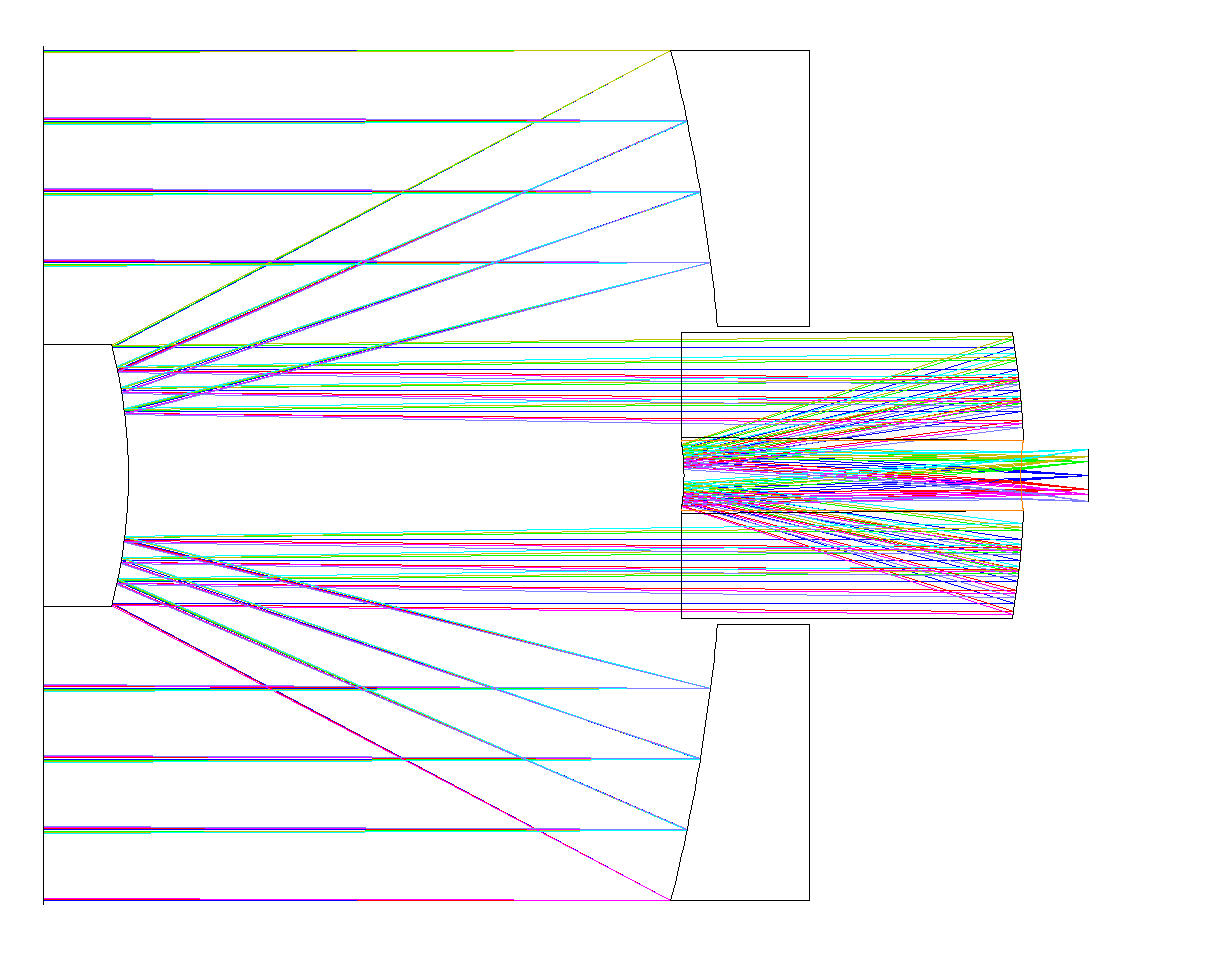}
			\caption{1U monolithic guider}\label{fig:design_monolith_guider}
		\end{subfigure}
		\begin{subfigure}{0.55
        \textwidth}
			\includegraphics[width=1\linewidth]{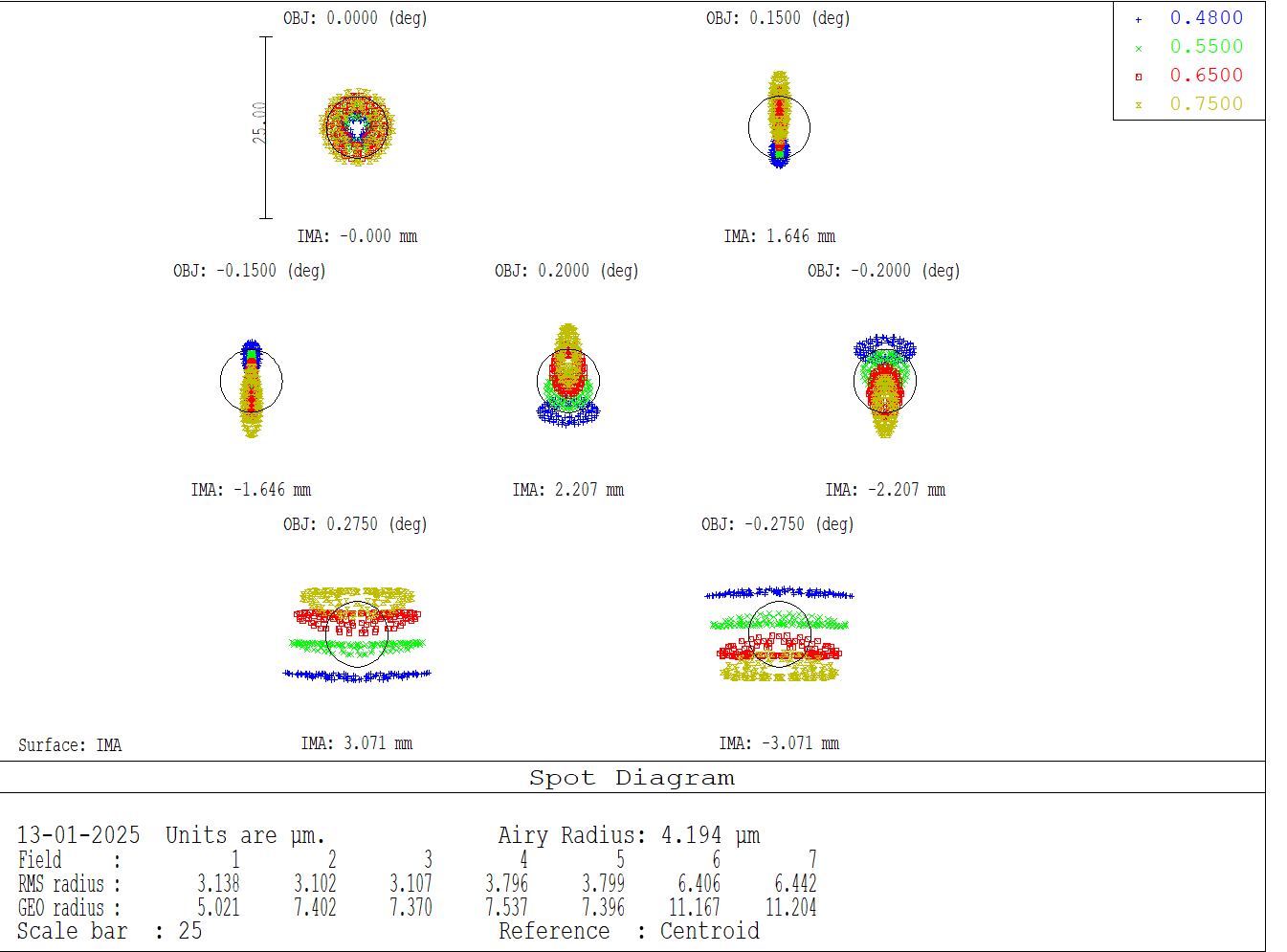}
			\caption{Spot for monolithic guider}\label{fig:spot_monolith_guider}
		\end{subfigure}
		\bigskip

        \begin{subfigure}{0.44\textwidth}
			\includegraphics[width=1\linewidth]{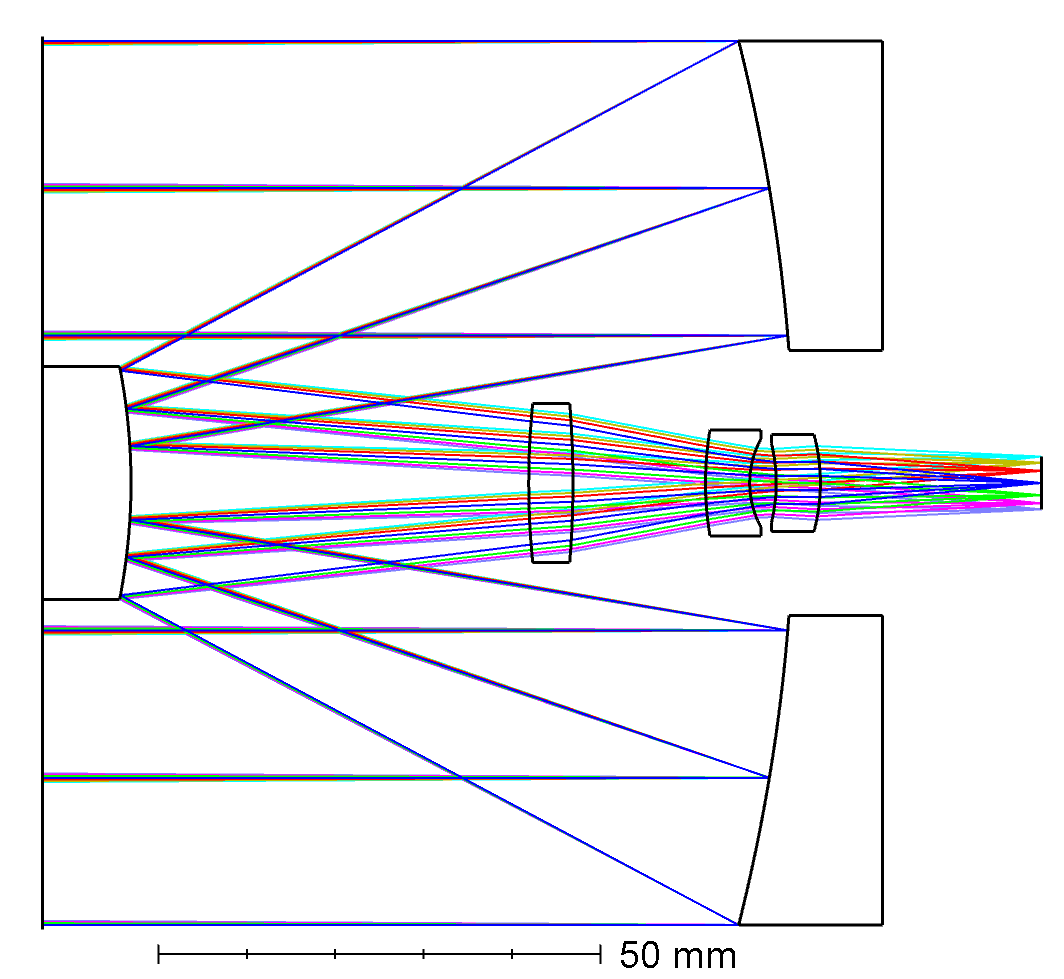}
			\caption{1U Guider using 3 lens elements}\label{fig:design_lens_guider}
		\end{subfigure}
		\begin{subfigure}{0.55
        \textwidth}
			\includegraphics[width=1\linewidth]{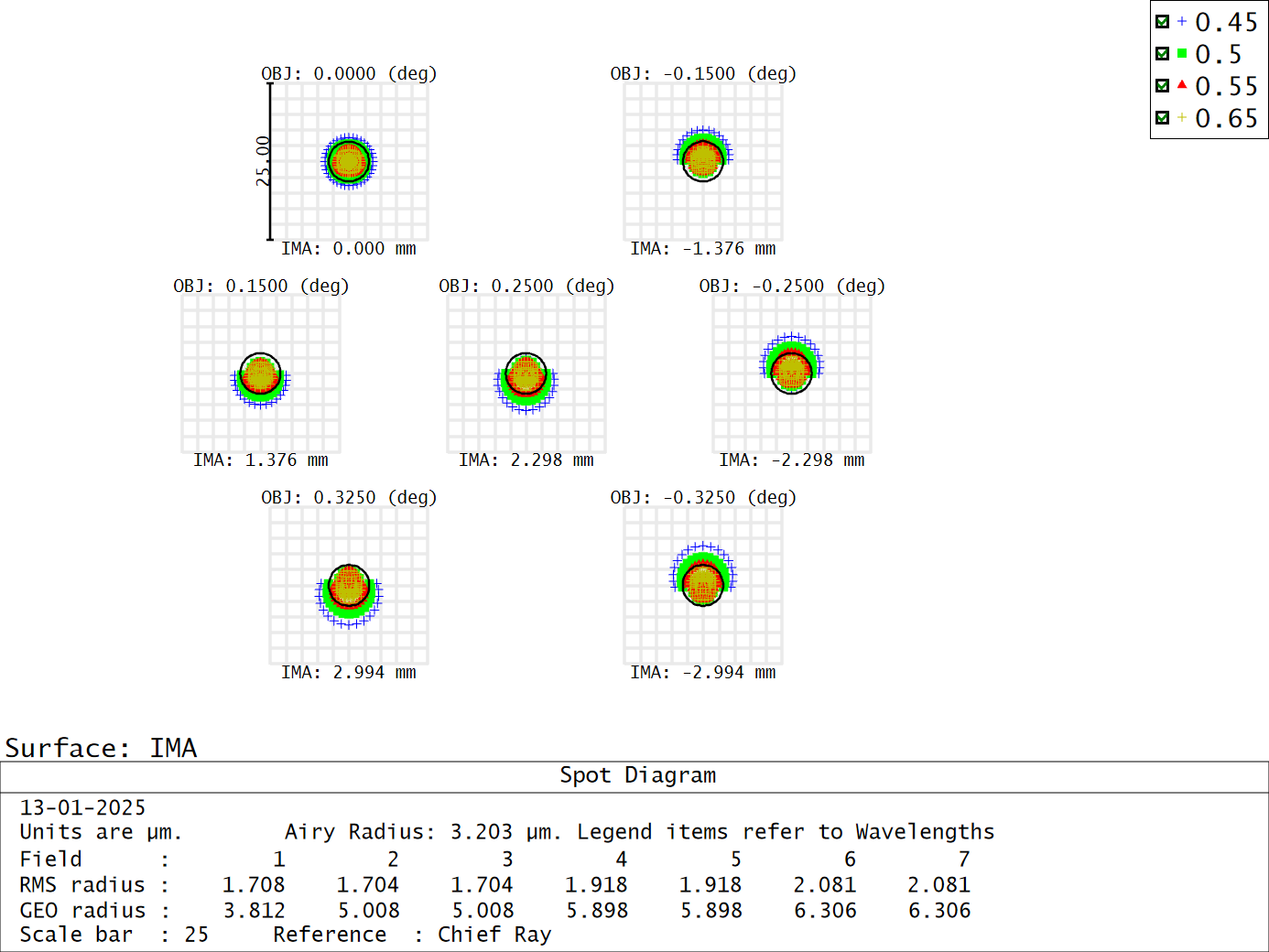}
			\caption{Spot for 1U guider}\label{fig:spot_lens_guider}
		\end{subfigure}
		\bigskip

     \caption{Comparison of  two guide camera designs: the layout in (a) is by means of a monolithic camera following an afocal telescope. This design has an output of F6.25 and a field of view of about 30 arcminutes. The corresponding image performance is shown in (b). Similarly the layout of a 3 lens corrector following a fast telescope is shown in (c). This design has an F5.25 output and is optimised for a field of 40 arcminutes. The imaging performance is shown in (d). The 3 lens catadioptric system is marginally better in terms field of view and image quality. }
		\label{fig:1U_guiders}  
	\end{figure*}

A complete observatory must include some provision for coarse and fine guiding as well as beam steering mechanisms to provide a stable image at the focal plane for long integration durations. It must also include necessary avionics components such as battery, computation, reaction wheels, magneto-toqruers, communication antennae etc. It is desirable to have a significant number of components/subsystems to be standardised and realized using modular and off-the-shelf components.  This approach reduces  cost, effort, and the time for practical implementation as well testing and characterisation of components. However, we also desire that our observatory framework must be flexible enough so as to target specific science cases. We present an optical framework based on afocal designs to accommodate both of these goals. 
We present two designs templates based on standard 12U Cubesat geometry. One template uses four instances of 10 cm aperture telescopes and the other uses a single 20 cm aperture telescope. We demonstrate that it is possible to realize a range of scientifically relevant designs using the same off-the-shelf primary mirror. The secondary mirror is also  kept identical across all afocal designs within a template. The optomechanical design of the afocal telescope is very similar to a conventional RC type telescope. Therefore, the telescope as the "light gathering element" can be standardised. The final output parameters such as F-number, wavelength range and field of view are mostly decided by the camera optics. Optical design of a number of such cameras along with relevant science cases is discussed. Also, both of these templates allow for some provision for a fine guiding camera (although in slightly different manners) and leave ample space for the avionics components and other vital subsystems. To allow for complete and reliable avionics components, a minimum of 4U volume (about 4000 cm$^3$) must be allocated for the avionics components. Therefore the optics must fit within a 8U volume. The templates are discussed as following:

\begin{figure}
		\centering

        \begin{subfigure}{0.5\textwidth}
			\includegraphics[width=1.\linewidth]{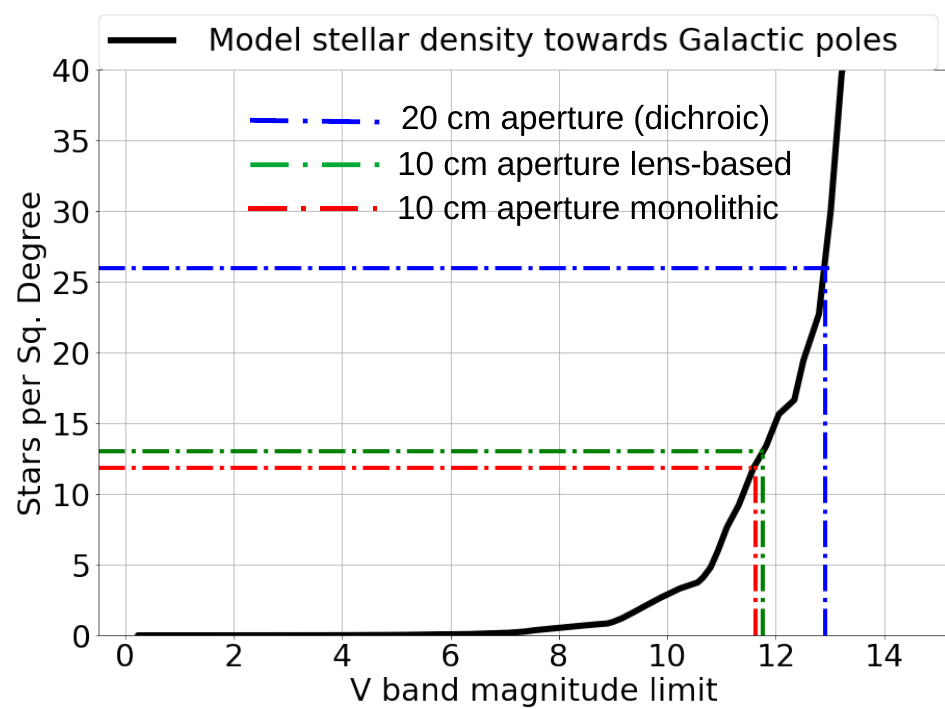}
			\caption{Guiding limits v stellar distribution}
		\end{subfigure}

     \caption{ A comparison of guiding limits of various cameras:
     the SNR limits for various guiding cameras is plotted along with  model stellar distribution from \cite{douglas2021practical}. The green and red line are for 1U guiders (monolithic and lens based respectively) in figure \ref{fig:1U_guiders}, and the blue line is for the 2U guider shown in figure \ref{fig:2U_guiders}. The corresponding stellar density in terms of stars per square degree is used to estimate the probability of finding at least one star in the field of view (table \ref{guideTable}).}
		\label{fig:guiding_limits}  
\end{figure}

\begin{table}
%% use tabular font for a smaller size font
\tabularfont
\caption{Three different guiding cameras are evaluated, the field of view as well as limiting magnitudes for 0.5 Second exposure (SNR =10) is presented. The quantity P is the probability of finding at least one guide star within the field of view. This probability is calculated for stellar density models toward the Galactic poles (figure \ref{fig:guiding_limits})and hence is a conservative lower bound.}\label{guideTable} %%10/12
\renewcommand{\arraystretch}{1.25}
\begin{tabular}{lcccc}
\topline
Design & FOV & Pixelscale & Vmag & P \\\midline
1U(lens) & 40'$\times$ 30'& 0.6"& 11.62& 0.97\\
1U(mono) & 30'$\times$ 22'& 0.5"& 11.76&0.92 \\
2U(lens) & 21'$\times$ 15.7'& 0.35"& 12.9& 0.91 \\
\hline
\end{tabular}
%%use \tablenotes{footnote} to get the table foot note
% \tablenotes{Sample table footnote}%%9/11
\end{table}

\begin{table}[htb]
%% use tabular font for a smaller size font
\tabularfont
\caption{Commercial piezo-driven X-Y stages for fine steering of 1U telescopes, required range and precision is listed for an initial (coarse) pointing stability of 150 arcseconds and desired stability of 0.25 arcseconds.}\label{xyTable} %%10/12
\renewcommand{\arraystretch}{1.25}
\begin{tabular}{lccccc}
\topline
F\#& Range & Accuracy & Model\\\midline

F5 &0.36mm& 0.6$\mu$m & Cedrat XY500M \\
F8 &0.58mm &0.96$\mu$m& PI P-625.2CL  \\
F12 & 0.87mm& 1.5$\mu$m & P-628.2CL \\
F15& 1.1mm & 1.8$\mu$m& Cedrat APA1500L*\\
F20& 1.45mm& 2.4$\mu$m& P-629.2CL \\

\hline
\end{tabular}
%%use \tablenotes{footnote} to get the table foot note
\tablenotes{*:APA1500L is a single actuator, at least two must be used to implement an X-Y stage}%%9/11
\end{table}

\begin{figure*}
		\centering

        \begin{subfigure}{0.44\textwidth}
			\includegraphics[width=1.\linewidth]{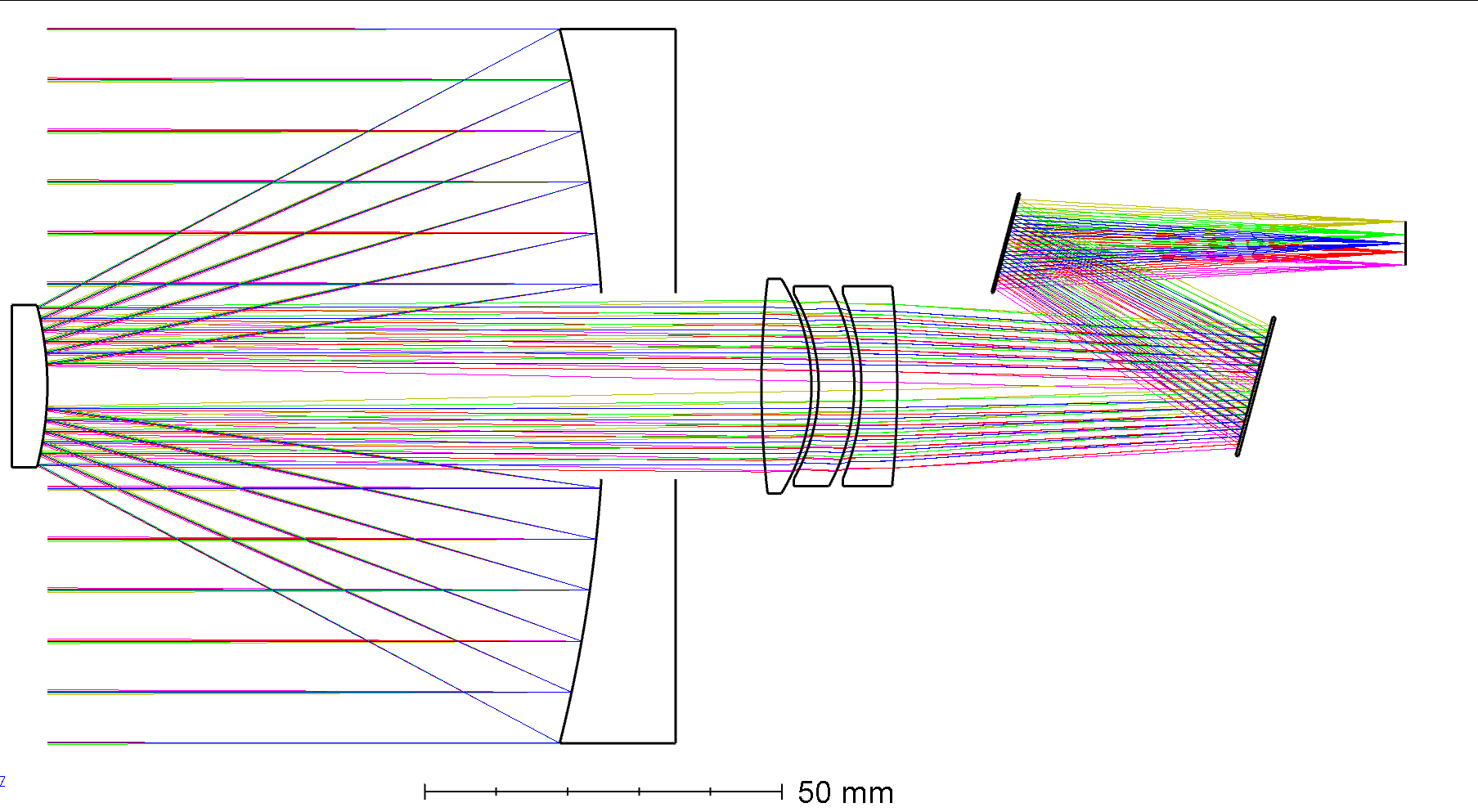}
			\caption{1U visible camera}\label{fig:1U_VIS}
		\end{subfigure}
		\begin{subfigure}{0.55
        \textwidth}
			\includegraphics[width=1\linewidth]{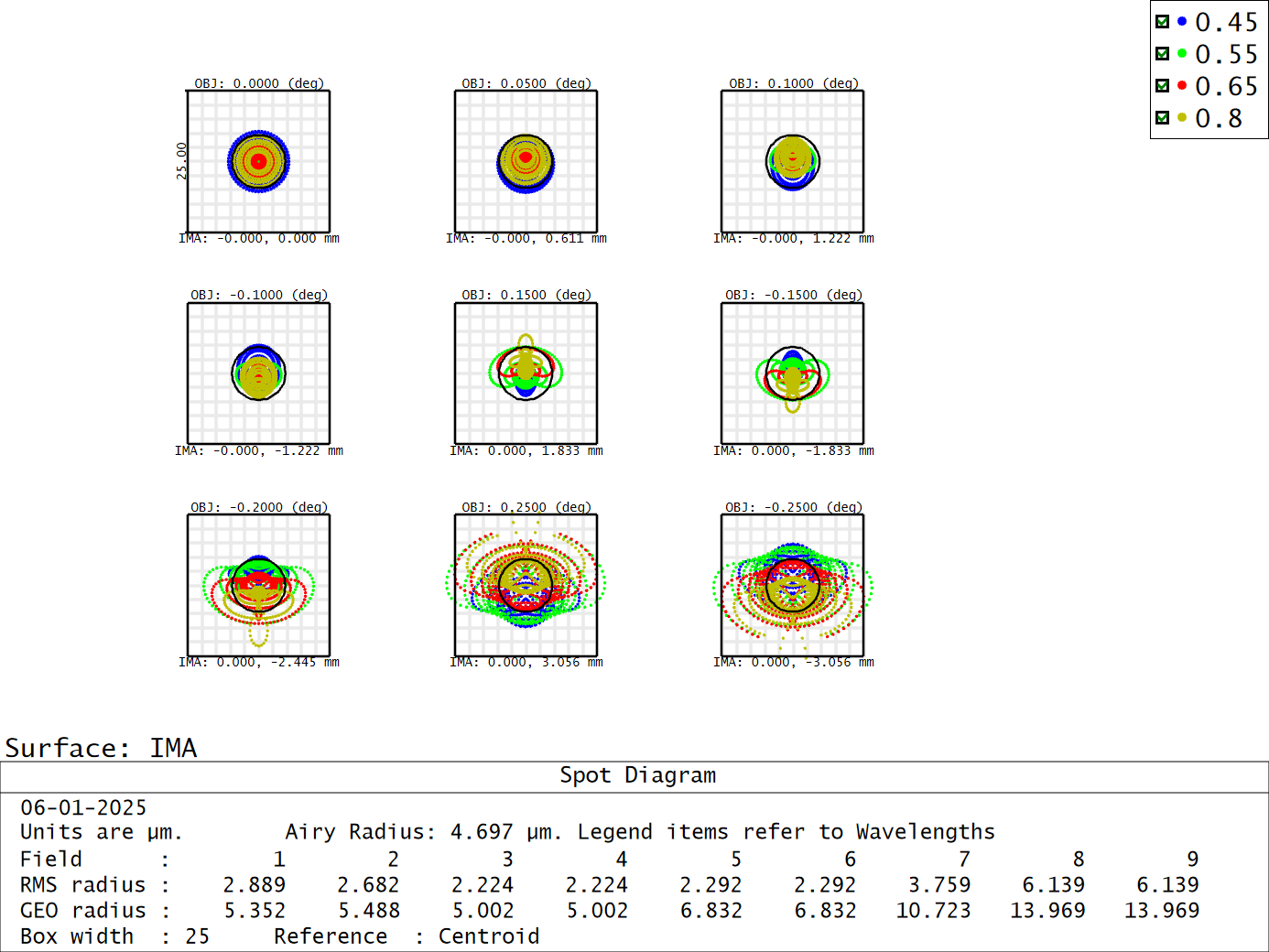}
			\caption{Spot for 1U camera}\label{fig:1U_VIS_spot}
		\end{subfigure}
		\bigskip

        \begin{subfigure}{0.44\textwidth}
			\includegraphics[width=1\linewidth]{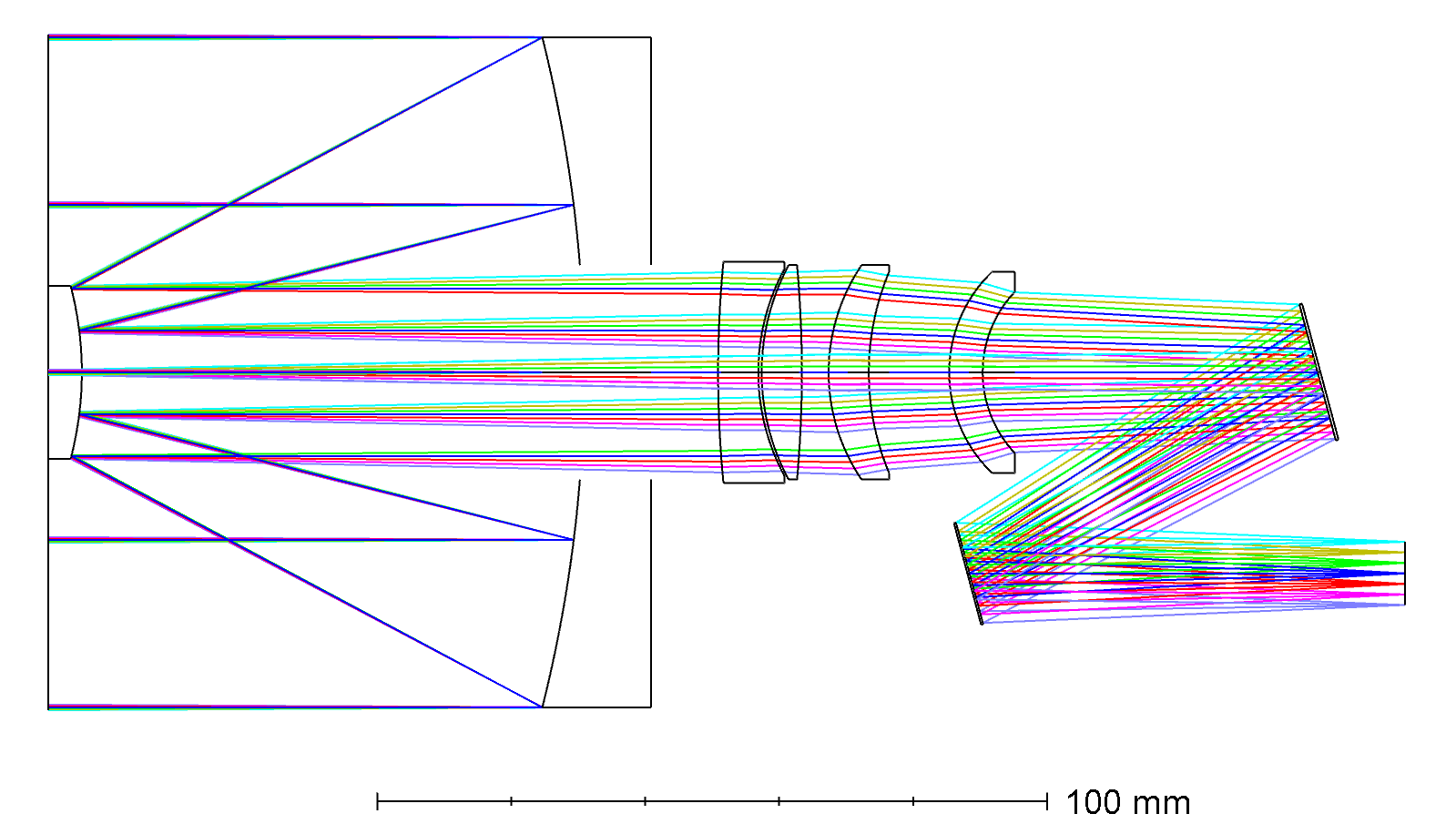}
			\caption{1U NIR camera}\label{fig:1U_NIR_dio}
		\end{subfigure}
		\begin{subfigure}{0.55
        \textwidth}
			\includegraphics[width=1\linewidth]{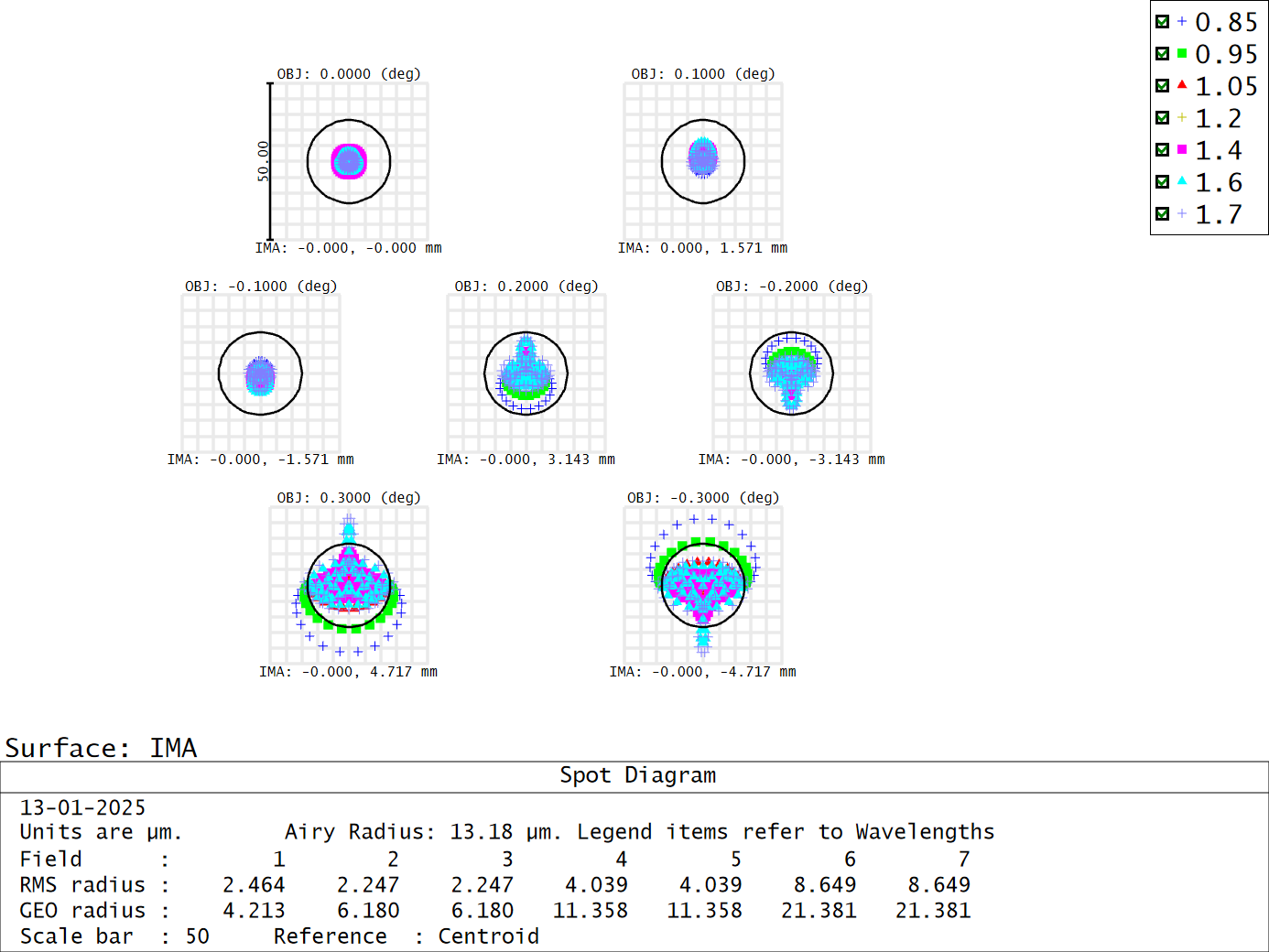}
			\caption{Spot for 1U NIR camera}\label{fig:1U_NIR_dio_spot}
		\end{subfigure}
		\bigskip

     \caption{Review of 1U designs based on dioptric cameras: figure (a) shows a visible camera with a fairly fast output (F7) and wide field of view. The imaging performance of this design is shown in (b); illustrating reasonably well optimised performance for up to 30 arcminutes. An NIR camera is shown in (c) and corresponding spot performance is shown in (d). The design is optimised for 30 arcminutes. Both visible and NIR designs are seen to be optimised close to the diffraction limit. }
		\label{fig:1U refractive designs}  
	\end{figure*}

\begin{figure*}
		\centering

       \begin{subfigure}{0.44\textwidth}
			\includegraphics[width=1\linewidth]{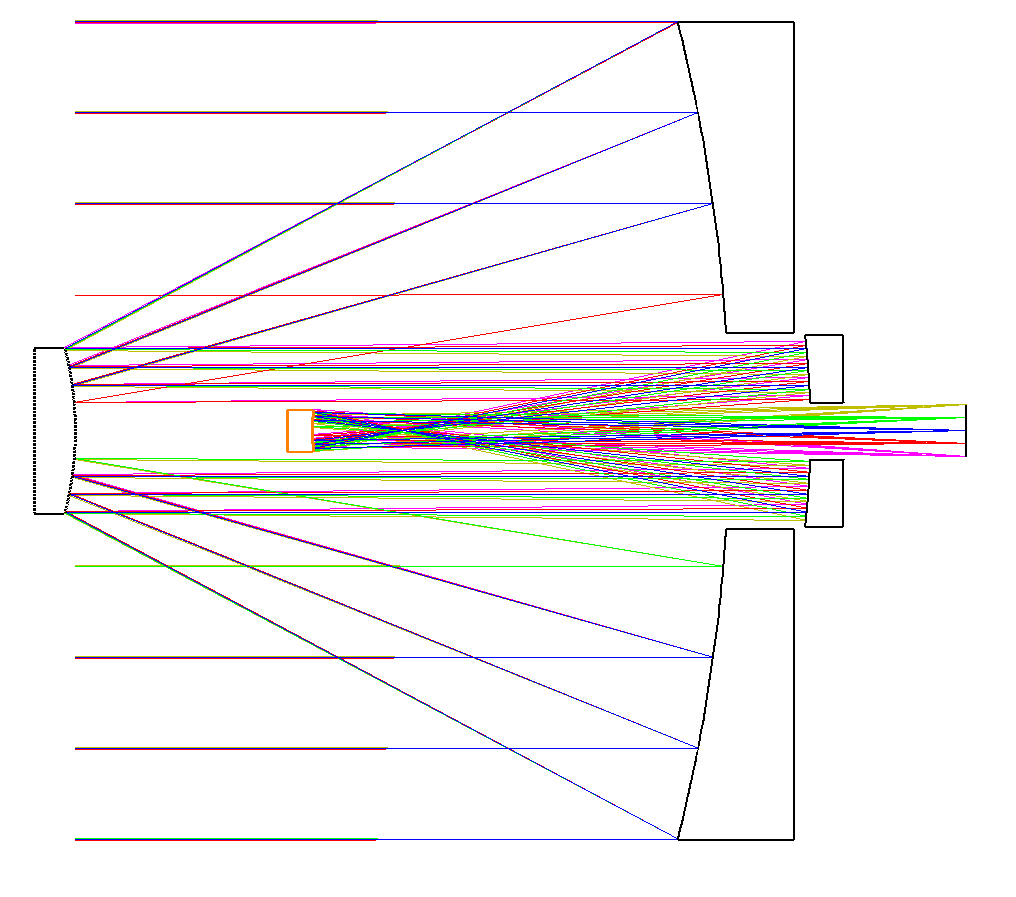}
			\caption{1U imager with a Gregorian camera}\label{fig:1U_greg}
		\end{subfigure}
		\begin{subfigure}{0.55
        \textwidth}
			\includegraphics[width=1\linewidth]{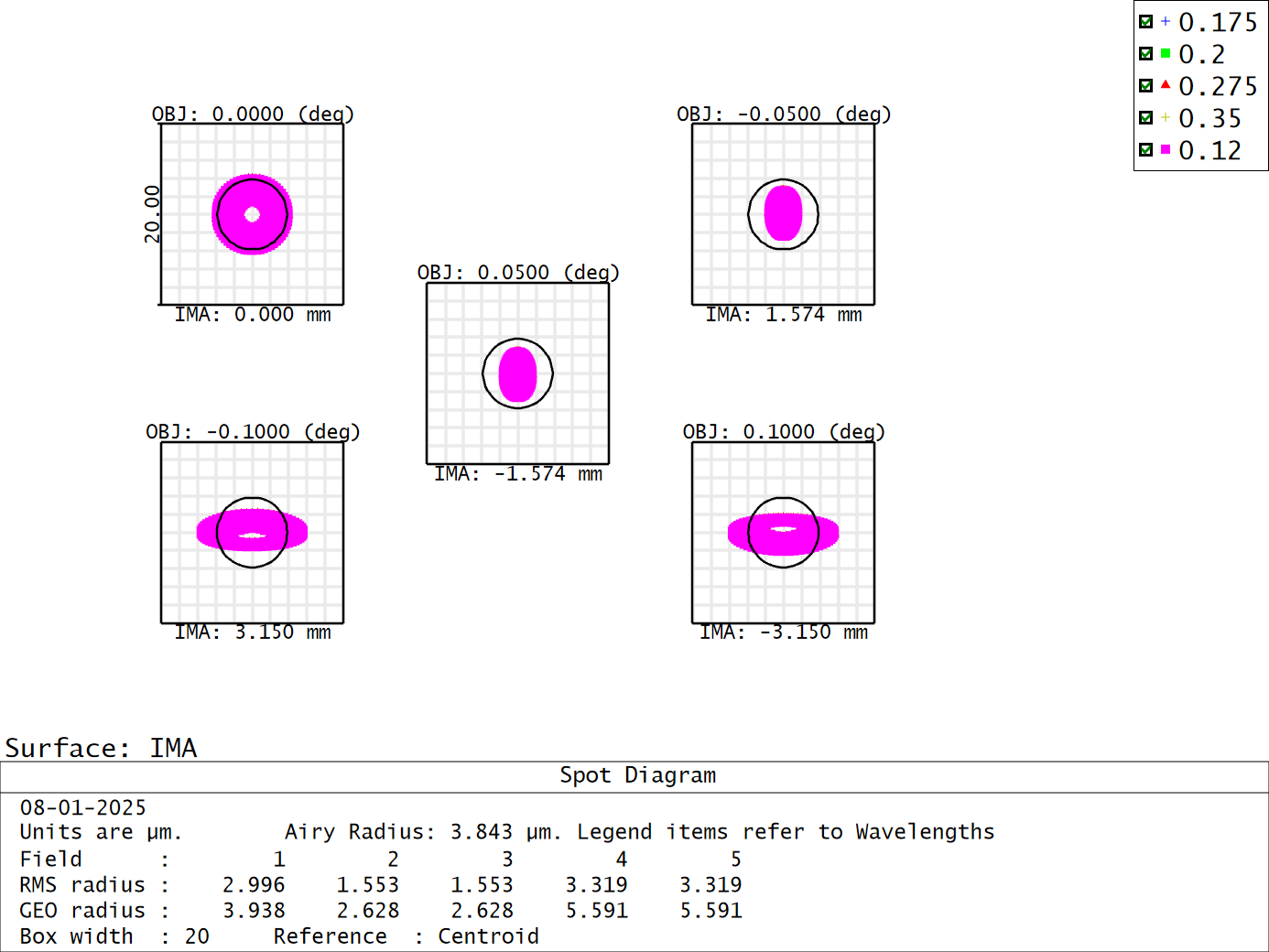}
			\caption{Spot for the Gregorian camera}\label{fig:1U_greg_spot}
		\end{subfigure}
		\bigskip

        \begin{subfigure}{0.44\textwidth}
			\includegraphics[width=1\linewidth]{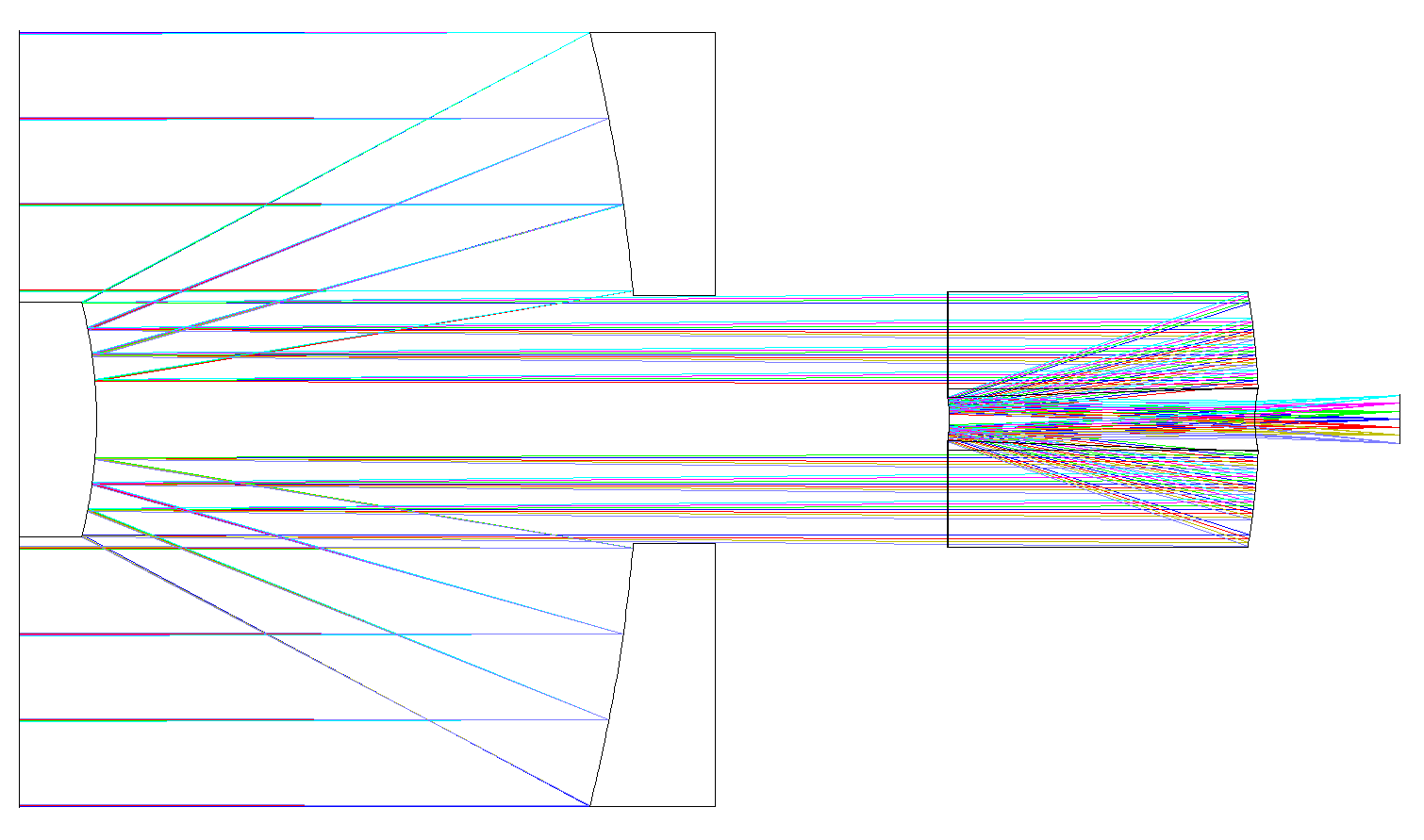}
			\caption{1U monolithic camera}\label{fig:1U_monolith}
		\end{subfigure}
		\begin{subfigure}{0.55
        \textwidth}
			\includegraphics[width=1\linewidth]{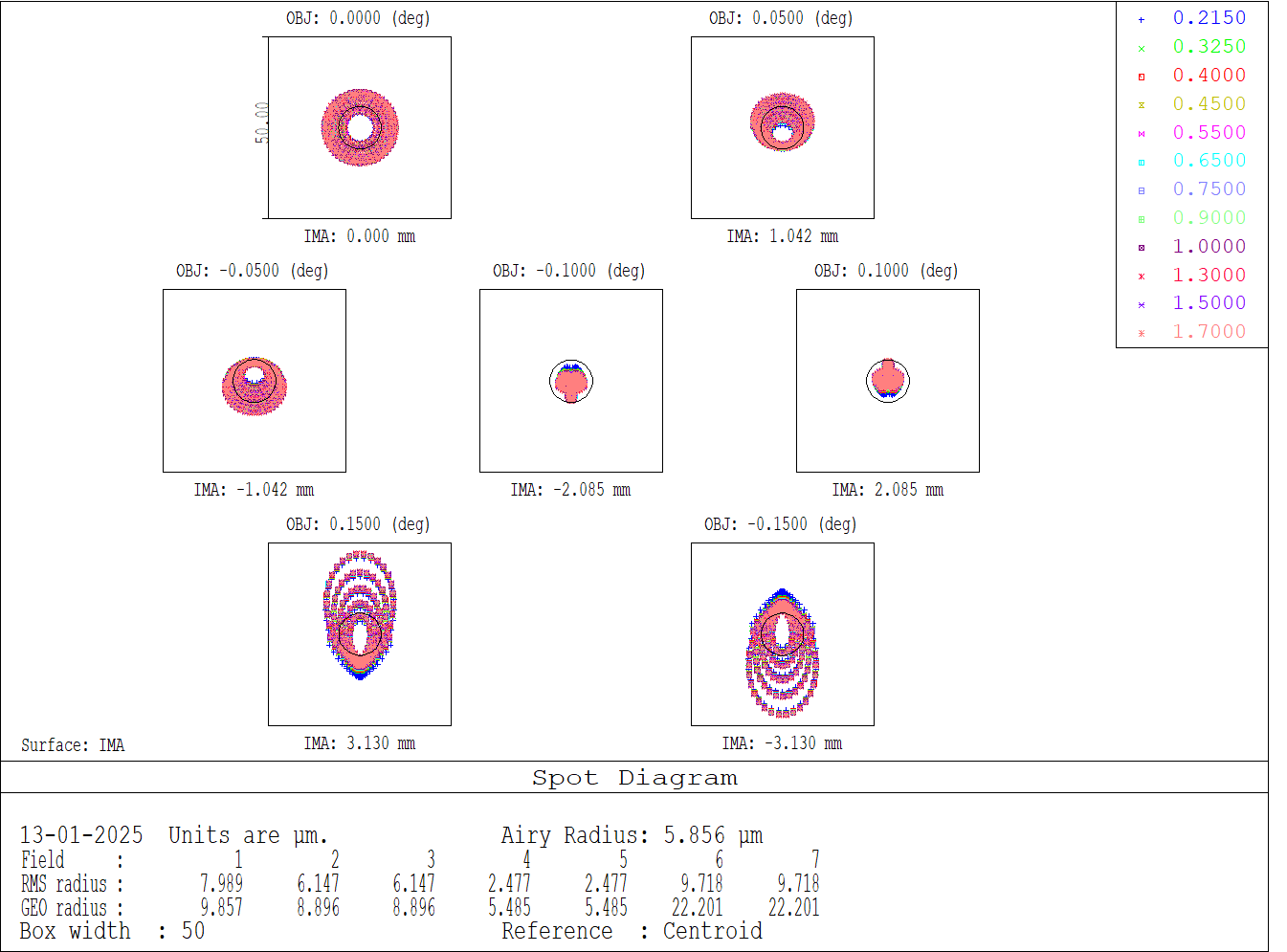}
			\caption{Spot for monolithic camera}\label{fig:1U_monolith_spot}
		\end{subfigure}
		\bigskip

     \caption{Review of 1U reflective cameras:  In (a) a "nested" Gregorian camera is shown following the afocal telescope.  This design can achieve fairly high F-numbers (as high as F18) and reasonable imaging performance (b) from a compact size. An example of a monolithic camera is shown in (c) and its corresponding imaging performance is shown in (d). All designs in this figure and in figure \ref{fig:1U refractive designs} use the same afocal telescope. }
		\label{fig:1U_reflective designs}  
	\end{figure*}

\begin{figure}
		\centering
    
        \begin{subfigure}{1\linewidth}
			\includegraphics[width=0.9\linewidth]{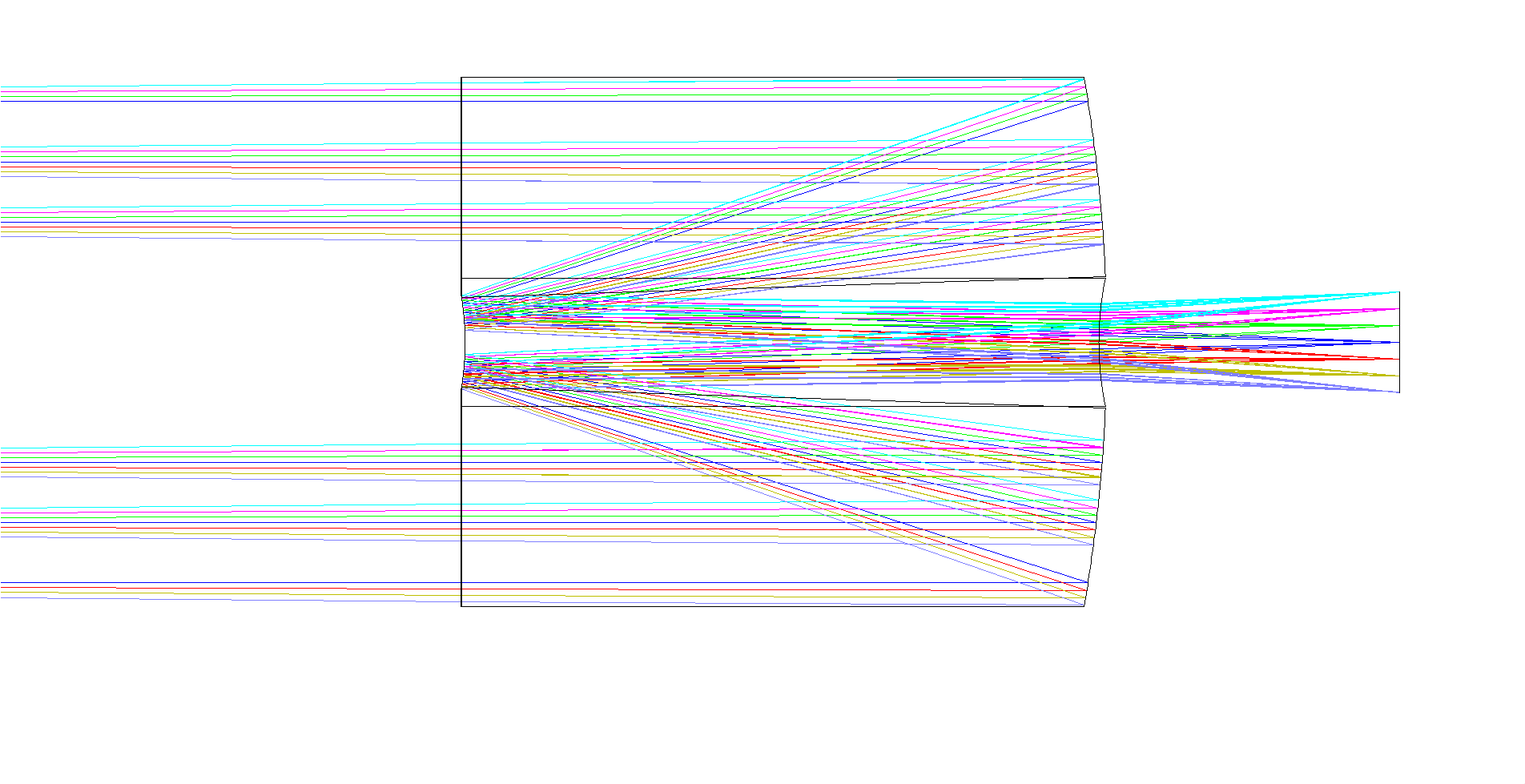}
			\caption{Monolithic camera}\label{fig:monolith_detail}
		\end{subfigure}

        \begin{subfigure}{1\linewidth}
			\includegraphics[width=0.9\linewidth]{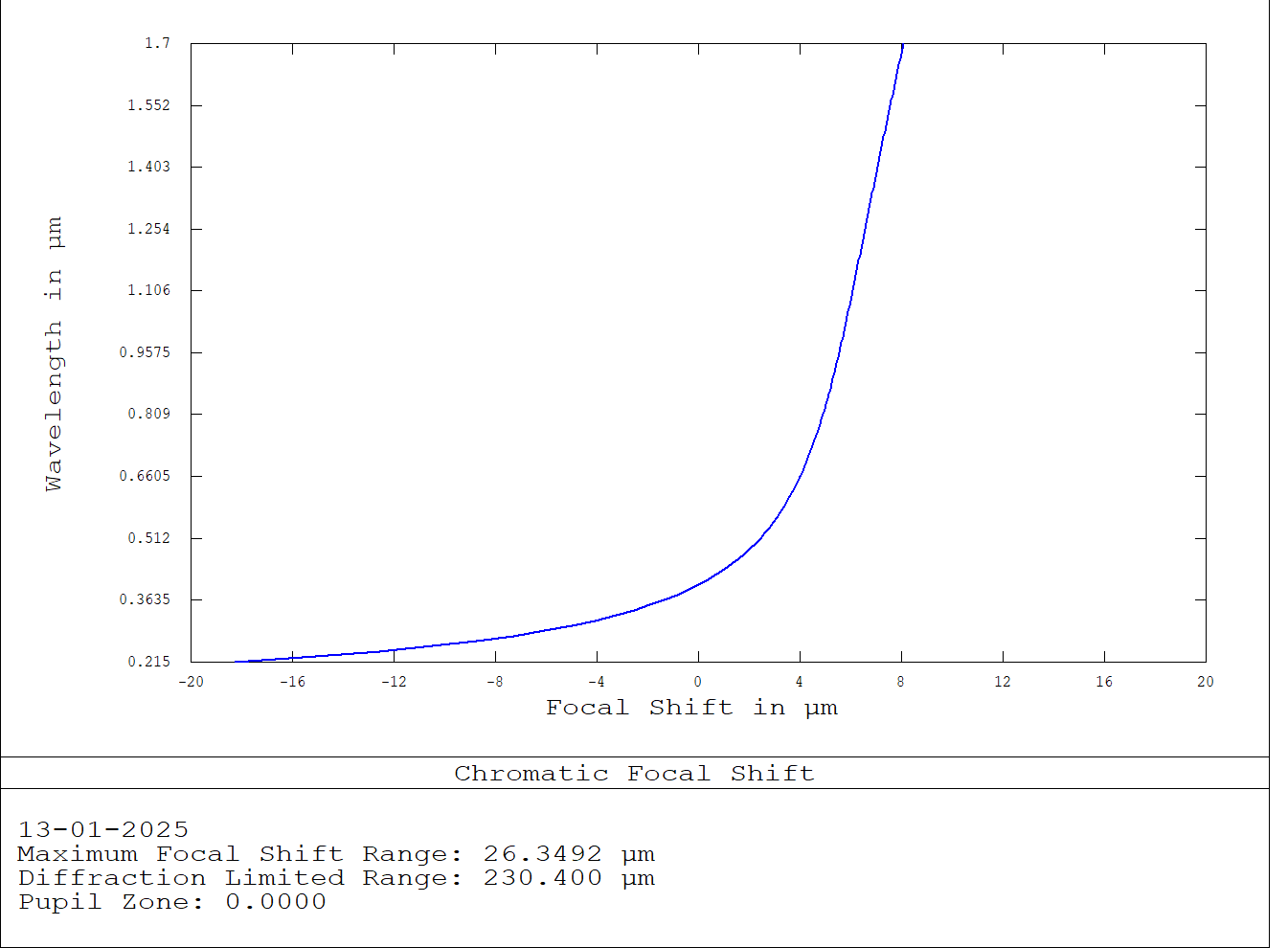}
			\caption{Chromatic shift in the monolith design}\label{fig:focal_shift_monolith}
		\end{subfigure}
        
		\begin{subfigure}{1
        \linewidth}
			\includegraphics[width=0.9\textwidth]{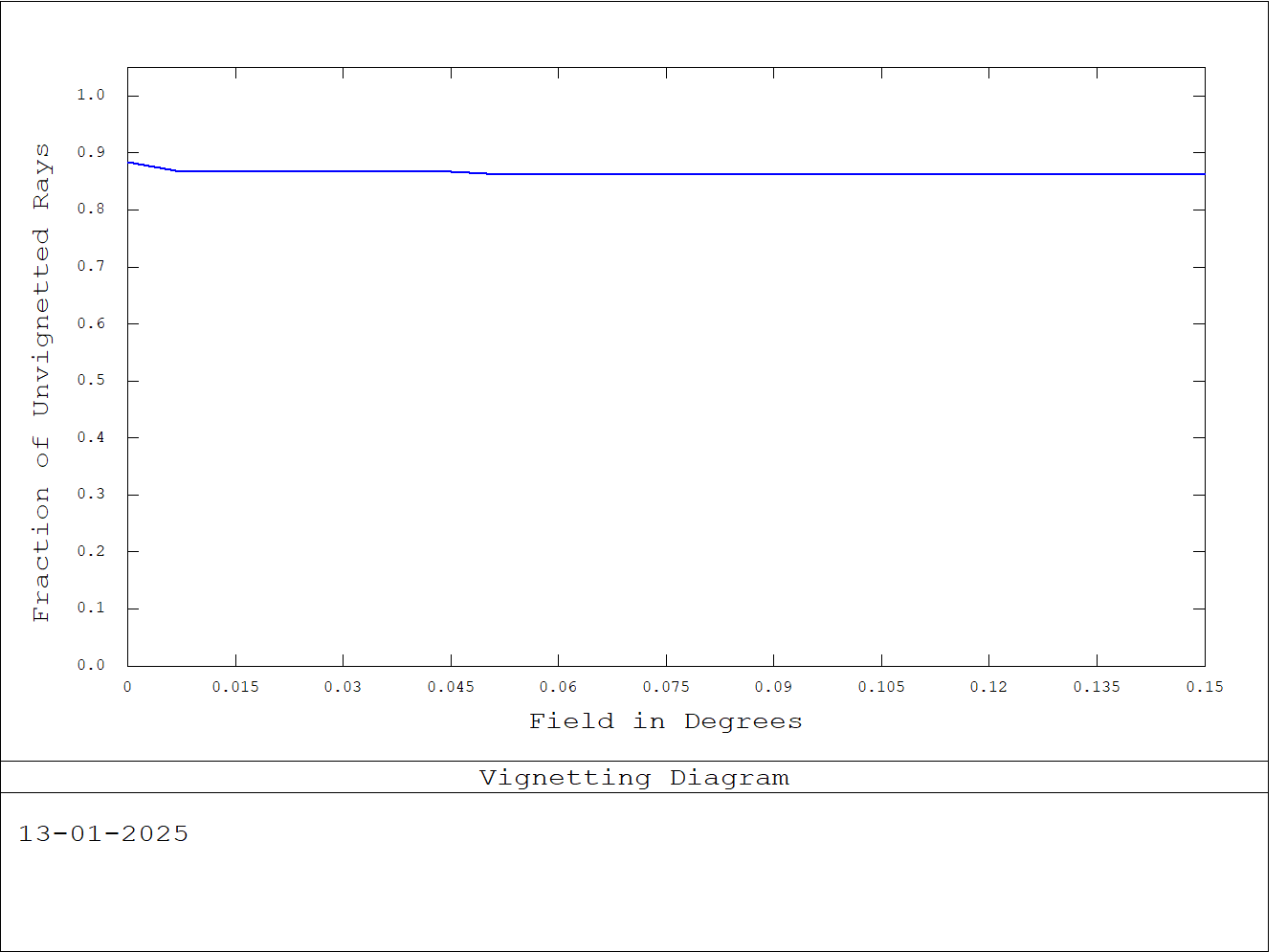}
			\caption{Field dependent vignetting}\label{fig:vignetting_monolith}
		\end{subfigure}
		\bigskip

     \caption{Monolithic cameras for afocal designs: the detail layout of a monolithic camera is shown in (a). The camera consists of an RC telescope "encased" in fused silica. This camera is capable of achieving fairly large focal lengths while being fairly compact. Since, most of the optical power is generated by reflection the design is also relatively free from chromatic aberration (b) and for a well optimised design the field dependent vignetting is also fairly low (c). }
		\label{fig:monolith}  
	\end{figure}

\subsection{Observatory template using 1U aperture telescopes}

A schematic plan for implementing a Cubesat based observatory from 1U aperture telescopes is shown in figure \ref{fig:1U_observatory}. The complete system is to be implemented in a standard 12U  Cubesat. It is of note that the official 12U standard is actually somewhat larger than 2UX2UX3U format and is actually allowed to be of size 226 mm X 226 mm X 366 mm. The template follows this size guideline \footnote{https://www.cubesat.org/}. Figure \ref{fig:1U_topview} shows the top view which describes the allocation of available aperture towards guiding and science instruments. Similarly, figure \ref{fig:1u_3d} and \ref{fig:1u_sideview} illustrate the space allocated for science instruments and for avionics. The complete aperture is divided into a total of 5 sub-apertures. Two of these apertures are reserved for coarse and fine guiding; the rest three aperture are for science focused instruments. Out of the three available science apertures, 2 can be made into 1U X 2U volume while the other one can be only a 1U by 1U to allow for some space for avionic components. The monolithic camera as in figure \ref{fig:design_monolith_guider} is a good match for the 1U camera.

The coarse guiding aperture is allocated to a star-sensor (similar to ones already developed and space qualified at IIA, (\cite{chandra2024low})). This camera working in tandem with the spacecraft Attitude Determination and Control System (ADCS) module must provide initial pointing stability within 2-3 arcminutes. The fine pointing stability is provided by a closed loop control involving a 1U aperture fine guiding camera and individual X-Y translation stages on all apertures. As discussed earlier (section 3), X-Y stage actuators work well for optical designs that have fairly short focal lengths. A number of suitable X-Y stages are shown in table \ref{xyTable} corresponding to different focal lengths of the cameras. It is of note if the required precision is met, then a longer travel stage may be used for a shorter focal length but not vice versa. The fine guiding is to be provided by a 1U guiding camera. Two competing designs are shown in figure \ref{fig:1U_guiders}; one of them uses an afocal telescope followed by a monolithic camera the other uses a catadioptric system consisting of an RC telescope and 3 lens elements.  Both of these designs use the same off-the-shelf primary mirror and are intended for use with the IMX477 sensor from SONY. The monolithic telescope is optimised for a field of view of 30 by 22.5 arcminutes and at an output of F6.5 produces 0.49 arcseconds per pixel on the 1.55 micron pixel of IMX477. The 3 lens catadioptric system is optimised for 40 by 30 arcminutes and produces a F5.25 output and 0.6 arcseconds on the same detector. The pixelscale is an important parameter as this largely defines the fine guiding resolution of these cameras. Comparing the performance of both of these designs, it is seen that the monolithic guider falls marginally short in terms image quality of the 3 lens design. The sensitivity limits of both of these designs in the V band (closely approximated by the G filter of the Bayer pattern) is shown in figure \ref{fig:guiding_limits} and the probability of finding a guide star is listed in table \ref{guideTable} illustrating that these designs are fairly suitable as "all-sky" guiding cameras.

\subsection*{1U template: afocal design examples}

A number of optical designs are presented to illustrate the versatility and flexibility of the afocal template. A few refractive camera designs are shown in figure \ref{fig:1U refractive designs} and fully reflective camera designs are shown in figure \ref{fig:1U_reflective designs}. All the designs presented are of the same outline; an afocal telescope is followed by a suitable camera. The afocal telescope (i.e. the combination of the primary and secondary) itself is fixed across all the designs. Further, the primary mirror is actually an off-the-shelf (from Optisurf\footnote{https://optisurf.com/}) 10 cm aperture F1 primary mirror which is chosen for its fast output in order to save space. 

 Whenever possible, each optical design is also paired with a tentative detector. A list of useful small form factor detector are listed out in table \ref{detectorTable}. In particular a number of sCMOS type detectors from SONY are highlighted. A few of these have interesting wavelength coverages such as IMX487 in NUV and IMX990 in NIR (\cite{smilo2024deep}). Additionally larger arrays of the same family have been evaluated for scientific use and have been known for relatively low read and dark noise and acceptable uniformity across the sensor (\cite{alarcon2023scientific}, \cite{betoule2023stardice}).
 Typically these arrays achieve better than 0.1$e^-/S$ in dark current for modest cooling requirements (about 5$^oC$). Modern CMOS readout circuits are also fairly compact\footnote{https://www.framos.com/en/products/sony-imx541aamj-evb-kit-slvs-ec-c-mount-24063} and are based on standardised interfaces such as SLVS-EC\footnote{https://www.sony-semicon.com/en/technology/is/slvsec.html} which directly connect the detector to an FPGA or controller. A flexible PCB\footnote{https://www.qhyccd.com/qhy-space-camera-series/} is particularly useful as the interface allowing for freedom in detector mount and operation. 
 
 Similarly, a number of compact MCPs as well as suitable CCD/CMOS sensors are also included in the table for the NUV and FUV wavelength range. These MCPs are fairly compact and can be paired with compact high voltage power supplies (e.g. C10940 and C11152 from Hamamatsu). There has been progress on readout of MCPs by using simple Raspberry Pi based cameras at IIA (\cite{chandra2024development}). However, MCPs typically have low quantum efficiency and one may consider replacing them with some modern CCD/CMOS type detectors for certain wavelength regions. The short wavelength limit of these detectors is typically decided by the window material. Hamamatsu offers some of the CCDs without a window\footnote{https://www.hamamatsu.com/content/dam/hamamatsu-photonics/sites/documents/99\_SALES\_LIBRARY/ssd/uv\_koth0022e.pdf}, and can achieve close to 30-40 \% efficiency at 130 nm and about 60-70 \% at 150 nm. \\

The design in figure \ref{fig:1U_VIS} is a visible camera. The design is based on a 3 lens system. The output is  F7 and is fairly well optimised for a field of view of 30 arcminutes and a wavelength range of 450 to 800 nm. The primary use of such a camera would be as a general purpose wide-field camera. The output matches well with sensors such as IMX477 \footnote{www.raspberrypi.com/products/raspberry-pi-high-quality-camera/} or IMX226. A nominal pixel scale of 0.45 arcseconds per pixel is achieved for the IMX477 sensor. The spot diagram of this design is shown in figure \ref{fig:1U_VIS_spot}. This design is optimised to be close to the diffraction limit for the entire field of view as well as the full wavelength range of sensitivity.\\

Similarly, in figure \ref{fig:1U_NIR_dio} an NIR imager is presented. This NIR imager is designed with an InGaAs sensor such as SONY-IMX992 in mind. This detector has a pixel size of 3.45 microns and is available in formats up to 2.5K by 2K. The wavelength sensitivity range is typically from 600-1750 nm while good quantum efficiency is achieved from 900 nm to 1700 nm. Modern InGaAs arrays are quite capable for astronomical and scientific applications (\cite{lourie2020wide}, \cite{smilo2024deep}, \cite{yang2025first}). The outstanding feature of InGaAs arrays when compared to H2RG detectors is that InGaAs arrays typically can be operated with minimal cooling requirement.  Simple Peltier cooing is often sufficient and the weight and complexity of a cryogenic system is avoided. This makes them a simple and affordable alternative (\cite{batty2022laboratory}). Additionally, these detectors also do not have issues such as ghost images and saturation. Although these arrays typically have relatively higher dark current (about 10-20 $e^-/S$ at about $-30^o C$\footnote{https://www.zwoastro.com/product/asi990mm-pro-asi991mm-pro/}), they are quite suitable for imaging and photometry applications(\cite{mishra2022filters}) of brighter sources. The optical design for a 1U NIR imager is implemented in a similar fashion as the visible imager, however, the pixel size of IMX992 requires for an output of F9. The layout and spot diagram are shown in figure \ref{fig:1U_NIR_dio} and figure \ref{fig:1U_NIR_dio_spot}. The pixel-scale is 0.8 arcseconds which is tuned to diffraction limit of the 1U aperture for the NIR wavelength range. The design is optimised for a field of view of 30 arc-minutes and for a wavelength range of 850 nm to 1700 nm. \\

\begin{table*}[htb]
\tabularfont
\caption{A sample of detectors that would be suitable for Cubesat based implementations is presented. The sensors either have relatively small pixels or a small format. The specifications of the detector as well as some target use cases are listed.  It should be kept in mind these are selected for the utility and opportunity they provide for a Cubesat format -- particularly within the lower radiation levels of the low-earth orbit (\cite{kimura2015optical})-- and are not necessarily radiation hardened for deep space applications.}\label{detectorTable}
\renewcommand{\arraystretch}{2}
\begin{tabular}{lcc|ccc|r}
\topline

Sl.no&Model&Manufacturer & Pixel size and Format& Wavelength &Read-noise&Features and Usage\\
\midline
1&IMX226&SONY & 1.85$\mu$m; 4K$\times$3K &400-900nm & 3.7$e^-$& VIS imager   \\
2&IMX487&SONY & 2.75$\mu$m; 2.8K$\times$2.8K &200-900nm & 3.7$e^-$& NUV   \\
3&IMX992&SONY & 3.45$\mu$m; 2.6K$\times$2K &600-1700nm & 20$e^-$& NIR imager  \\
4&IMX990&SONY & 3.45$\mu$m; 1.3K$\times$1K &600-1700nm & 20$e^-$& NIR imager  \\

5&S7170-0909&Hamamatsu & 24$\mu$m; 0.5K$\times$0.5K&200-1000nm & 8$e^-$&FUV and NUV  \\
6&S7031-0907S&Hamamatsu &24$\mu$m; 0.5K$\times$0.1K & 200-900nm&8$e^-$ & FUV spectroscopy \\
7&S10141-1107S&Hamamatsu & 12$\mu$m; 2K$\times$0.1K& 200-1000nm&4$e^-$ & NUV spectroscopy \\
8&2020BSI&GSENSE& 6.5$\mu$m; 2K$\times$2K &200-900nm & 1.2$e^-$&NUV-VIS imager
\\
9&F1552-04&Hamamatsu&7.5$\mu$m;14mm dia &-- &-- & MCP  \\
10&CIS115&e2V& 7$\mu$m; 2K$\times$1.5K &400-900nm & 5$e^-$& VIS imager  \\
11&IMX477 &SONY & 1.55$\mu$m; 4K$\times$3K &400-900nm & $\approx$3$e^-$& Fine guiding  \\
12&MT9P031& On Semi& 2.22$\mu$m; 2.5K$\times$1.9K &400-900nm & --& Fine guiding  \\
\hline
\end{tabular}
\tablenotes{IMX477 is commonly available as the Raspberry PI HQ camera module.}
\tablenotes{IMX* series sensors have a different readout noise depending on the exact conversion gain used, the mentioned read noise is for high gain count mode.}
\end{table*}

The design in figure \ref{fig:1U_greg} is  an all reflective camera implemented by a classical Gregorian type telescope. This results in a very compact layout that produces a fairly slow output of F18 while being compact enough to fit in a total volume slightly larger than 1U. This systems achieves a focal length  as high as 1800 mm and hence allows for the use of larger pixel size detectors. If the supports for the nested Gregorian can be hidden in the shadow of the afocal secondary, then this design is a useful general purpose imager. Additionally, this designs remains fully rotationally symmetric. This feature is particularly desirable for polarimetric instruments, this will be briefly discussed at a later stage (section 5). The spot diagram for this design is shown in figure \ref{fig:1U_greg_spot}.\\

 Among the designs, figure \ref{fig:1U_monolith}  is particularly special as it implements the complete camera optic chain within one single piece of glass and thereby presents a unique solution. The "monolithic telescope as introduced by Rik Ter Horst (\cite{ter2024monolithic}) consists of a Schmidt corrector, a classical RC telescope as well as a field flattener all within a single block of fused silica. The version shown in figure \ref{fig:1U_monolith} (and shown in detail in figure \ref{fig:monolith_detail}) is a slightly simplified without the Schmidt corrector. The exit surface that is utilised as a field flattener is also spherical. The example shown in figure \ref{fig:1U_monolith} produces a F12 output and an image size of 6.25 mm for a field of view of 18 arc-minutes. The combination spot diagram for the same is shown in figure \ref{fig:1U_monolith_spot}. Although the spot performance is slightly degraded from the original monolithic telescope described by \cite{ter2024monolithic} the performance is still acceptable for the considered field of view. Additionally, since most of the optical power is generated from the reflection at the mirrored surfaces, the chromatic aberration is very small within the whole NUV-VIS-NIR range. Chromatic focal shift for the  design is shown in figure \ref{fig:focal_shift_monolith}.  There is some amount of field dependent vignetting  as shown in figure \ref{fig:vignetting_monolith}. The un-vignetted fraction of rays for the central field is about 89\% -- the 11\% drop is largely from the secondary obscuration -- and it drops to about 85\% for a field of about 18 arc-minutes. This is a particularly promising implementation of a camera in a rugged, thermally insensitive and compact package. At present, some manufacturing concerns remain for such an unconventional design, however, the benefits of the monolithic approach is apparent for Cubesats and correspondingly there have been efforts to realize these designs (\cite{smilo2024deep}).\\

\begin{figure}[!ht]
\centering

  \begin{subfigure}{0.49\textwidth}
			\includegraphics[width=1\linewidth]{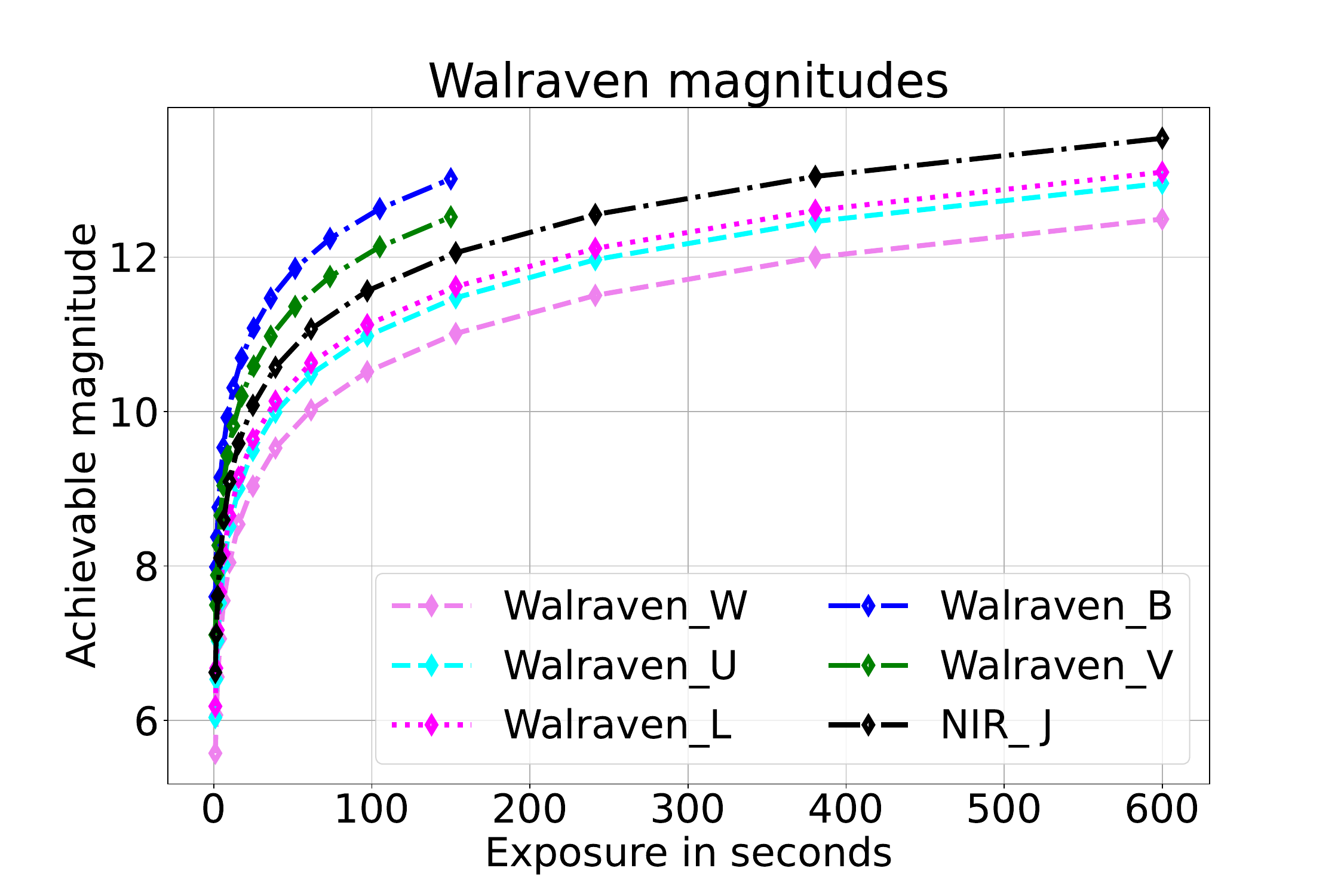}
			\caption{Walraven SNR limits}\label{fig:SNR_walraven1}
  \end{subfigure}

		\begin{subfigure}{0.47
        \textwidth}
			\includegraphics[width=1\linewidth]{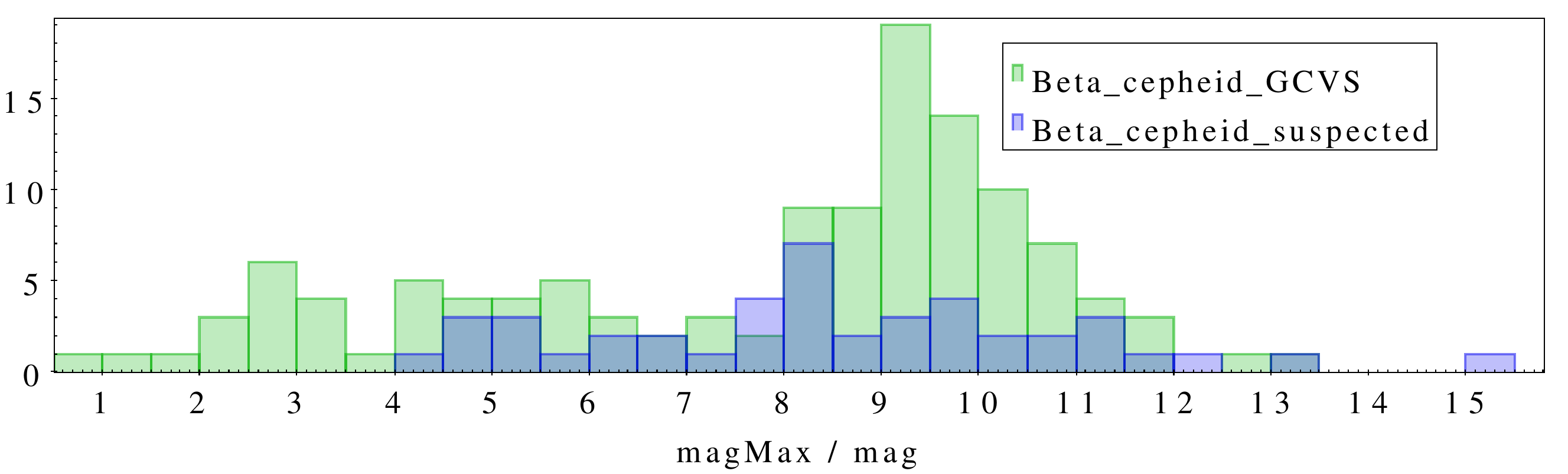}
			\caption{Brightness distribution of Beta Cepheids}\label{fig:dist_Bcep}
		\end{subfigure}
		\bigskip
        
       \begin{subfigure}{0.47
        \textwidth}
			\includegraphics[width=1\linewidth]{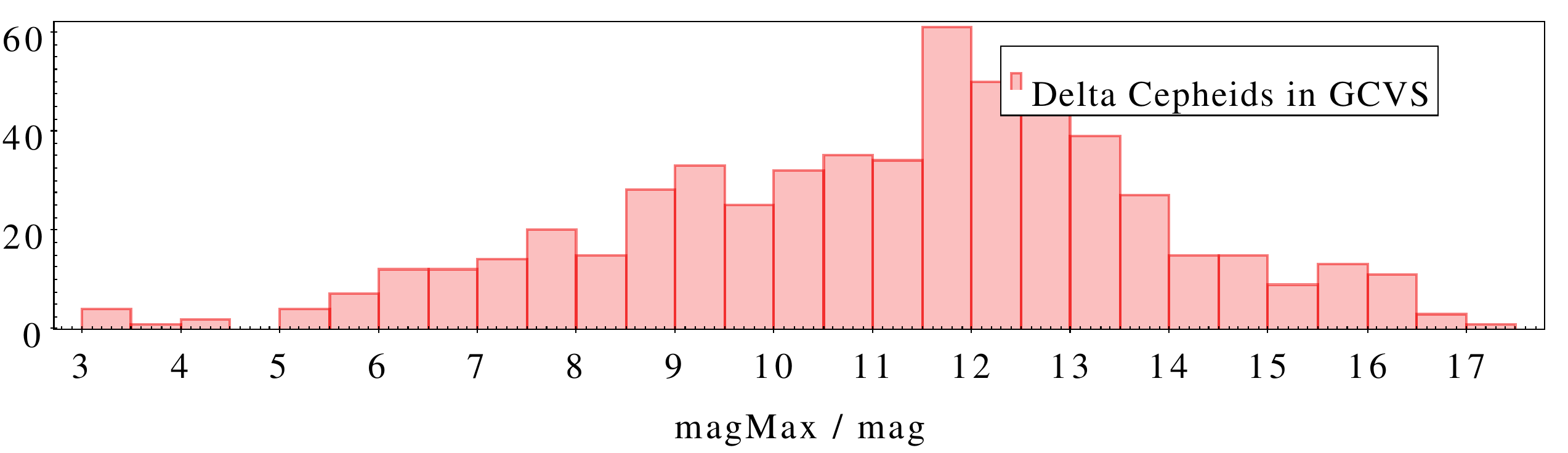}
			\caption{Brightness distribution of $\delta$Cepheids}\label{fig:dist_dcep}
		\end{subfigure}
		\bigskip

     \caption{Limiting magnitudes versus integration time in the Walraven photometric system from a 1U aperture is shown in (a). Luminosity distribution (from GCVS, \cite{samus2017general}) of a few $\beta$cepheids is shown in (c) and of classical cepheids is shown in (d). The magnitude limits are for an photon noise limited SNR of 100 and are calculated by assuming nominal values of throughput (typically 35-40\%), quantum efficiency (50\% in NUV and 70-80\% in VIS-NIR). SED distributions are from (\cite{fitzpatrick2010uv})for NUV wavelength range and from (\cite{zombeck2006handbook}) for VIS-NIR wavelength range. }
		\label{fig:SNR_walraven}  
\end{figure}

\subsection*{Example science case: Walraven photometry}
\begin{figure}[!ht]
		\centering

		\begin{subfigure}{0.45
        \textwidth}
			\includegraphics[width=1\linewidth]{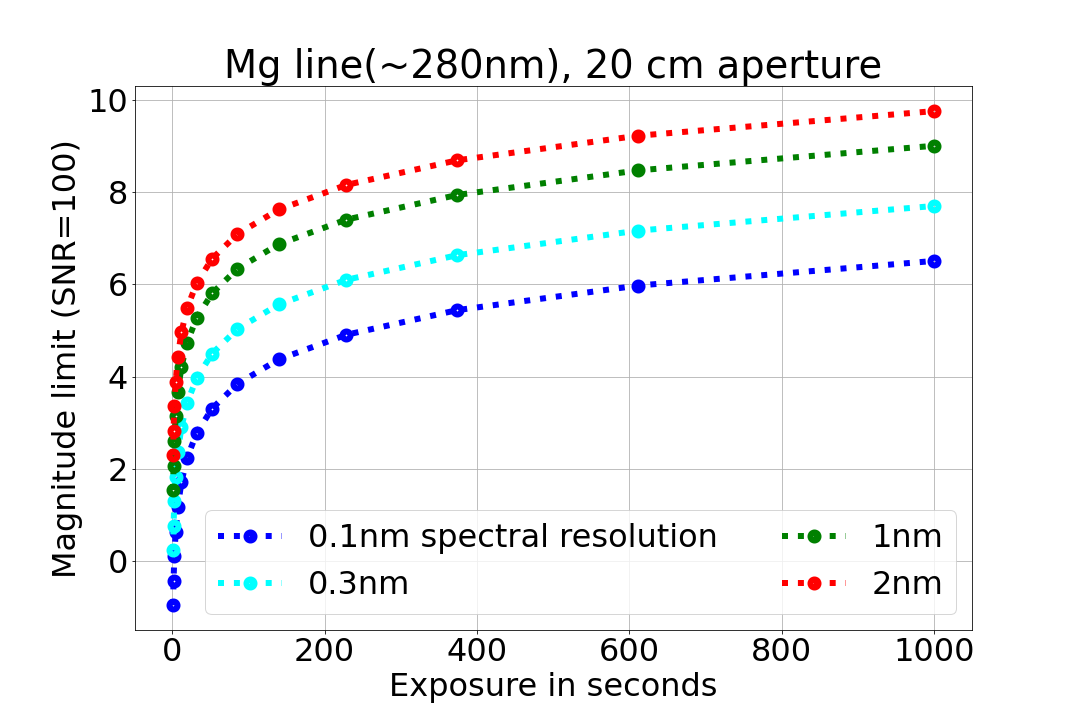}
			\caption{Achievable magnitude limits for 2U aperture}\label{fig:SNR_resolution}
		\end{subfigure}
		\bigskip

        \begin{subfigure}{0.45\textwidth}
			\includegraphics[width=1\linewidth]{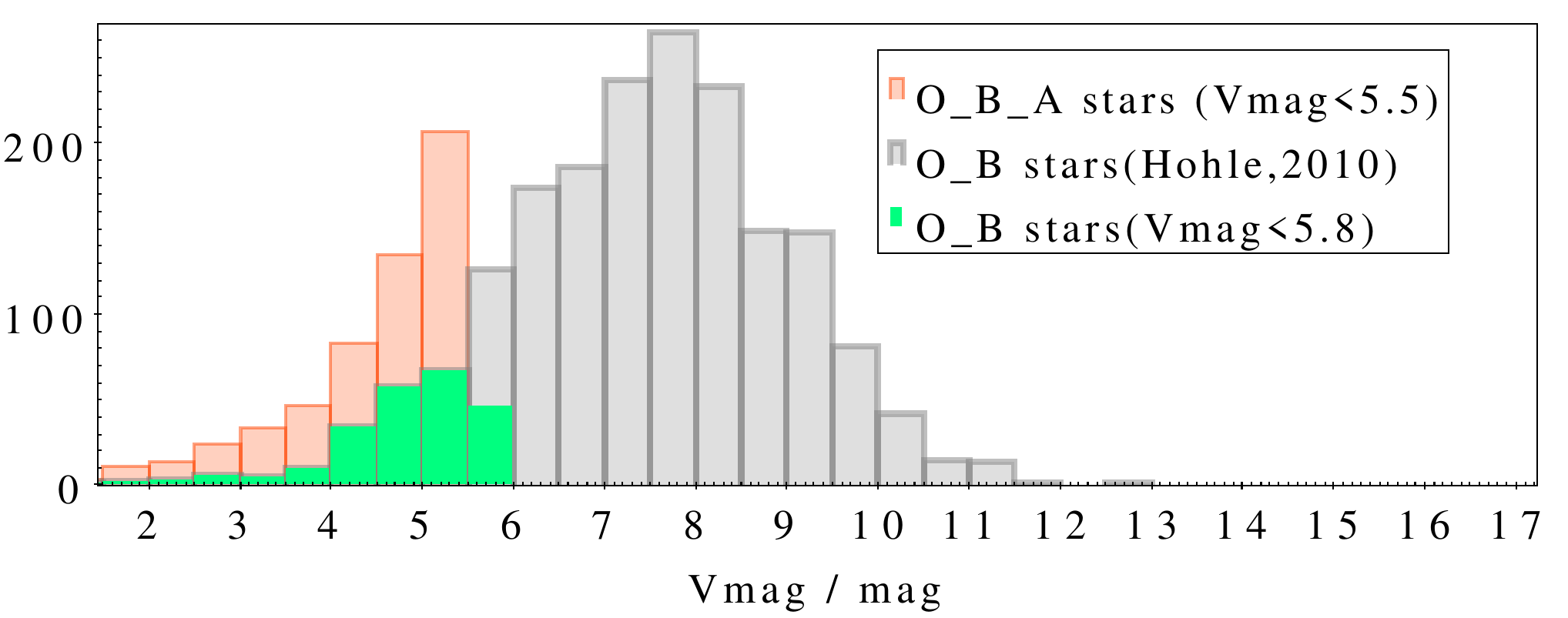}
			\caption{Brightness distribution of O and B type stars.}\label{fig:dist_O_B}
		\end{subfigure}

        \begin{subfigure}{0.45
        \textwidth}
			\includegraphics[width=1\linewidth]{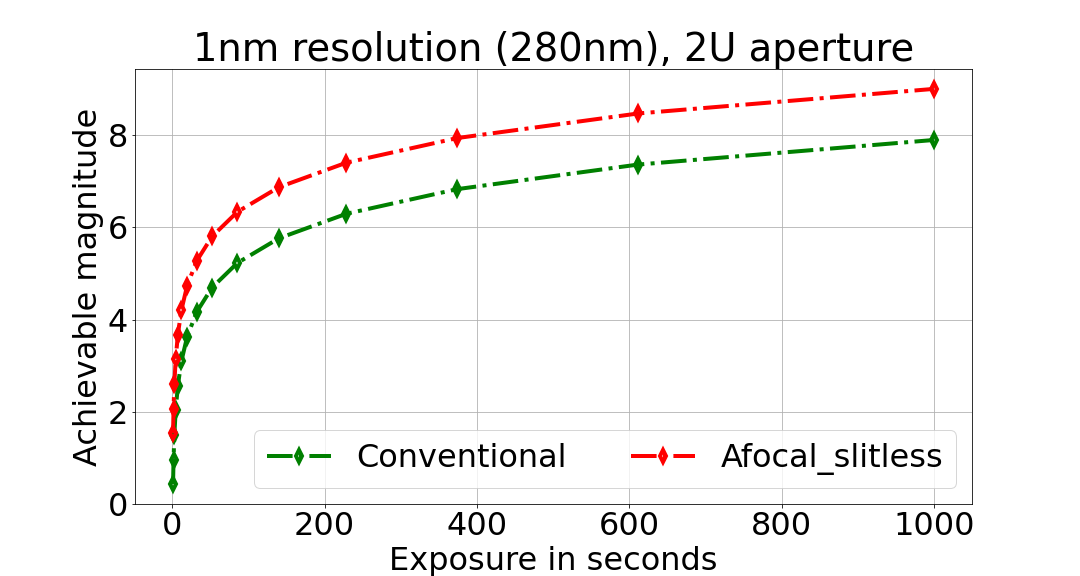}
			\caption{SNR advantage with an Afocal slit-less spectrograph}\label{fig:SNR_afocal}
		\end{subfigure}

     \caption{The limiting magnitudes shown in (a) are for an SNR of 100. The estimations are for nominal efficiencies such as 50\% detector quantum efficiency, 60\% grating efficiency and 25\% optical throughput. The Spectral energy distribution is collected from \cite{fitzpatrick2010uv}. In (b) the luminosity distribution of early type stars (from \cite{hohle2010masses} and \cite{warren1987bright}) is presented. Finally in (c) the throughput advantage of an afocal design when compared to a conventional design is shown.
     }
		\label{fig:SNR_spec}  
\end{figure}

The Walraven photometric system is an unique ground based system that expands from visible into ultraviolet wavelength ranges. Typically it consists of 5 photometric filters ( 542 nm (V), 427 nm (B), 385 nm (L), 362 nm (U) and 323 nm (W)). This photometric system (\cite{walraven1952new}) has been known to be excellent in determining cepheid mass and radius  \cite{sollazzo1981cepheid}). Colour-colour loops in U-B versus W-U (or similar) can be a systematic as well as convenient way of determining physical parameter for pulsating variables (\cite{onnembo1985importance}). Similarly, the Baade-Wasselink method (\cite{molinaro2011cors}) in the Walraven photometric system is also particularly suitable for period-luminosity and period-luminosity-colour relations. As the NUV wavelength range is particularly sensitive to pulsation driven shocks and heating of gas, observations in this wavelength range is highly sought after. However, ground based Walraven photometry is typically limited to fairly bright sources (3-5th magnitude) only, as atmospheric transmission is fairly poor in U and W passbands. Additionally, the Walraven photometer has been retired as of 1991\footnote{https://www.eso.org/public/teles-instr/lasilla/09metre/walraven/}. A Cubesat based platform can extend the capability of these systems to 8-10th magnitudes. For example, limiting magnitudes for SNR 100 in figure \ref{fig:SNR_walraven1}, along with luminosity distribution of $\beta$ cepheids (figure \ref{fig:dist_Bcep}) and classical cepheids(figure \ref{fig:dist_dcep}). This allows for dedicated study of a significant number of such variables providing for an independent estimate of the mass and radius. This photometric system is also particularly suited for studying early type stars (\cite{van1986five}), cataclysmic variables (\cite{van1987multiwavelength}) as well as novae and supernovae. Magnitude distributions of  some bright early type stars is shown in figure \ref{fig:dist_O_B}.

\begin{figure*}
		\centering
    
        \begin{subfigure}{0.52\linewidth}
			\includegraphics[width=1\linewidth]{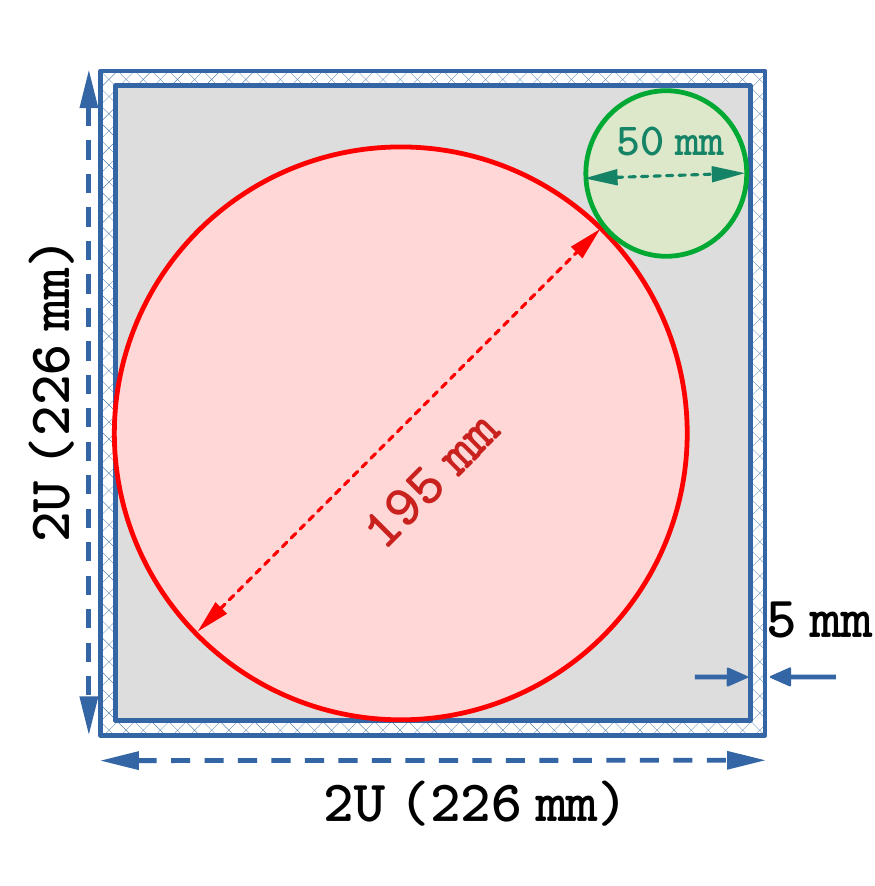}
			\caption{Top view}\label{fig:2U_topview}
		\end{subfigure} 
        \begin{subfigure}{0.46\linewidth}
			\includegraphics[width=1\linewidth]{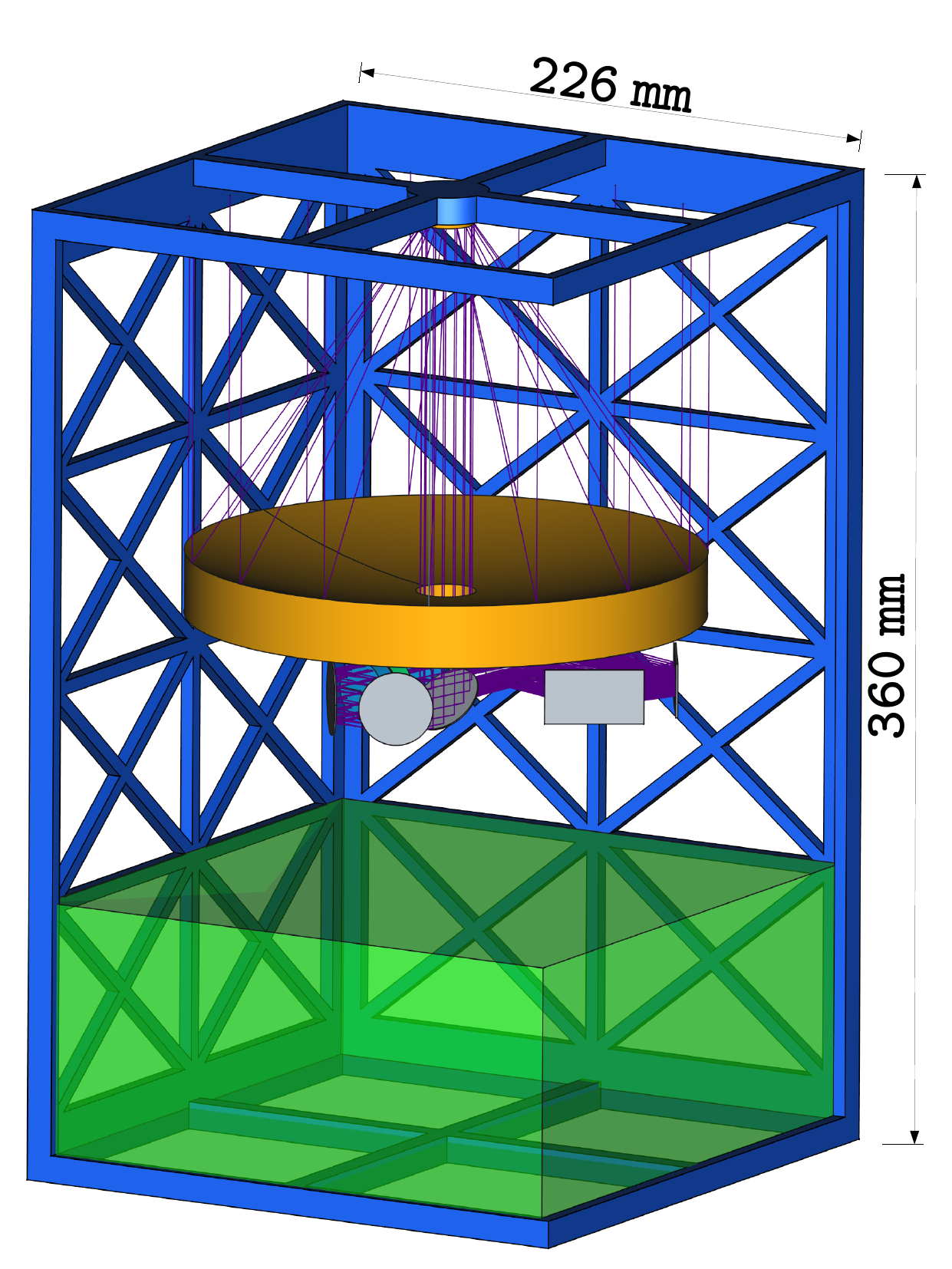}
			\caption{Side view }\label{fig:2u_sideview}
		\end{subfigure}

      \begin{subfigure}{0.49\linewidth}
			\includegraphics[width=1\linewidth]{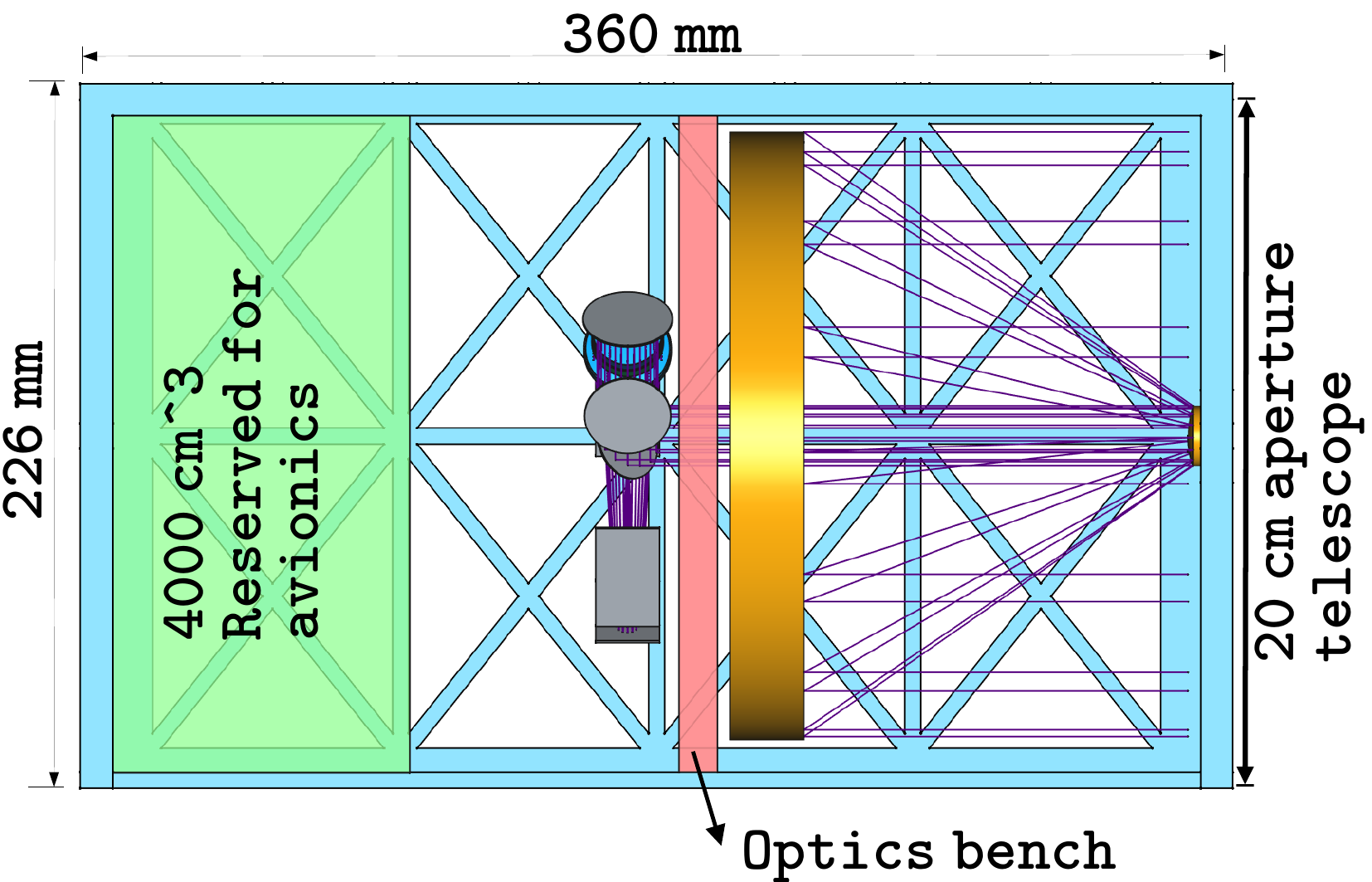}
			\caption{Side view }\label{fig:2u_bench}
		\end{subfigure}
        \begin{subfigure}{0.49\linewidth}
			\includegraphics[width=1\linewidth]{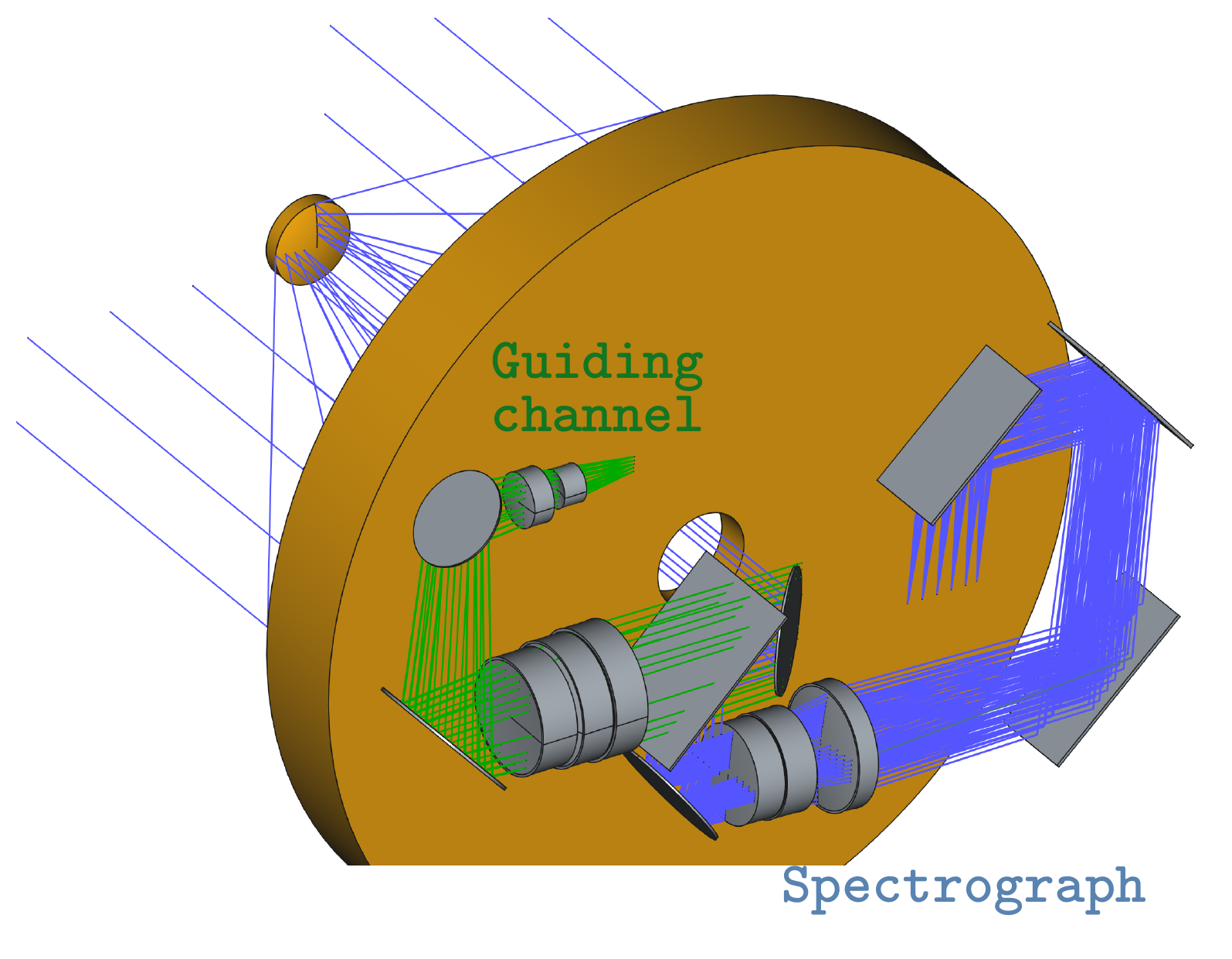}
			\caption{Side view }\label{fig:2u_layout}
		\end{subfigure}
     \caption{ Plan for a Cubesat observatory utilising a 20 cm (practically 195 mm) aperture telescope. The entire observatory fits within a 12U volume, of which about 8U is reserved for the optical telescope and components. Aperture distribution is shown in (a). Space distribution within the 12U volume is shown in (b) and (c). An illustrative combination of a guiding channel and a science channel (NUV spectrometer)  is shown in (d). These optical designs are discussed in detail in figure \ref{fig:2U_guiders} and figure \ref{fig:design4b}. }
		\label{fig:2U_observatory}  
	\end{figure*}

\begin{figure*}
		\centering
        \begin{subfigure}{0.44\textwidth}
			\includegraphics[width=1.\linewidth]{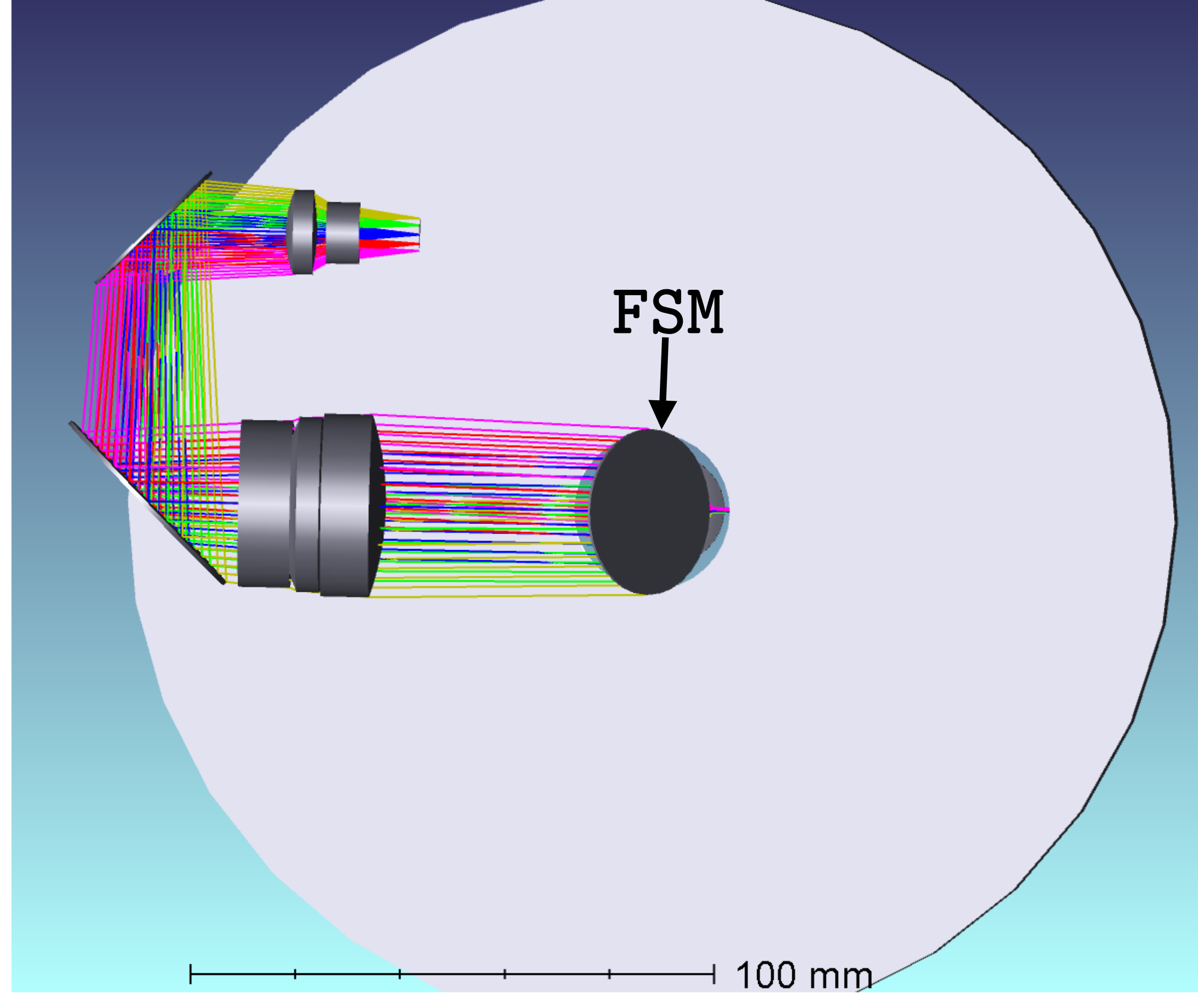}
			\caption{2U guiding camera}\label{fig:design4a}
		\end{subfigure}
		\begin{subfigure}{0.55
        \textwidth}
			\includegraphics[width=1\linewidth]{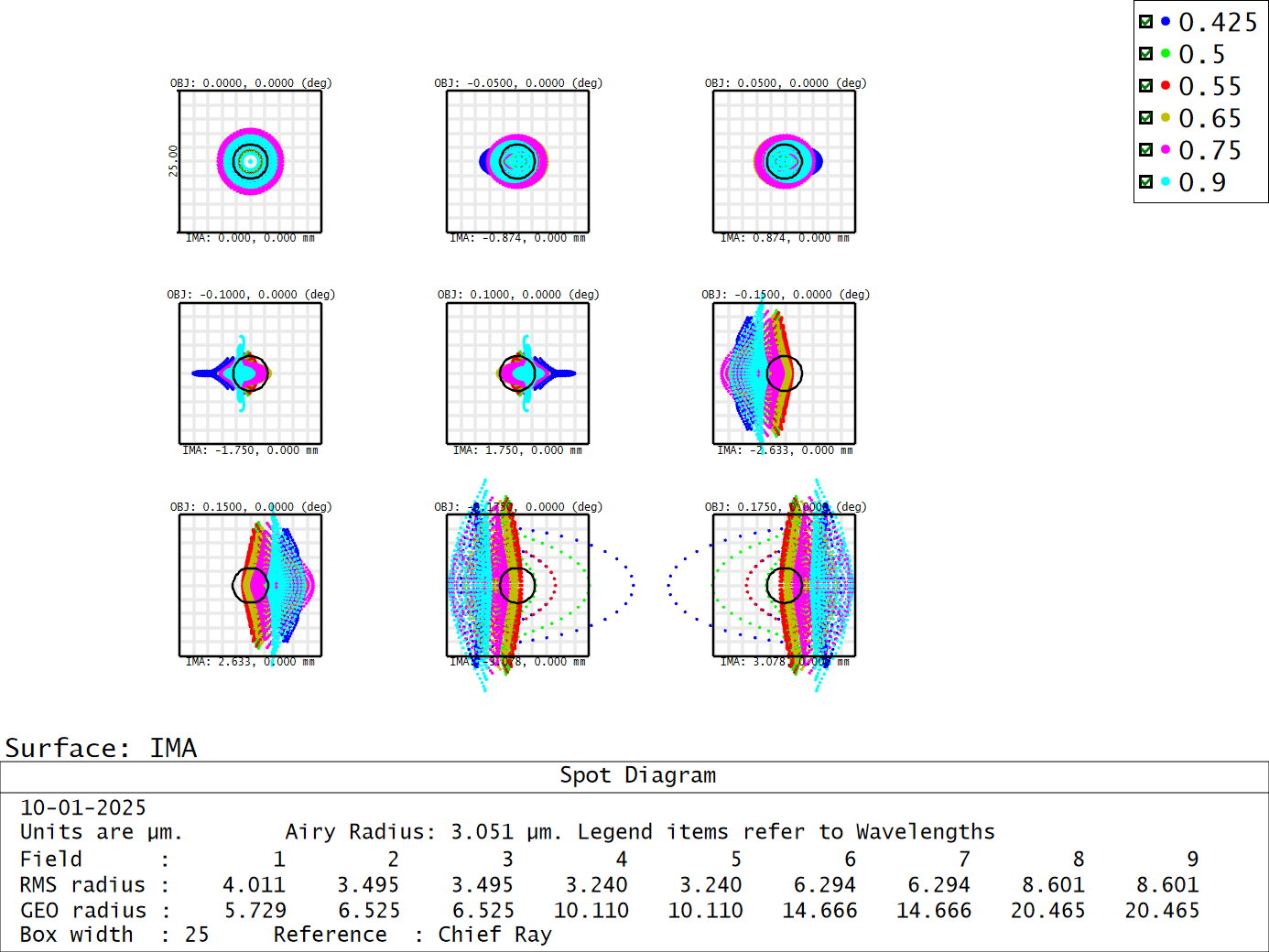}
			\caption{Spot for 2U guiding camera}\label{fig:spot4a}
		\end{subfigure}
		\bigskip

     \caption{2U guider camera : layout for a guide camera for the 2U design template is shown in (a). The guide camera is dichroically split from the main science camera. The camera design is done keeping the IMX477 sensor (similar to the 1U guiders) and has a fairly fast F5 output to accommodate the small 1.55 micron pixels of the sensor. The field of view is about 24 arcminutes. The imaging performance of the camera is shown in (b) showing reasonably good optimisation for the required field of view. The sensitivity limit of this camera is shown in figure \ref{fig:guiding_limits}and the probability of finding a guide star is listed in table \ref{guideTable}.}
		\label{fig:2U_guiders}  
	\end{figure*}

\begin{figure*}
		\centering

        \begin{subfigure}{0.44\textwidth}
			\includegraphics[width=1\linewidth]{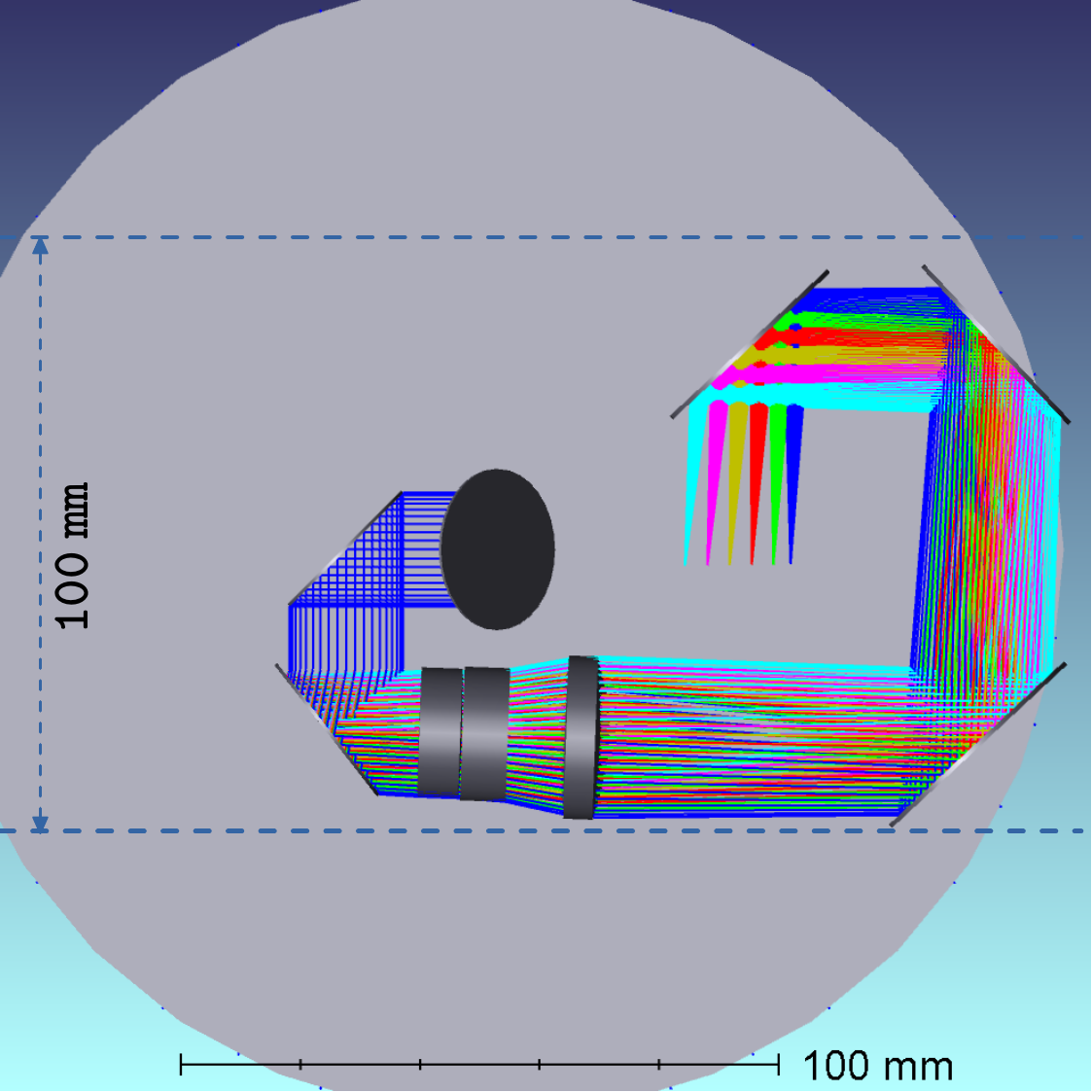}
			\caption{2U NUV spectrograph}\label{fig:design4b}
		\end{subfigure}
		\begin{subfigure}{0.55
        \textwidth}
			\includegraphics[width=1\linewidth]{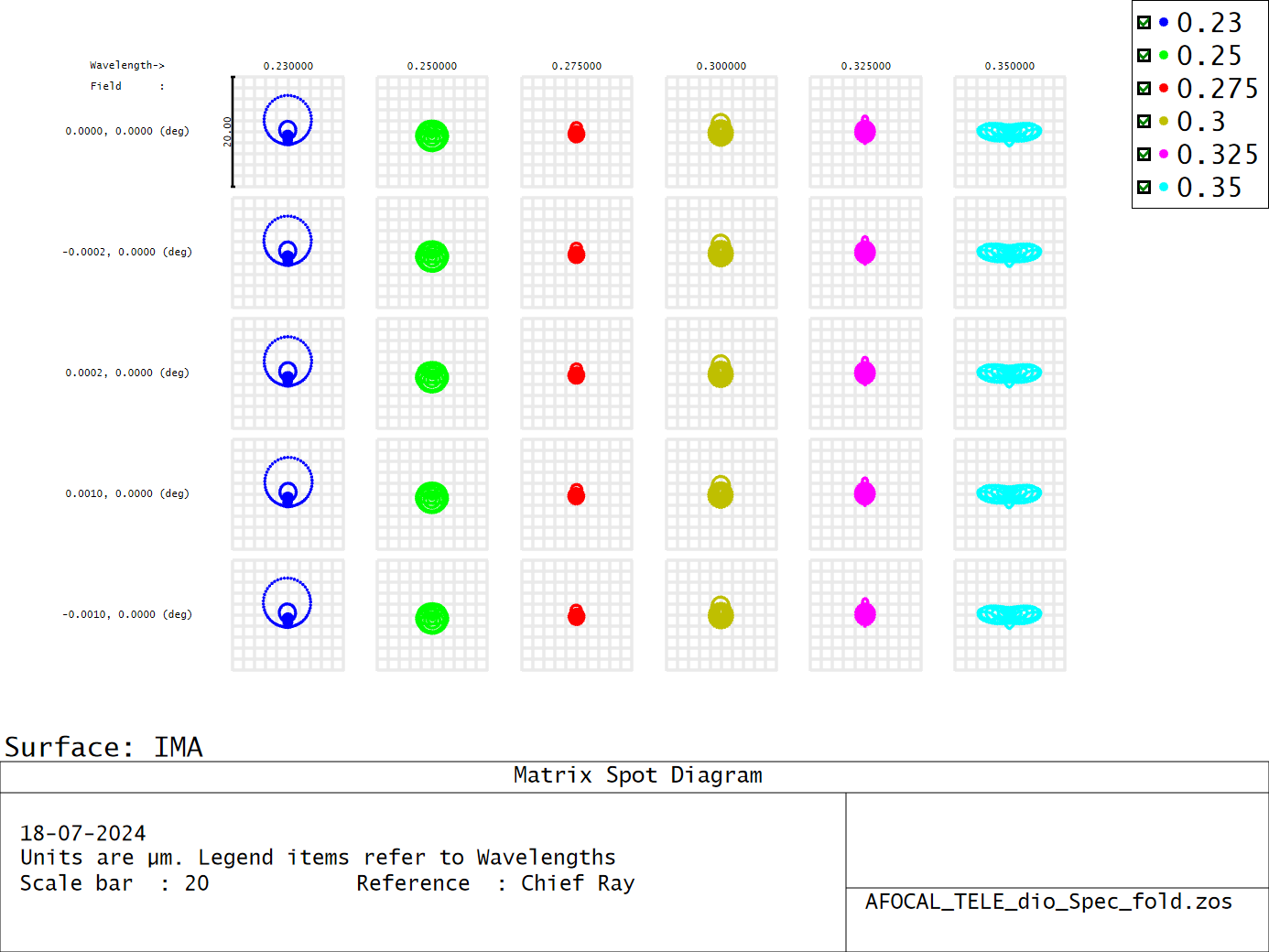}
			\caption{Spot for 2U NUV spectrograph}\label{fig:spot4b}
		\end{subfigure}
		\bigskip

        \begin{subfigure}{0.44\textwidth}
			\includegraphics[width=1\linewidth]{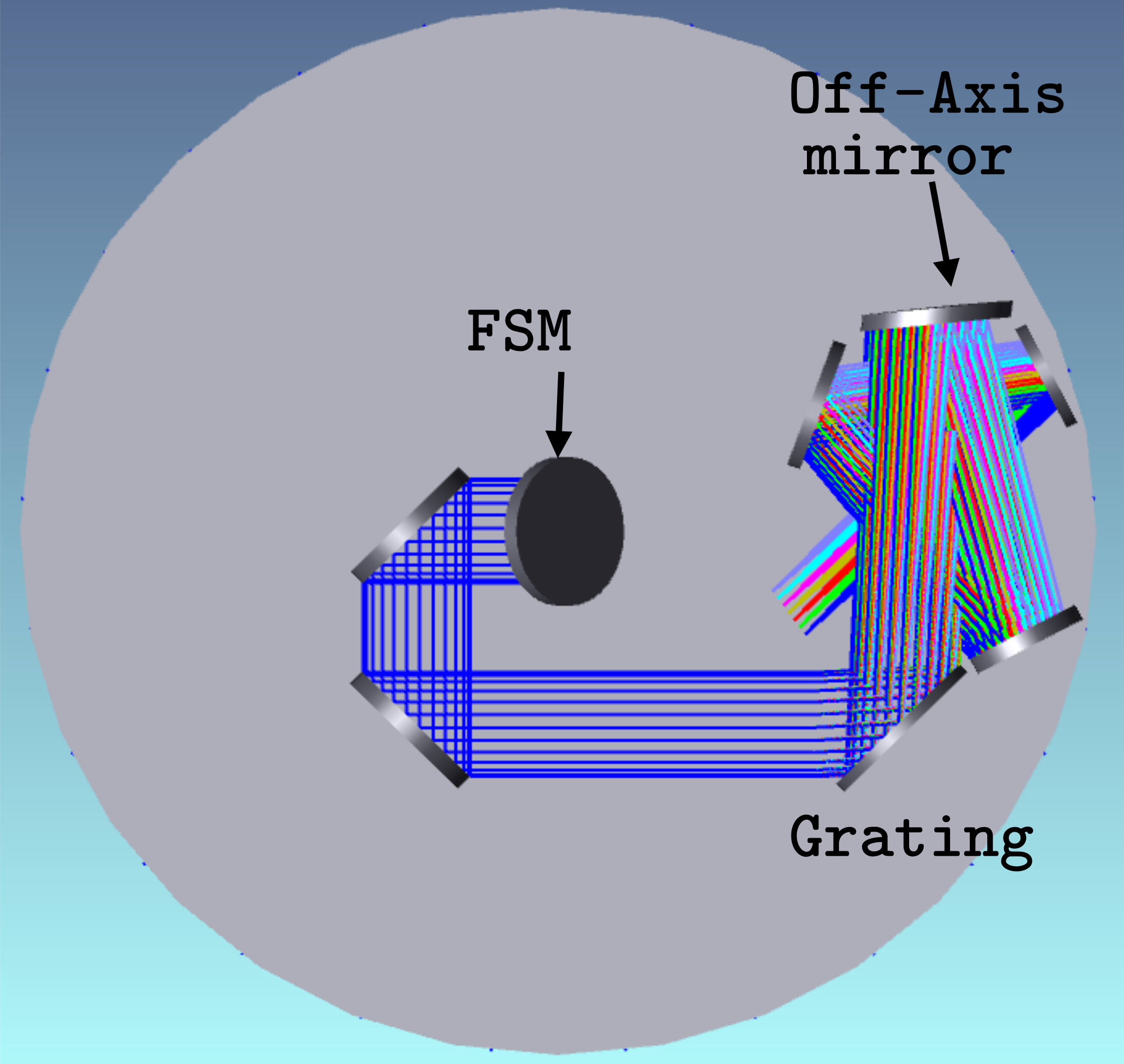}
			\caption{2U FUV spectrograph}\label{fig:design4c}
		\end{subfigure}
		\begin{subfigure}{0.55
        \textwidth}
			\includegraphics[width=1\linewidth]{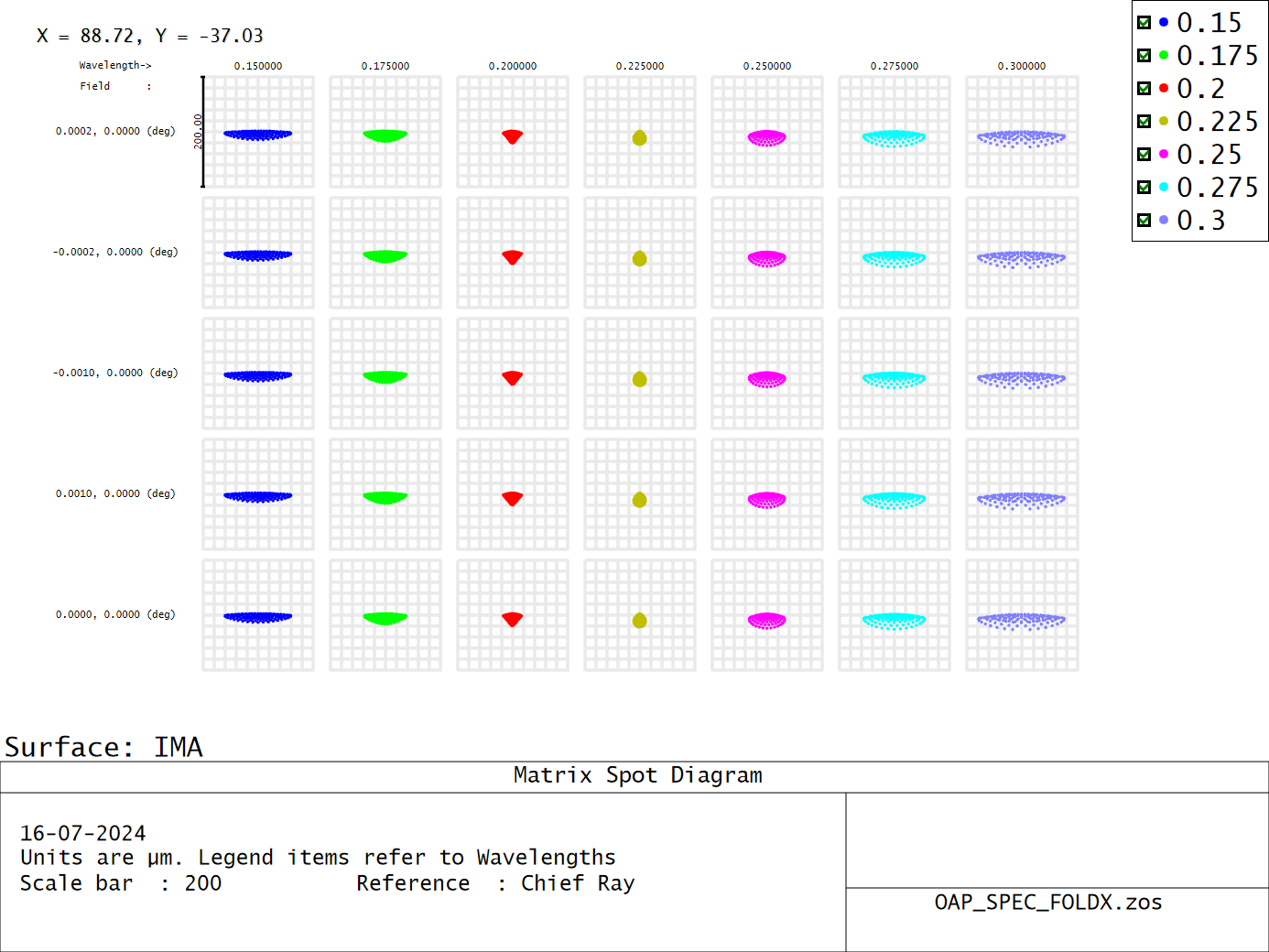}
			\caption{Spot for 2U FUV spectrograph}\label{fig:spot4c}
		\end{subfigure}
		\bigskip

     \caption{Review of 2U spectrographs: Layout of a NUV and a FUV spectrograph is shown in (a) and (C). The NUV spectrometer covers 230 nm to 350 nm and is based on three lens design. The spot diagram for the same is shown in (b); illustrating typical line spread within 15-20 microns. Similarly the FUV spectrograph covers 150 to 300 nm in wavelength range has reasonable line spread as shown in (d). Both of these designs are to be used with the guider design shown in figure \ref{fig:2U_guiders} and use the same afocal telescope i.e. the same primary and secondary combination.}
		\label{fig:2U_spectrographs}  
	\end{figure*}

\subsection{Observatory template using 2U aperture telescopes:}

 Another conceivable template within the 12U format would be one that makes use of one single 20 cm aperture.  This version has more light gathering power allocated to one single instrument and is more suitable for observing astronomical sources in one specific manner. The available aperture for a 12U type optical system is shown in figure \ref{fig:2U_topview}. The aperture is mostly utilised by a "2U" aperture telescope as well as a coarse guiding camera. All designs under this template share the same afocal telescope. The primary mirror is an off-the-shelf 20 cm aperture and F0.8 mirror from Optisurf. The telescope and scientific instrumentation are allowed to take up about 8U of the total volume while 4U is reserved for the avionics; as illustrated in figure \ref{fig:2u_sideview} and \ref{fig:2u_bench}. The fine guiding is done by dichroically splitting the telescope output into a guiding channel and a science channel (figure \ref{fig:2u_layout}). A simple guiding camera for this format is shown in figure \ref{fig:2U_guiders}. This guiding camera is also based around the Raspberry PI HQ camera module (IMX477 sensor). This camera is realised by means of 5 lens elements and produces a fast F5 output; the layout for the same is shown in \ref{fig:design4a}. The camera is optimised for a field of view of about 24 arcminutes by 18 arcminutes. The spot performance of this camera is shown in \ref{fig:spot4a} and as shown in table \ref{guideTable} this camera has a probability of over 0.9 to find at least one guiding star within the field of view. The SNR estimates are shown for 500 millisecond integrations and for fairly sparse fields (e.g. towards the Galactic poles) and hence this probability can be taken as a conservative lower bound.\\
   
 The fine pointing stability is achieved by means of Fine Steering Mirror (FSM) as shown in figure \ref{fig:design4a}. The FSM is responsible for correcting the pointing of both the guide camera as well as the science camera. It should be theoretically possible to utilise off-the-shelf tip tilt mirrors such as the V-931 from PI\footnote{https://www.pi-usa.us/en/} or DTT10M-SG-SV from CEDRAT\footnote{https://cedrat-technologies.com/} as the FSM in this scenario. This is again made possible by the afocal implementation which allows for beam steering from within a collimated beam section and hence the offset rotation of the mirrors is not a significant problem; this has been illustrated in figure \ref{fig:steer_offset}. However, the occupied  volume of these off-the-shelf modules is still a concern and the same goes for the required electronic drivers as well sensor conditioning components. There is a need for miniaturization of these components as well as space-qualification and ruggedisation of the complete system. In this regard, significant amount of progress is being  such as by (\cite{de2024high}). While there are still some challenges in terms of "cross-talk" of rotation axes as well as deformation of mirrors under load, it is expected that a Cubesat optimised FSM will be available for the astronomical community soon.\\

\begin{figure}
		\centering

        \begin{subfigure}{0.5\textwidth}
			\includegraphics[width=1.\linewidth]{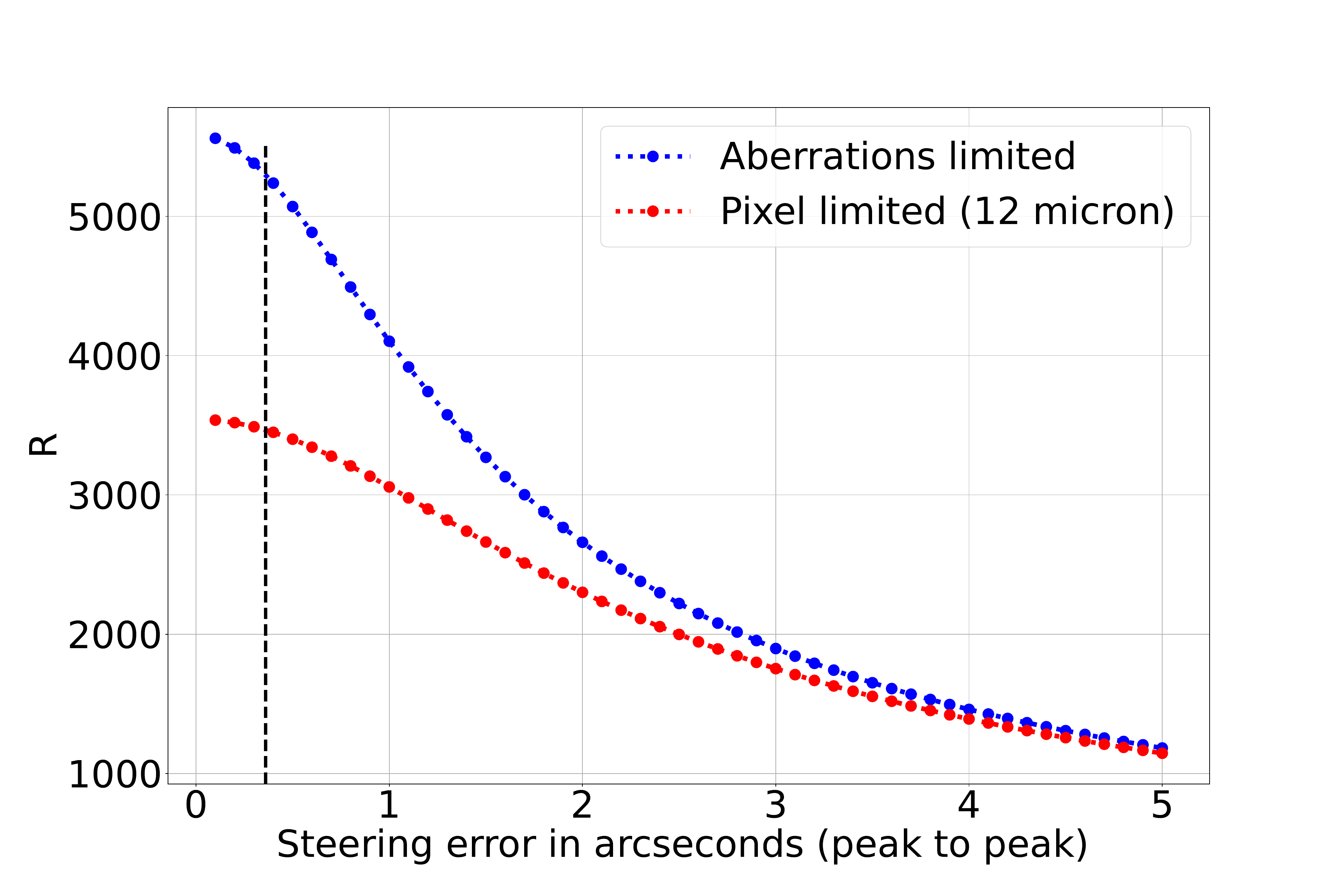}
			\caption{Effect of steering error on resolution}
		\end{subfigure}

     \caption{ For slit-less spectroscopy, the final resolution is dependent on the steering error. Expected degradation in resolution is shown. It is to be noted that the plot is inclusive of other effects such as pixel size, optical aberrations etc. The discussed NUV spectroscope is actually pixel size limited and can maintain this resolution if the steering error is kept less than 1 arc-second. The black vertical line shows the pixel scale of the guiding camera.}
		\label{fig:steering_R}  
\end{figure}

  Optical design for spectrographs utilising the 2U template are presented in figure \ref{fig:2U_spectrographs}. A spectrograph focusing on the NUV wavelength range  is shown in figure \ref{fig:design4b}. This spectrograph is implemented using a camera consisting of 3 lenses, produces an output of F9.3 and a dispersion of about 6.8 nm per mm at the image plane. The dispersion is by means of a 800 lines per mm grating ( similar to models available from \footnote{https://www.shimadzu.com/opt/}). The spot diagram corresponding to different wavelengths is shown in figure \ref{fig:spot4b}. The design is optimised to produce minimum spot size along the dispersion direction and the spot width is less that 14 microns for most wavelengths. When paired with detectors such as S10141-1107S from Hamamatsu (table-1), the achievable spectral resolution is about R=3000 ($\Delta \lambda = 0.9$\r{A}) at 275 nm. An FUV spectrograph is shown in figure \ref{fig:design4c}. The optics for the same is based on a single off-axis mirror (the rest are flat folding mirrors). The design is intended for use in the 150-300 nm wavelength range. The design makes use of a 300 lines per mm grating and produces a dispersion of about 15.6 nm per mm at the image plane. The spot diagram corresponding to different wavelengths is shown in figure \ref{fig:spot4c}. The resulting spectral resolution is about R=500 ($\Delta \lambda = 3.6$\r{A}) at the wavelength of 180 nm. The spectrographs discussed above will need to be wavelength and flux wavelength calibrated at ground before launch. Mercury and Cadmium based spectral lamps\footnote{https://spectrolamps.com/spectral-lamps/} can be used for this purpose. However, for on-orbit calibration standard stars have to be used for wavelength calibration as well as flux calibration. In this regard, work done by \cite{sreejith2022autonomous} will be quite useful. The spectrographs are intended for observing bright sources in relatively sparse fields. In case the field is more crowded, e.g. up to 2-3 stars within an arcminute, the complete CUBESAT may be rotated (along the "roll" or line of sight axis) to position the dispersion axis along the least crowded direction of the field.

 An analysis from first principles can be done to evaluate the relation between pointing stability and achievable resolution. The resolution of a spectrograph is:

\begin{equation}
    R = \frac{\lambda_c}{\Delta \lambda}
\end{equation}
Where, $\lambda_c$ is the central wavelength and $\Delta \lambda$ is the smallest resolvable wavelength element. Which can be expressed as:
\begin{equation}
    \Delta \lambda = [\frac{\lambda_{span}}{d_t}] \times \Delta S
\end{equation}
Where $\lambda_{span}$ is the total wavelength span which is mapped to a detector of size $d_t$ and $\Delta S$ is the physical spot size along the direction of dispersion. 

\begin{equation}
    \Delta S = \sqrt{(\Delta a)^2 +(\Delta d)^2+ (\Delta x)^2+(\Delta p)^2}
\end{equation}
The physical size of the spot includes contributions from optical aberrations ($\Delta a$), diffraction ($\Delta d$), steering errors scaled by the platescale ($\Delta x$) and detector pixel size ($\Delta p$). Thus, the final resolution can then be written as:
\begin{equation}
    R= \frac{\lambda_c d_t}{\lambda_{span} \sqrt{(\Delta a)^2 +(\Delta d)^2+ (\Delta x)^2+(\Delta p)^2} }
\end{equation}

A relation between steering error and resolution is plotted in figure \ref{fig:steering_R} by taking an example of Hamamatsu S10141-1107S matched to the NUV spectrograph described in figure \ref{fig:design4b}. A nominal 1 arcsecond of steering stability is desired to retain the original resolution at 3000. The corresponding guide camera (figure \ref{fig:2U_guiders}) has a pixel-scale of 0.36 arcseconds, therefore the total guiding error must be within 3 pixels in order to maintain this nominal resolution.

\subsection{Example science case: FUV/NUV spectroscopy}
For observing pulsating variables, FUV/NUV spectroscopy has regained recent interest. The variable star $\delta$ Cepheid has been observed by (\cite{engle2014secret}) to show strong  emission lines in FUV which also vary in correlation with the pulsation phase. This definitely hints at a pulsation driven shock mechanism as the source. Such spectroscopic studies can be extended to a number of other variable stars if sufficient observing time can be allocated -- such as from a Cubesat based platform. Of particular interest will be intermediate period Type-II cepheids such as W Virginis which have been reported to show signatures of pulsation driven shocks in the form of emission lines in H-$\alpha$ (\cite{kovtyukh2011neutral}). Similarly non-radial pulsators such as $\beta$ Cepheids which are inherently bright in shorter wavelengths will also be of interest for such studies. An estimate of achievable magnitude limits for a Cubesat based platform is shown in figure \ref{fig:SNR_resolution} and a magnitude distribution of $\beta$ cepheids is shown in figure \ref{fig:dist_Bcep}; demonstrating that a significant number of such sources should be within the capabilities of a 2U Cubesat. 

We have assumed a Vega like Spectral Energy Distribution in our SNR analysis. In this regard, (\cite{fitzpatrick2010uv}) have reported slight anomalies (of the order of 5\%) in the SED of Vega when compared to best fit theoretical models. They have recommended follow-up observation of a large number of similar A type stars to constrain whether the anomalies are unique to Vega or extend to other stars as well. This is also an important area where Cubesats can provide much sought after observations.

Since, Cubesat based platforms can only have limited aperture, it is absolutely essential to be efficient with the gathered photons. In this regard, the afocal slit-less spectrograph has a unique benefit. For low and medium resolution (R$<$5000), the slit-less design has "perfect" slit transmission -- 100\% compared to slit based systems where the typical slit efficiency can be around 60\% -- as well as better throughput( about 20\% higher than conventional) as the collimator group is not required. A comparison of afocal slit-less design and a conventional slit based design is presented in figure \ref{fig:SNR_afocal}. Improvement of about 0.5-0.7 magnitudes in terms of limiting magnitude may be expected for an afocal design. Throughput improvements resulting from afocal slit-less design is complemented well by the good quantum efficiency of modern CCD detectors (typically about 50-60\% compared to only 20-30\% of MCP type detectors) as well as specialized coatings in the UV (e.g. \cite{woodruff2019optical}); thus allowing a broader scope for Cubesat based spectrographs.

\begin{figure*}
		\centering

        \begin{subfigure}{0.44\textwidth}
			\includegraphics[width=1\linewidth]{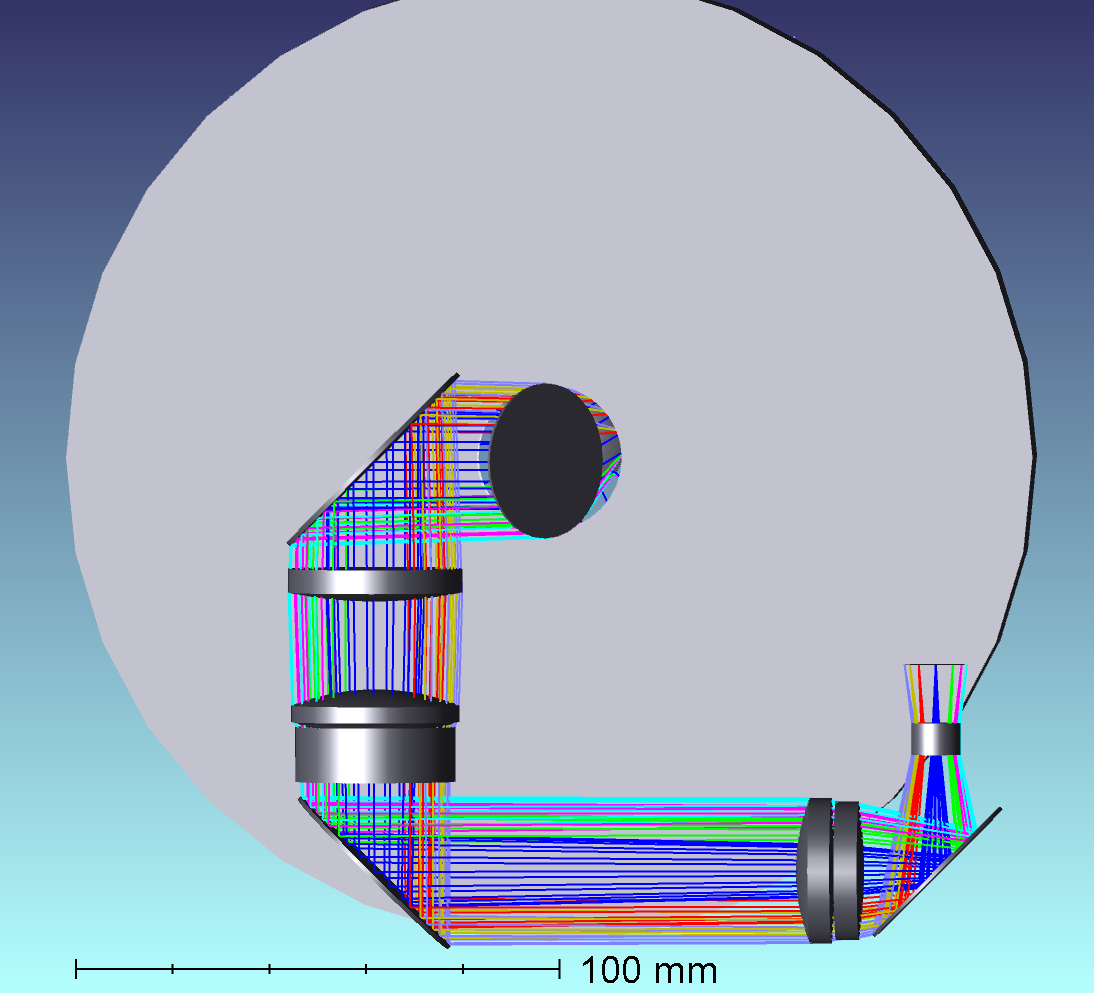}
			\caption{2U NUV imager}\label{fig:2U_nuv_layout}
		\end{subfigure}
		\begin{subfigure}{0.55
        \textwidth}
			\includegraphics[width=1\linewidth]{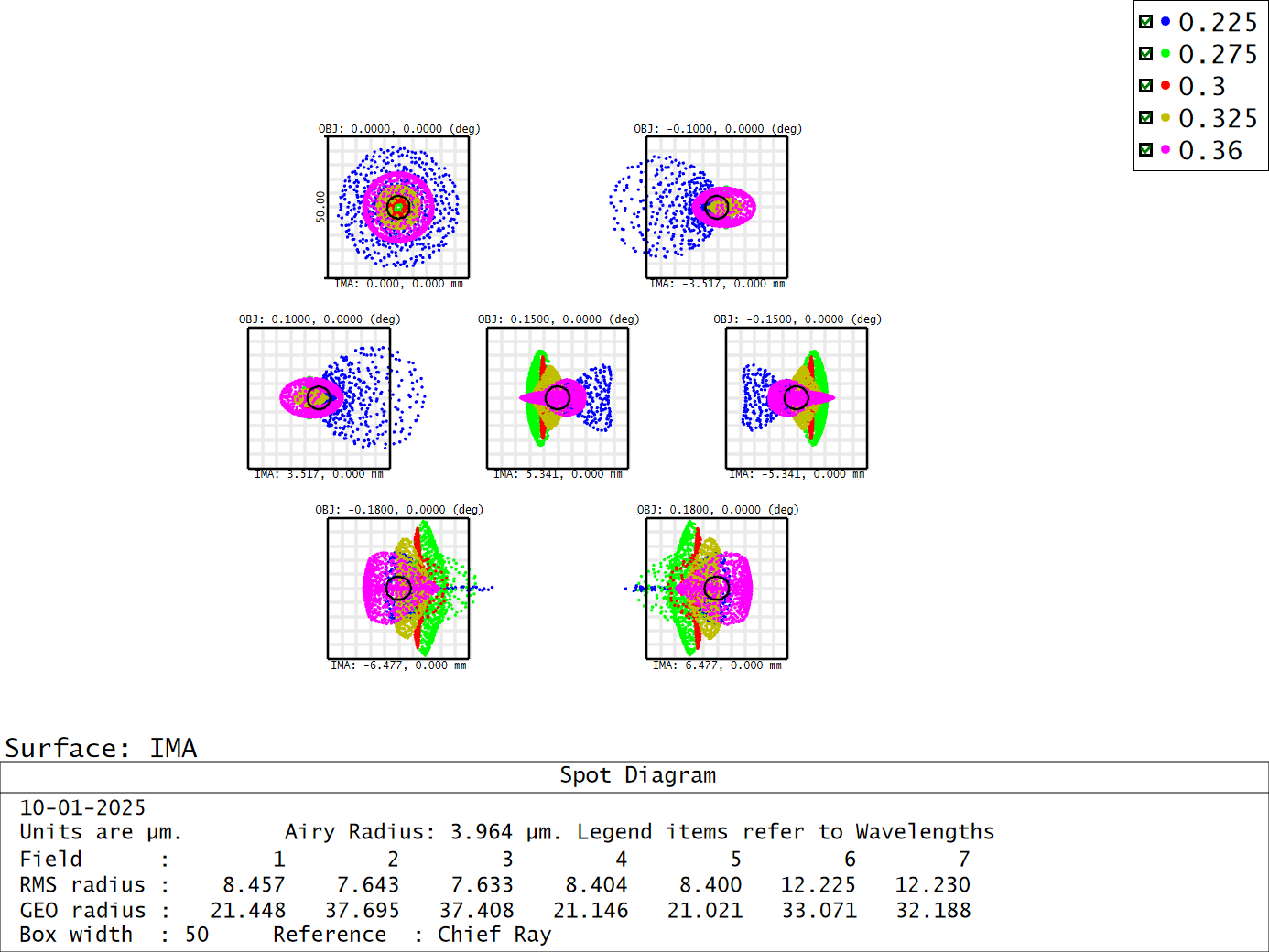}
			\caption{Spot for 2U NUV imager}\label{fig:2U_nuv_spot}
		\end{subfigure}
		\bigskip

		\begin{subfigure}{0.44\textwidth}
			\includegraphics[width=1\linewidth]{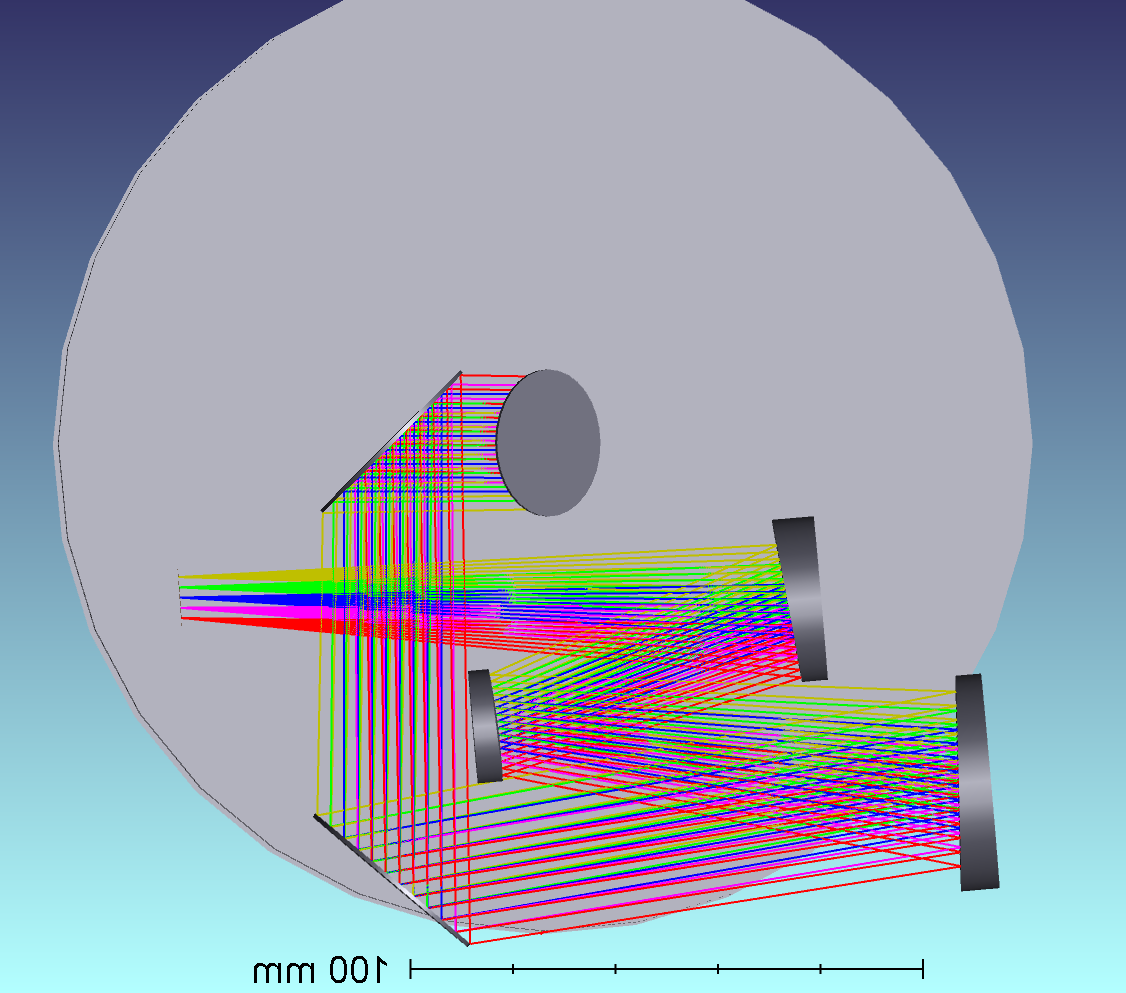}
			\caption{2U TMA imager imager}\label{fig:2u_tma_layout}
		\end{subfigure}
		\begin{subfigure}{0.55
        \textwidth}
			\includegraphics[width=1\linewidth]{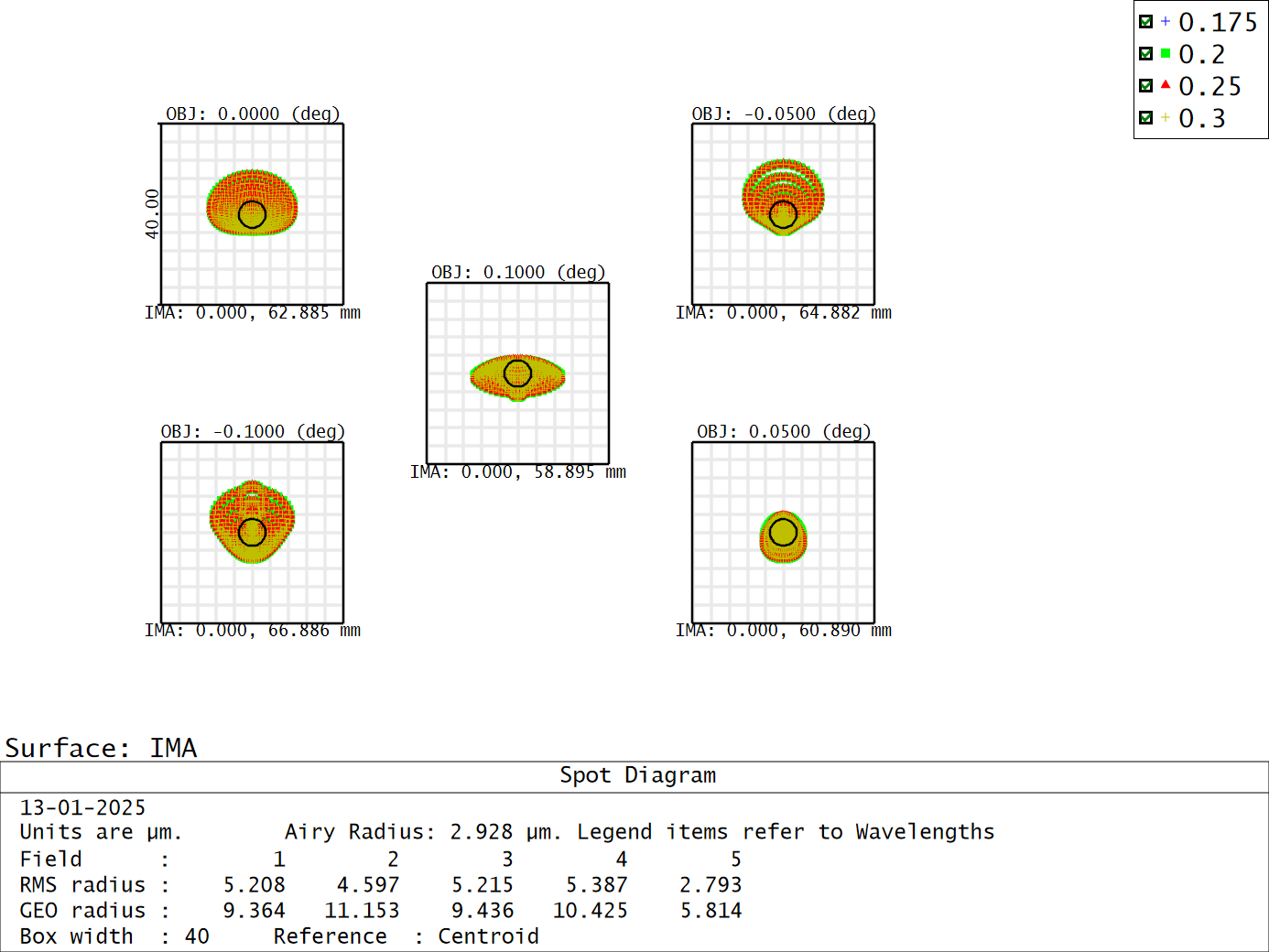}
			\caption{Spot for 2U TMA imager}\label{fig:2u_tma_spot}
		\end{subfigure}
		\bigskip
     \caption{Review of 2U imagers: layout of a NUV imager is shown in (a). The NUV imager has a F8 output and a field of view of about 21 arcminutes. The imaging performance of the same is illustrated in (b). In (c) a fully reflective camera based on TMA design is shown, and the corresponding spot diagram for 12 arcminutes field of view is shown in (d). It is seen that the TMA type imager is close to diffraction limited whereas the refractive NUV imager is not. This is partly due to the presence of strong chromatic aberration in the NUV imager. } 
		\label{fig:2U_imagers}  
	\end{figure*}

\subsection*{Imaging cameras in the 2U template} In figure \ref{fig:2U_imagers}, a number of imaging cameras based on the 2U template are shown. Again, all designs share the same afocal telescope. Figure \ref{fig:2U_nuv_layout} presents an NUV imager in the wavelength range of 225-360 nm. The design is implemented with MgF2 and Fused silica glass; a total of 6 lenses are used. The imager is optimised for 21 arcminutes field of view. The corresponding spot diagram is shown in \ref{fig:2U_nuv_spot}. The imager has an F10 output and produces an image circle of about 13 mm. Detectors such as the IMX487 or GSENSE2020BSI will be a good match for this. With IMX487, the pixel scale is about 0.3 arcseconds. The design in \ref{fig:2u_tma_layout} is a fully reflective camera based on the Three Mirror Anastigmat (TMA) design. As the name suggests three off-axis mirrors are used to realize camera. The design is optimised  for a field of view of about 12 arcminutes and produces an images size of 14 mm. The spot diagram for this design is shown in  \ref{fig:2u_tma_spot}. As the designs only uses reflective components, this camera can be used for wide range of wavelengths. In particular, this design will be useful as an FUV camera suited for detectors like S7170-0909 from table \ref{detectorTable}.

\section{A few illustrative science cases}
\begin{figure*}
		\centering

        \begin{subfigure}{1\textwidth}
			\includegraphics[width=1.\linewidth]{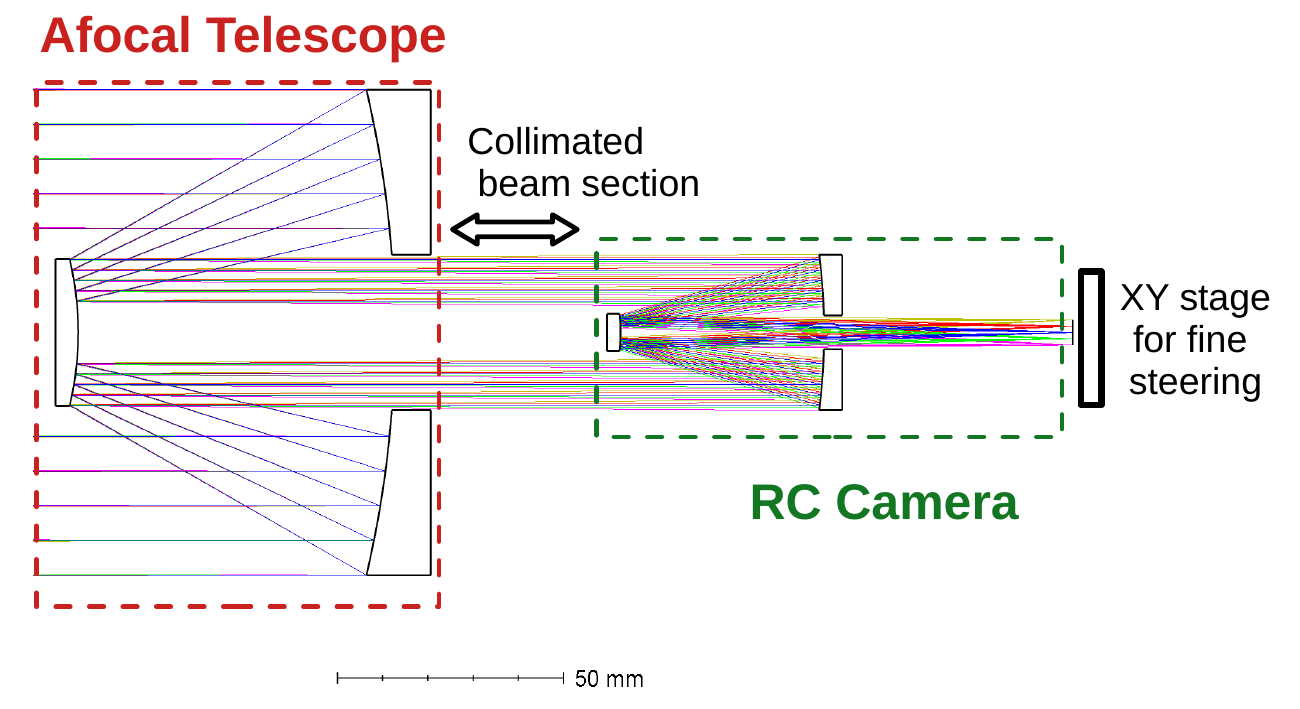}
			\caption{Afocal spectropolarimeter implementation by an RC type camera.}\label{fig:design_SP}
		\end{subfigure}

        \begin{subfigure}{0.45\textwidth}
			\includegraphics[width=1\linewidth]{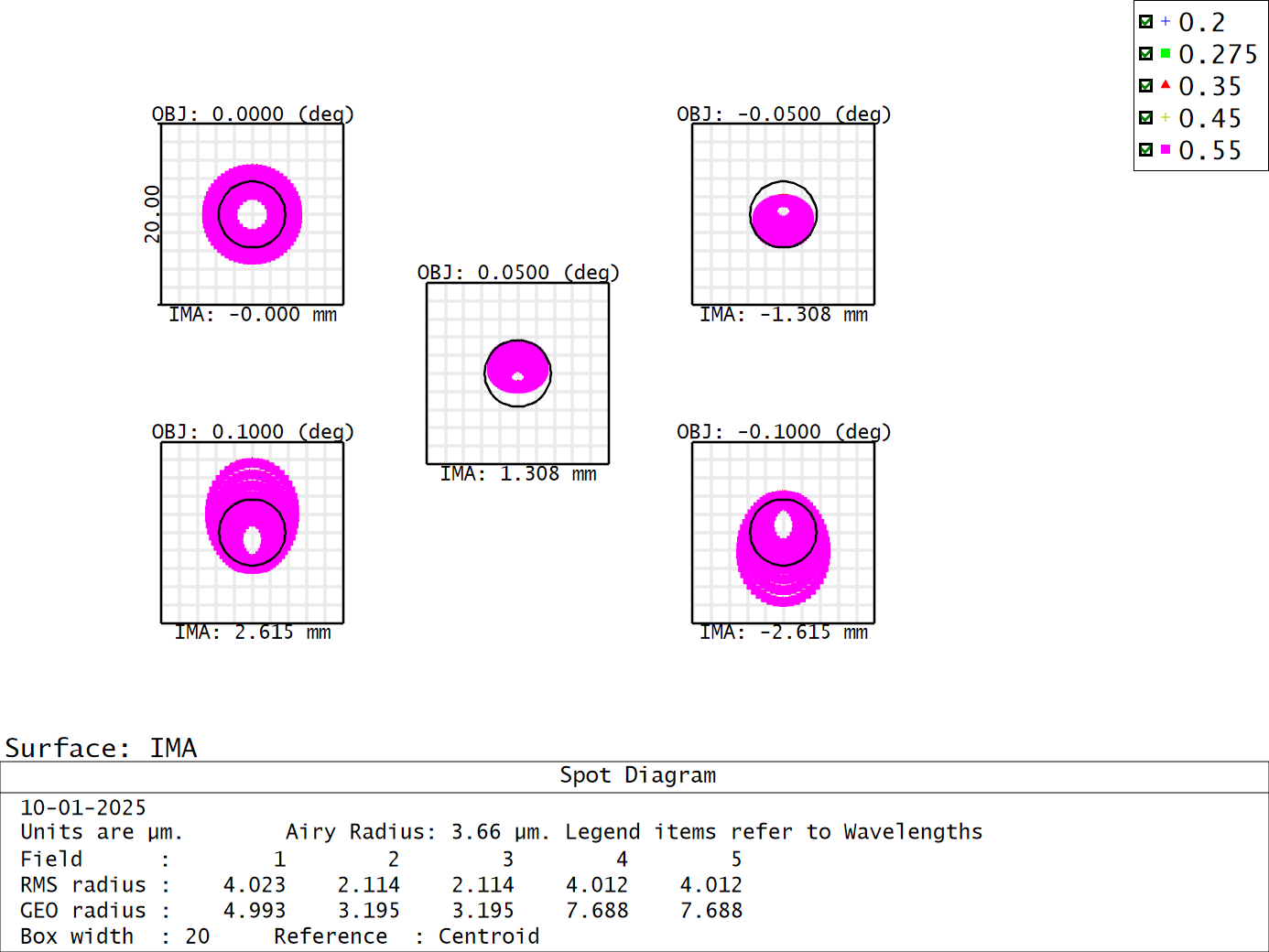}
			\caption{Spectropolarimeter spot}\label{fig:spot_SP}
		\end{subfigure}
		\begin{subfigure}{0.5
        \textwidth}
			\includegraphics[width=1\linewidth]{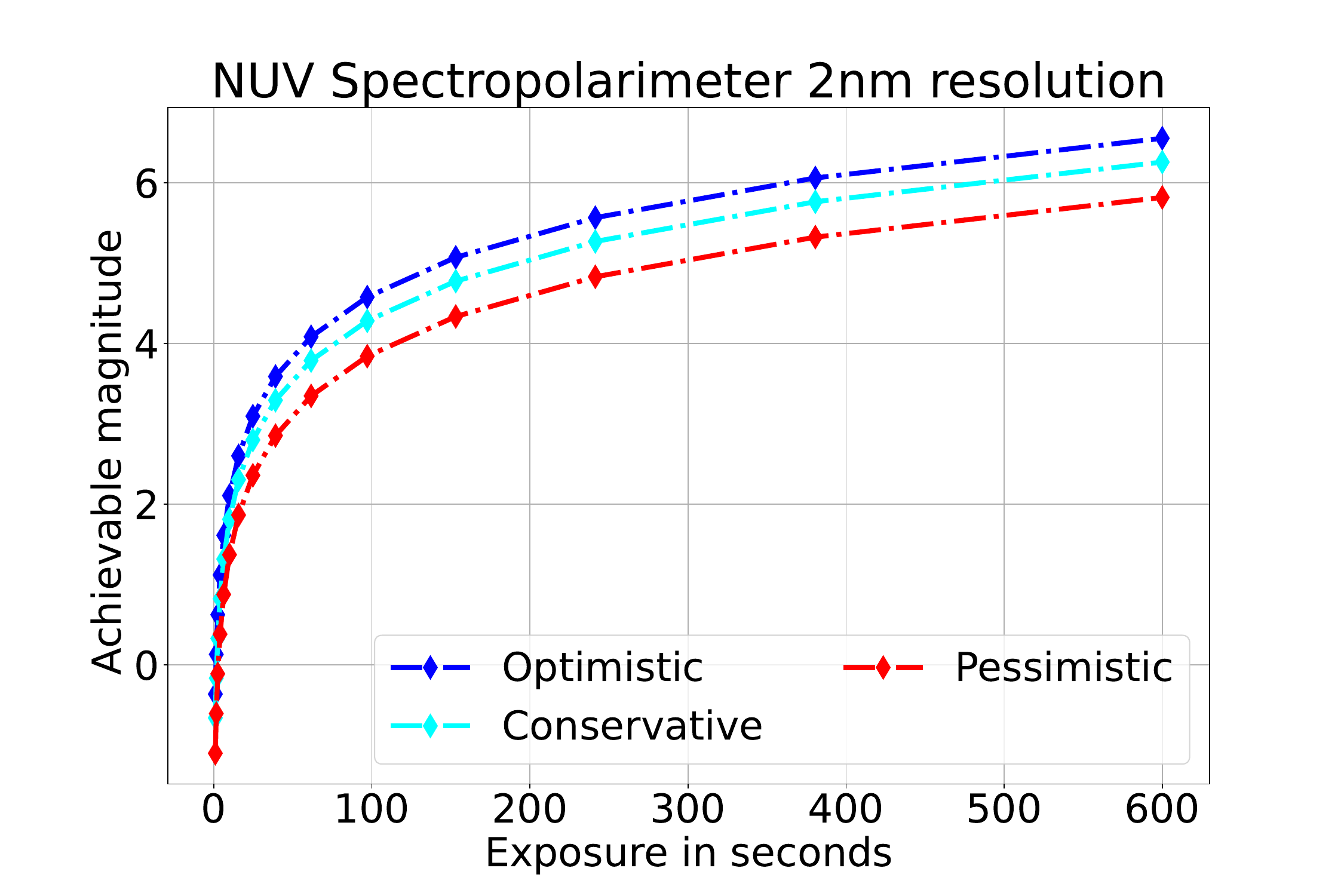}
			\caption{SNR limits}\label{fig:SNR_SP}
		\end{subfigure}
		\bigskip

     \caption{ A fully rotationally symmetric afocal implementation : an interesting implementation utilising an afocal telescope is shown in (a). The complete imaging chain is fully rotationally symmetric so as to reduce instrumental polarisation. The fine pointing stability is achieved by using a X-Y stage such as from table \ref{xyTable} to match the exact F-number. The design achieves fairly good imaging performance as shown in (b) while implemented with all reflective components. The desired characteristics of the Wollaston as well as the dispersing element can be estimated by using equations 17-19. Expected magnitude limits from a 10 cm Cubesat for NUV wavelength range is shown in (c). Throughput and quantum efficiency assumptions remain the same as in figure \ref{fig:SNR_spec}. The Wollaston efficiency is further split into optimistic (0.4), conservative (0.30) and pessimistic (0.2).}  
		\label{fig:SP}  
	\end{figure*}

\subsection{Multi-wavelength Asteroseismology}
In recent times, significant insights has been made into stellar structure and evolution owing to progress in astero-seismology (\cite{aerts2021probing}). However, most of the asteroseismology has been done in missions  such as TESS  (\cite{ricker2015transiting}) where the observation is through broadband filters optimised for exoplanet detection. The concept of astero-seismological modes may naturally be extended to shorter wavelengths for early type stars. As the radiant flux of these stars is higher in the shorter wavelength it is also possible that the amplitude of the modes is higher in these wavelengths. The nature of observation (of continuous data over long durations) will also reveal nature of flaring and related activity of these stars.

Similarly, it will also be useful to see the behaviour of pulsating variables in the infrared wavelength range. Particularly, infrared light-curves are important as they are inherently a better indicator of the period-luminosity law of these sources. From a small aperture Cubesat it will be possible to cover a large number of bright infrared sources and their variability can be studied at good time resolution. Additionally, since pulsating variables are typically evolved stars their radiant flux increases as wavelength is increased and hence there is a significant scopes of detecting astero-seismological modes of these sources in infrared. Currently modes in the infrared range are not very well understood and some very unique opportunities remain in this type of detection(\cite{kallinger2005asteroseismology}).

\subsection{Emission line stars}
Emission line stars -- such as classical Be stars(\cite{rivinius2013classical}) and more exotic variants thereof(\cite{wang2022identification},\cite{bhattacharyya2021identification}) -- present opportunities for study of some very unique astrophysical phenomena such as circum-stellar disc formation, dissipation and many others(\cite{rivinius2013classical},\cite{porter2003classical}). Dedicated monitoring of these sources can reveal a significant amount of information regarding the characteristics of the disc (\cite{banerjee2022study},\cite{banerjee2024study}) and the evolution of the central massive star. However, ground based observations remain at the mercy of weather and geographical constraints. A Cubesat based platform can provide continuous observations of these sources. In this capacity, the 1U observatory template discussed previously (section 4.1) is of particular interest as it can provide simultaneous photometric observations in a wide wavelength range (NUV-NIR). This is crucial to understanding evolution of different components of the circum-stellar disc at different timescales. It is also expected that NUV photometry can be used as an indicator for activity in the disc and hence used as a trigger for ground based follow up observations .

 The numerous emission lines seen in the spectra of these stars (\cite{banerjee2021optical},\cite{mathew2011optical}) are also diagnostic tools for the circum-stellar disc as well as the central massive star. ln this regard, NUV spectroscopy will provide an additional and yet to explore tool to understand and characterise different line forming regions in Be stars and thus investigate the evolutionary status of such systems. It will be interesting to match the NUV spectral features to that of emission lines in terms of line profile and variability at different epochs. Overall, it is expected that an approach combining ground based and Cubesat facilities can provide new insights as well as broaden the scope to include other similar variables such as Ae stars (\cite{anusha2021identification}).

\subsection{Spectropolarimetry}

 Afocal designs allow for implementation of fully rotationally symmetric cameras resulting in low instrumental polarization. Additionally, the  compactness (by skipping the collimator group)  of these designs make them good candidates for Cubesat based polarimetric observation. An illustrative example is shown in figure \ref{fig:design_SP}. In the example, an RC telescope has been used as the camera, but it should be possible to implement this with a Gregorian telescope or even a monolithic telescope.
The fine pointing stability is achieved by using a X-Y stage such as from table 3 to match the exact F-number. There have been examples of upgrading an existing collimator camera design with polarimetric capabilities by introducing an Wollaston prism in the collimated beam-section (e.g. \cite{helhel2015double}). This design is usable in a wide wavelength range (200-1700 nm)and can utilise both GRISM or zero-deviation prism, (e.g. \cite{hagen2011compound}) for dispersion and a Wollaston prism (e.g. \cite{oliva1997wedged}) for separation of orthogonal polarimetric components. The required parameters of the the dispersing and polarization element can be inferred directly from the geometry of the afocal design:

Let us consider a primary diameter $D_P$, a beam compression ratio $C$ and a final f-number of $F_n$. Then the focal length of the camera optics is given as:
\begin{equation}
    F_{camera} = \frac{D_PF_n}{C}
\end{equation}
If a Wollaston split of $d_W$ (in mm) is to be achieved at the image plane, then the separation angle $W_S$ (in degrees) of the Wollaston prism must be:
\begin{equation}
    W_S = \frac{206265 \times d_W}{F_{camera}\times 3600} = \frac{57.3 \times d_W \times C}{F_n \times D_P} 
\end{equation}
Similarly, the desired dispersion (in degrees per nm) from the dispersing element is:
\begin{equation}
    Dispersion = \frac{57.3 \times C \times d_{\lambda}}{F_n \times D_P \times \lambda_S} 
\end{equation}

Where, $\lambda_{span}$ is the desired wavelength span (in nm) and $d_{\lambda}$ is the linear size of the dispersed spectra at the image plane. Necessary equations for estimating spectral resolution are already presented in section 4.2 (equations 13-16).

The expected limiting magnitudes are shown in figure \ref{fig:SNR_SP}. The estimations are based on relatively conservative estimations of throughput (with details in the figure caption). The magnitudes may be compared to the luminosity distribution of O and B type stars is shown in \ref{fig:dist_O_B}. It is exciting to note that there are sufficient number of bright O and B stars to probe the interstellar polarization even from a small aperture telescope. It may be possible to probe different regions of sky along with different distance ranges. The implementation of the camera with an all reflective designs (such as the RC, Gregorian ) or the monolithic camera allows for a wide range of wavelength range to be accessible within one instrument. This allows for low resolution spectro-polarimetry over a large enough spectral range to discuss the nature of Serkowksi curves (\cite{papoular2018new}) and perhaps its variability indicating composition of the interstellar medium at different directions. 

\begin{figure*}
		\centering

        \begin{subfigure}{0.6\textwidth}
			\includegraphics[width=0.99\linewidth]{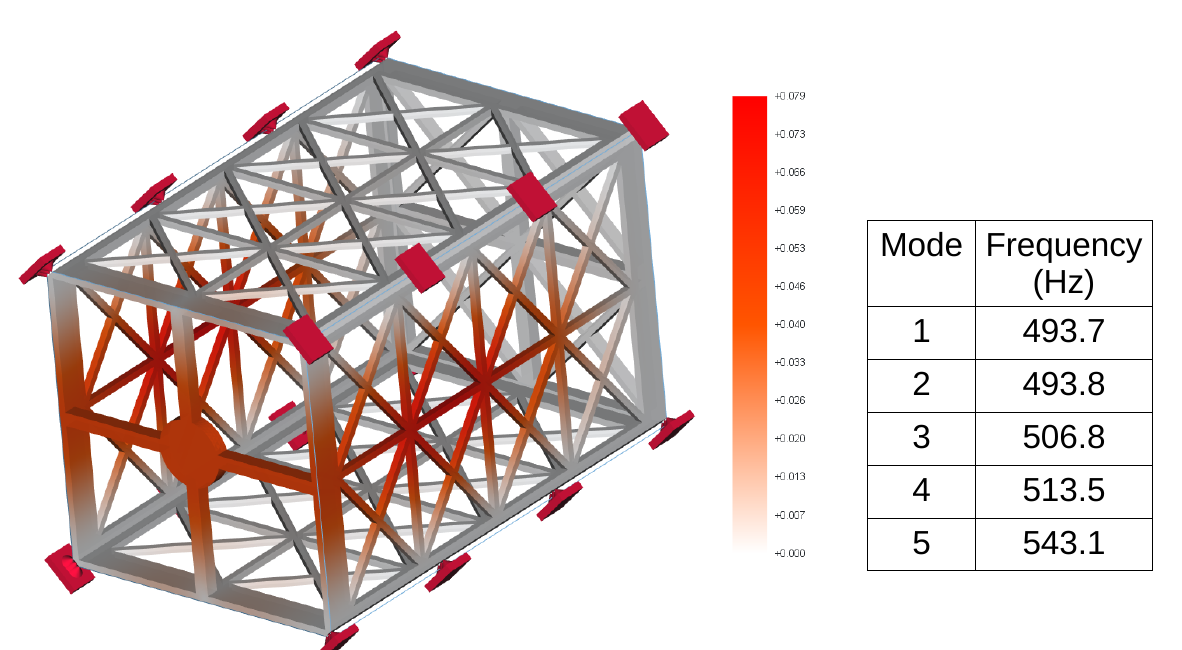}
			\caption{Illustrative mode-1 }\label{fig:12U_frame_freq}
		\end{subfigure}
		\begin{subfigure}{0.35
        \textwidth}
			\includegraphics[width=0.99\linewidth]{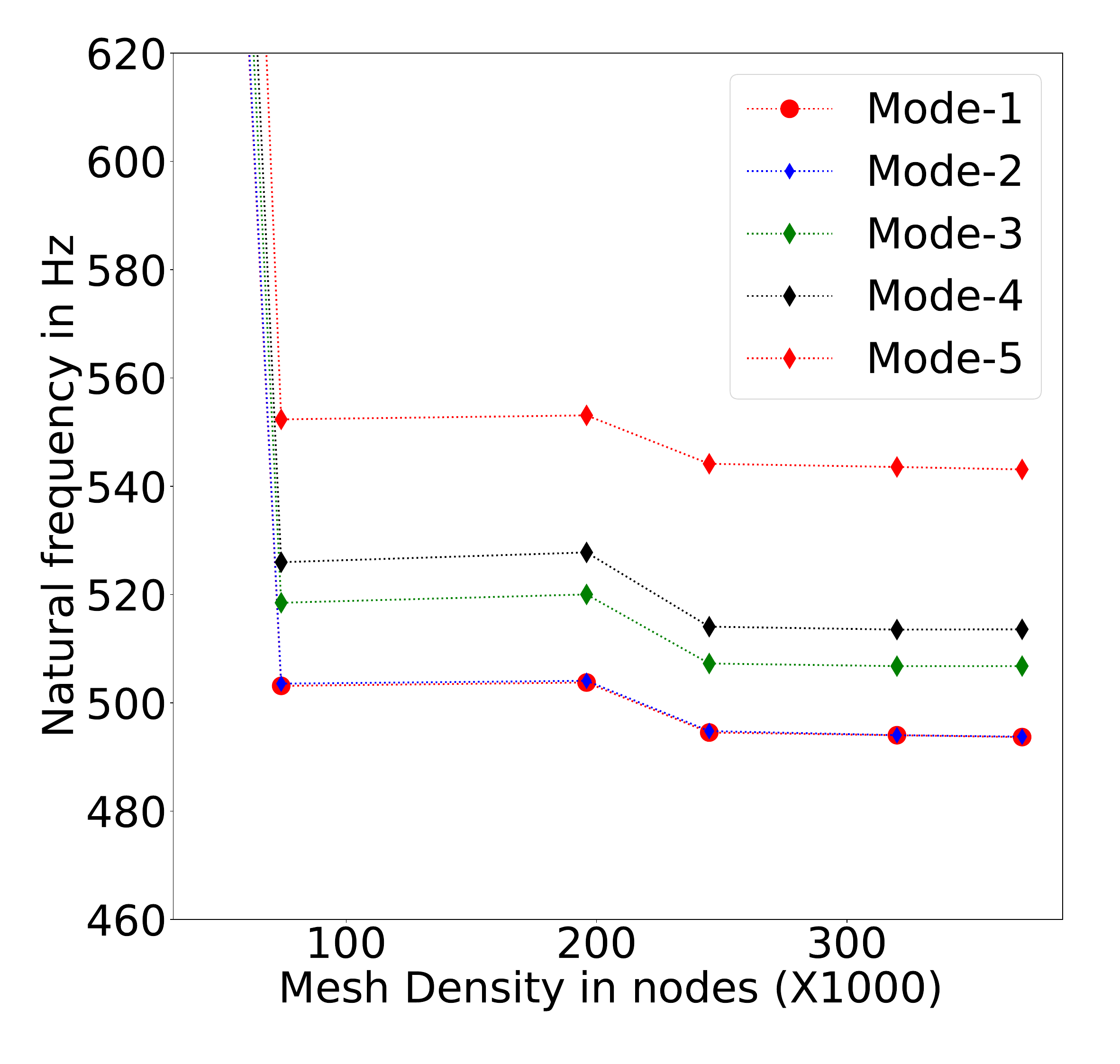}
			\caption{Mesh convergence}\label{fig:12U_frame_mesh}
		\end{subfigure}
		\bigskip

        \begin{subfigure}{0.99
        \textwidth}
			\includegraphics[width=0.99\linewidth]{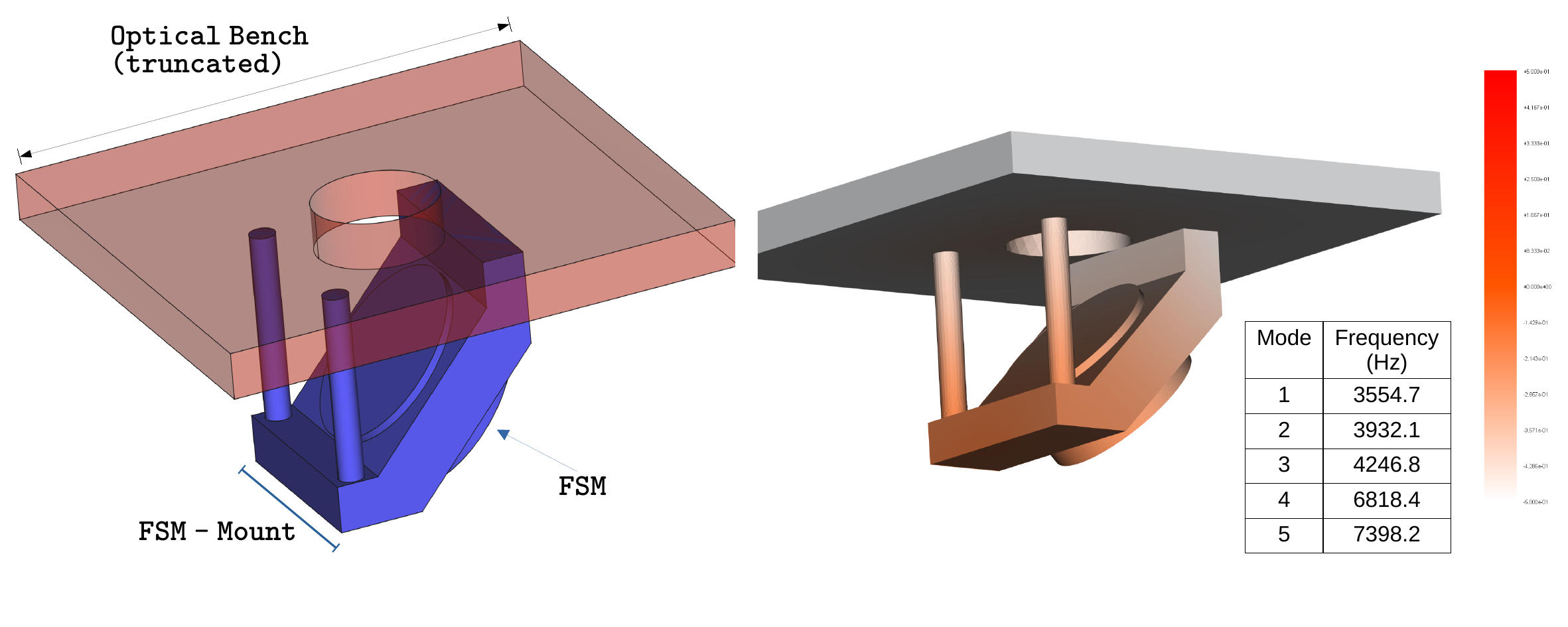}
			\caption{Frequency analysis of illustrative FSM mount}\label{fig:12U_fsm_mount}
		\end{subfigure}

        \begin{subfigure}{0.9
        \textwidth}
			\includegraphics[width=0.99\linewidth]{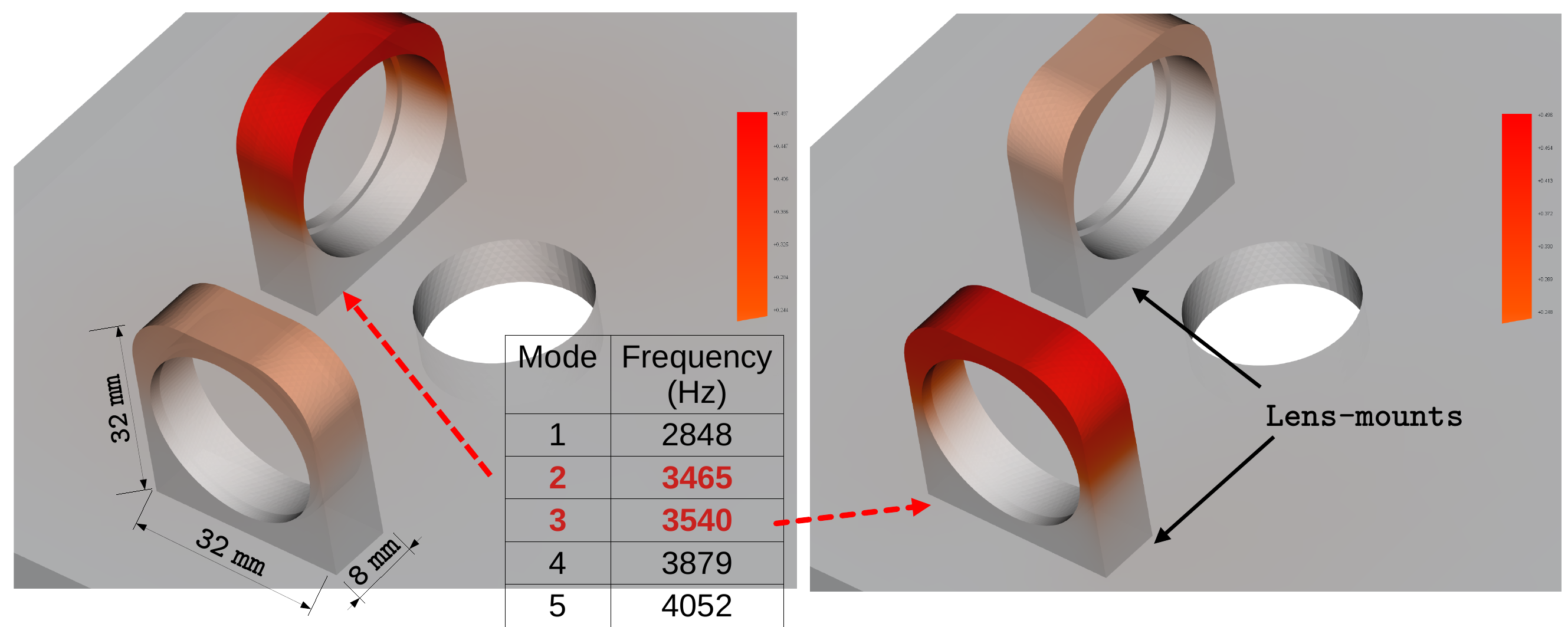}
			\caption{Frequency analysis of illustrative lens mount}\label{fig:12U_lens_mount}
		\end{subfigure}

     \caption{ Natural modes of vibration modes are identified for a few crucial structures. In (a) the lowest vibration mode of the 12U Cubesat frame is demonstrated. The mesh convergence of the same is shown in (b). The FSM mount (c) and lens mounts (d) are also analysed. All modes are well above 100 Hz, which is typically considered the minimum requirement for Cubesats. Instances of closely spaced modes are typically orthogonal modes in X and Y directions.}
		\label{fig:mode_fem}  
	\end{figure*}

\section{Practical aspects of afocal designs:}
A number of practical aspects related to realization of Cubesat based observatory templates are discussed:

\begin{figure}
		\centering
    
        \begin{subfigure}{1\linewidth}
			\includegraphics[width=1\linewidth]{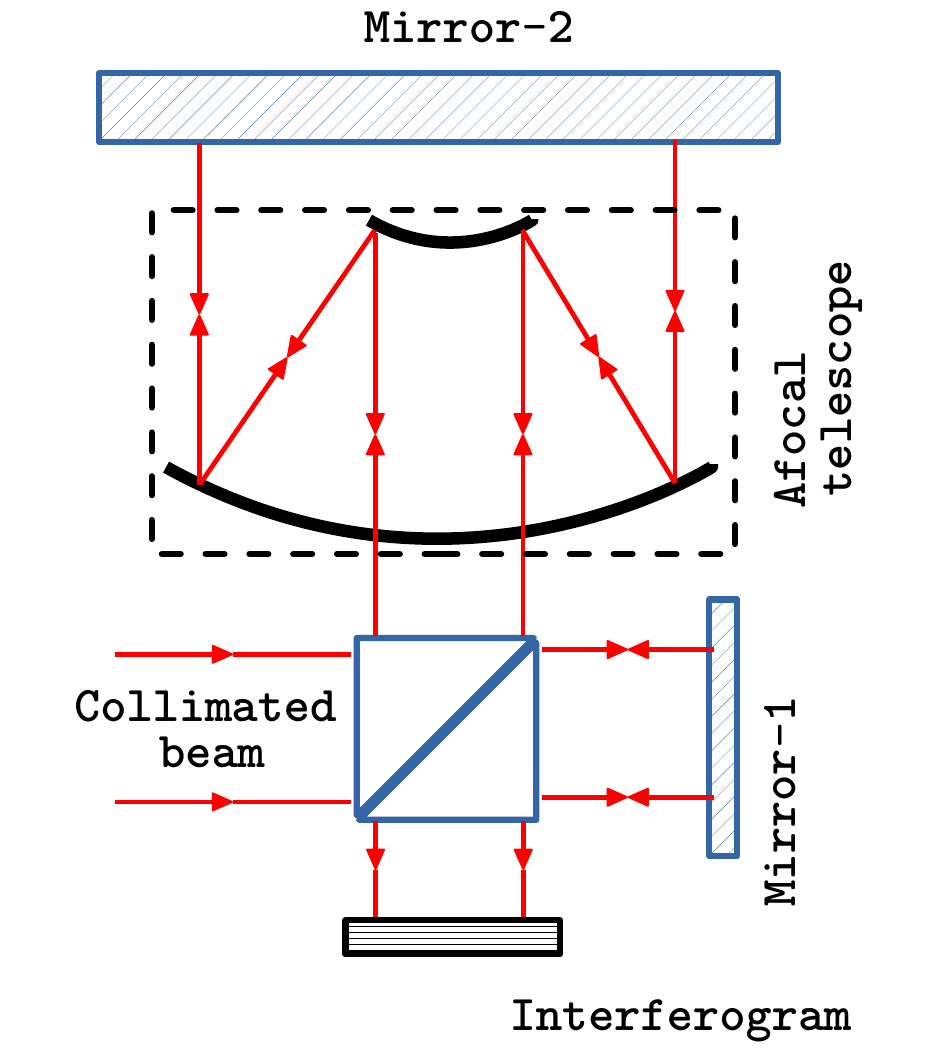}
			\caption{Interferometric test of afocal telescope}\label{fig:Lab_test_1}
		\end{subfigure}

        \begin{subfigure}{1\linewidth}
			\includegraphics[width=1\linewidth]{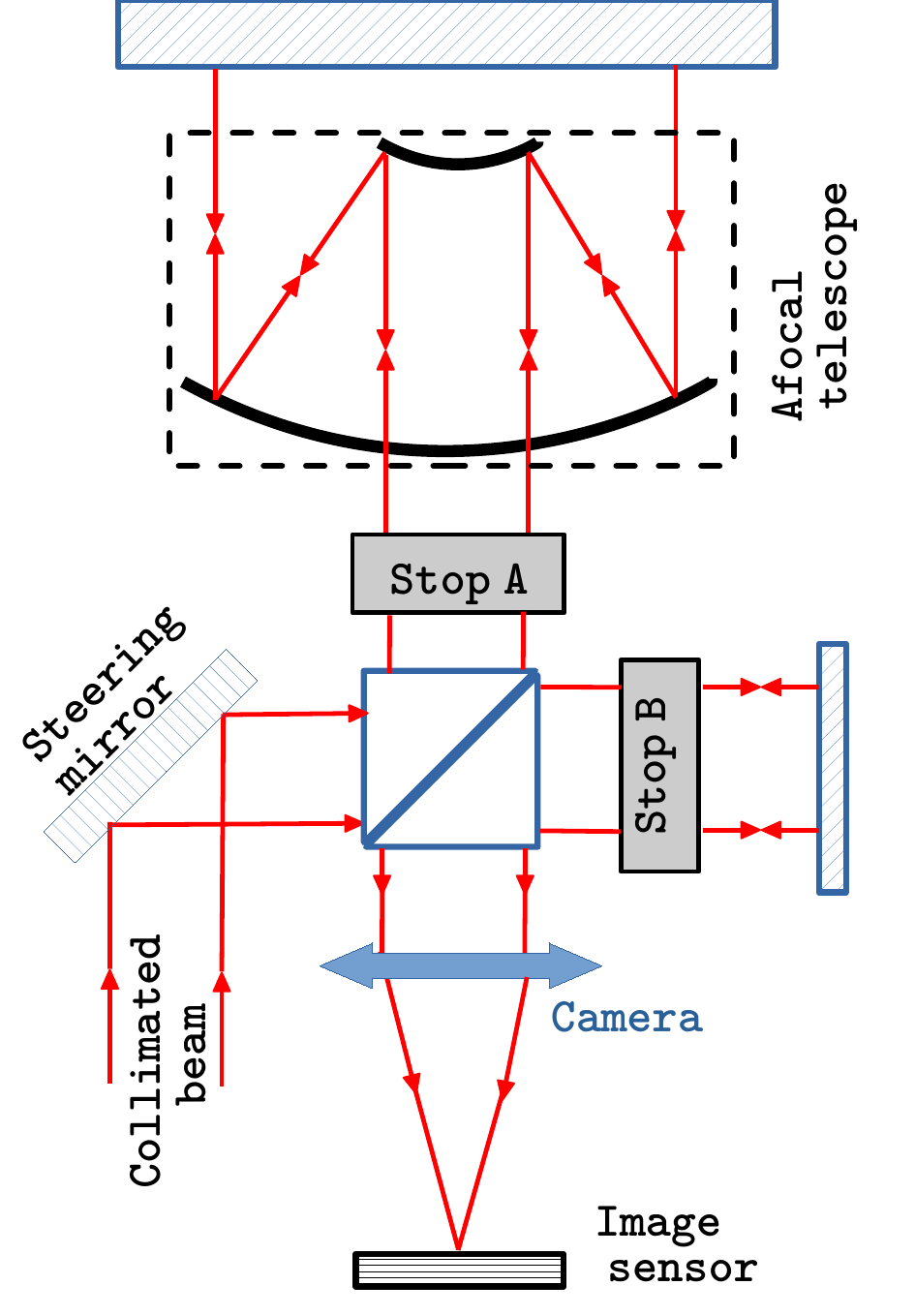}
			\caption{Test for telescope and camera}\label{fig:Lab_test_2}
		\end{subfigure}

     \caption{Laboratory test setups: see text for details}
		\label{fig:Lab_tests}  
	\end{figure}

\subsection{Tolerancing and athermalization:}

Cubesat designs typically require a tight tolerance limits (typically 5 $\mu$m or less; e.g. CUBESPEC). This is inadvertently related to using fast telescopes in a space constrained design  as well as the necessity to athermalise the design against thermally induced defocus. One aspect of afocal designs can help alleviate this issue: in an afocal design, the telescope and the camera are essentially decoupled from each other and can tolerate some amount of defocus (a few mm) and decenter (about 100 microns) without degrading the final image performance. This is a natural outcome since the coupling between the telescope and the camera happens within a collimated beam section. This allows for some amount of flexibility in the alignment of the complete system as the telescope and the camera can now be aligned separately and then assembled with the telescopes at a loose tolerance. The telescope itself still has to be aligned and athermalised. This can be done using standard opto-mechanics suitable for RC telescope. Typically the mechanical structure has to be of very low expansion material such as invar or zerodur to satisfy this requirement. 

\begin{table}
%% use tabular font for a smaller size font
\tabularfont
\caption{ Tolerancing estimates for various designs. According to Edmund Optics high-precision standard. Paraxial focus is used as the compensator.}\label{tolerance_Table} %%10/12
\renewcommand{\arraystretch}{1.25}

\begin{tabular}{lcccc}
\topline
Optical & Original & Tolerance & Compens. \\
Design &($\mu$m RMS) &  ($\mu$m RMS) &  (mm) \\ \midline

1U(Guider) & 1.6 & 2.7 & $\pm$ 0.35 \\
1U(Vis) &3.8 & 7.5& $\pm$ 0.7 \\
1U(NIR) & 4.8& 9.4 & $\pm$ 1.7  \\
1U(Greg.) & 5.5 & 9.3 & $\pm$ 4  \\
2U(Guider) & 5.1 & 9.7 & $\pm$ 0.7  \\
2U(NUV Spec.) & 3.1 & 8 & $\pm$ 2  \\
\hline
\end{tabular}
%%use \tablenotes{footnote} to get the table foot note
% \tablenotes{Sample table footnote}%%9/11
\end{table}

Specifically for the 2U template design discussed, the implementation of the FSM  allows for one additional convenience in terms of alignment. Some amount of initial tilt between the optical bench and the afocal telescope can be compensate by the FSM itself. This is possible  since the FSM is the coupling element between the telescope and the optical bench. This is of course subject to the maximum stroke of the FSM. However, this does allow for all the optical components on the optical bench to be tested before assembly onto the Cubesat platform without worrying too much about relative alignment between the telescope and the optical bench. A summary of tolerancing (closely following the Edmund Optics high precision standard\footnote{https://www.edmundoptics.in/knowledge-center/application-notes/optics/precision-tolerances-for-spherical-lenses}) related image degradation for various designs is presented in table \ref{tolerance_Table}.

It should also be possible to make use of commercial off-the-shelf cameras along with the afocal telescope. Typically cameras such as from Schneider\footnote{https://schneiderkreuznach.com/en} or Nikon\footnote{https://www.nikon.com/business/industrial-lenses/lineup/} (and many other manufacturers) which are designed for intermediate focal lengths(100-250 mm) and reasonably large apertures($>$35 mm) will be a good match. The exact implementation will involve detailed knowledge on specific camera models regarding exact mechanical construction, number of elements, vignetting effects and other factors. This is out of scope for the present discussion.

\subsection{Natural modes of a sensitive structures:}
It is of considerable importance that the natural vibrational modes within a space payload are identified. This is to ensure that no such modes are within the range of frequencies that may be excited during launch. Typically for Cubesats, modes below 100 Hz are best avoided (\cite{herrera2016cubesat}). Finite element analysis (using open-source Freecad\footnote{https://www.freecad.org/index.php} Calculix solver)  is presented in figure \ref{fig:mode_fem} for some of the critical components. The lowest excitation mode for a 12U frame is presented in figure \ref{fig:12U_frame_freq} and five of the lowest modes are listed. The frame is a structural component for the complete observatory. For simplicity, we have considered construction from 6061 Aluminium (with weight less than 3 Kg). The frame is constrained at the launch rails and is evaluated while "unfilled" (which is a worst case scenario ). The colour map identifies the mode (arbitrary units) and illustrates parts of the frame which are susceptible to this mode. A mesh convergence test is presented in figure \ref{fig:12U_frame_mesh} to demonstrate repeatability of the analysis with respect to mesh density. Similar illustrative analysis is done for FSM mount (figure \ref{fig:12U_fsm_mount}) and lens mounts (figure \ref{fig:12U_lens_mount}). The lens mount is tested without a lens; again a worst case scenario in terms of vibration. The preliminary analysis hints at a good margin of safety for these components. Of course, this type of analysis is more useful in highlighting problem areas and not a substitute for rigorous physical testing.

\subsection{Laboratory test and characterisation of afocal systems:}

Since the afocal telescope is also a beam expander/compressor with both input and output being collimated it can be conveniently tested by means of optical interferometry. A simple setup based on a Twymann-Green interferometer (\cite{malacara2007optical}) is shown in figure \ref{fig:Lab_test_1}. A reference collimated beam is given as input to the system. The input laser beam is required to be comparable in diameter to that of the secondary mirror only. The afocal telescope expands the beam further to fill its own primary mirror hence only a smaller collimated beam is required. This  can be conveniently generated from off-the-shelf He-Ne laser beam expanders(e.g. \footnote{https://www.newport.com/p/T81-30X}). One arm of the interferometer contains a reference mirror  and the other arm contains the telescope under test. A larger flat mirror is placed in front of the telescope to reflect Both reflected beams are combined by the beam-splitter and interfere at the output at position B in the diagram. Since the afocal telescopes is used in double pass its wavefront aberrations are doubled and hence easily measurable.

Another testing method that can be realized is shown in figure \ref{fig:Lab_test_2}. This method can test the camera of the afocal design by itself or the complete system consisting of the afocal telescope and the camera. Again, a reference collimator is required with a beam size comparable to that of the secondary mirror. This time however, a broadband light source may be used. This collimated beam is diverted by the beamsplitter to two different arms. Depending on which arm is blocked or allowed different parts of the system may be tested. If stop-A in the figure is kept in place and stop-B is removed then only the camera optics is tested. If stop-B is kept in place and stop-A is removed then the complete system is checked for image performance. The resulting image quality in either of the cases is  observed at the output of the camera optics and hence the system is mostly self-sufficient. The steering mirror as shown in figure \ref{fig:Lab_test_2} can be used to test the system for some amount of field of view. It should be also be possible to switch the simple steering mirror to a functional FSM and test for static characteristics such as stroke and precision of beam steering. It is of note that these test setups are realized mostly with "generic" flat mirrors and do not require external calibration components such as reference spheres. This is again one of the benefits of an afocal telescope towards reducing time and cost of their realization.

\section{Concluding remarks:}
We have presented optical designs based on afocal telescopes in the context of Cubesat based astronomical observations. The practical advantages of these systems such as standardisation and aberration free beam steering are discussed. It is expected that observatory templates based on afocal designs will result in quick turn-around time and a reduction in development cost and efforts. A few science cases that can be accessed from such Cubesat based platforms are also illustrated.\\

Of course, the goal of a standardised and accessible Cubesat platform also requires coordinated effort from other areas such as mechanical, electronic, communications as well as system design. We hope that our work paves way towards the ultimate aim  of extensive utilisation of Cubesat based platforms by the astronomical community.

\section{Acknowledgments:}
 Work at IIA is funded by Department of Science and Technology, Government of India.

\section*{Appendix-A : prescription table for guiding camera}

\begin{table}
%% use tabular font for a smaller size font
\tabularfont

\caption{Prescription data for the guiding camera, all glass types are from Schott catalog. }\label{prescriptionTable} %%10/12
\renewcommand{\arraystretch}{1.25}

\begin{tabular}{lcccc}
\topline
Element & Glass & RoC & Thickness&Semidia\\
 & &  (mm) &  (mm)& (mm) \\ \midline

Primary & mirror & -200.0  & -75&50\\
Secondary &mirror & -66.6048 & -45&13.1 \\
Lens-1 &Fluorine& 90.0909 & 5& 9 \\
 & crown& -111.3851&15 & 9 \\
Lens-2 & Boron &35.6488 &5  &6 \\
 & crown &10.4333 & 3 & 5\\
Lens-3 & Dense &-18.8786 &5 &4  \\
 & flint &-21.0445 &25 &5  \\
\hline
\end{tabular}
%%use \tablenotes{footnote} to get the table foot note
% \tablenotes{Sample table footnote}%%9/11
\end{table}
Specifications of the guiding camera is attached (design in figure \ref{fig:design_lens_guider}). The primary and secondary mirror have a conic constant of -1.0 and -2.5 respectively. All lenses are spherical. This design serves as a common guiding camera optimised in the visible wavelength range. The design has a FOV of 40 arcminutes and is matched to the commonly available Raspberry Pi HQ camera.

\balance

\bibliography{references}
% \begin{theunbibliography}{}

% \end{theunbibliography}

\end{document}